\documentclass{aa}

\usepackage{graphicx}
\usepackage[varg]{txfonts}

\usepackage{natbib}
\bibpunct{(}{)}{;}{a}{}{,} 
\usepackage[colorlinks=true,linkcolor=blue,citecolor=blue]{hyperref}%
\usepackage{orcidlink}

\usepackage{multirow}

\newcommand{\jotwelve}{\textrm{J}1240\mbox{--}2309}
\newcommand{\Msun}{\hbox{$\rm\thinspace M_{\odot}$}}
\newcommand{\ecgs}{erg~cm$^{-2}$~s$^{-1}$}
\newcommand{\ecgsA}{erg~cm$^{-2}$~s$^{-1}$~\AA$^{-1}$}

\newcommand{\lboledd}{$L_{\rm Bol}/L_{\rm Edd}$}
\newcommand{\mdotedd}{$\dot{m}_{\rm Edd}$}
\newcommand{\mdotEdd}{$\dot{m}_{\rm Edd}$}

\begin{document}

   \title{A changing-look Seyfert discovered by eROSITA reveals a two-component broad-line region} 

\author{
Alex Markowitz\orcidlink{0000-0002-2173-0673},\inst{1}
Mirko Krumpe,\inst{2}  
David Homan\orcidlink{0000-0002-3243-874X},\inst{2,3}
Bo\.zena Czerny\orcidlink{0000-0001-5848-4333},\inst{4}
Mariusz Gromazdki\orcidlink{0000-0002-1650-1518},\inst{5}
Hartmut Winkler\orcidlink{0000-0003-2662-0526},\inst{6} 
Joern Wilms\orcidlink{0000-0003-2065-5410},\inst{7}
Steven H\"ammerich\orcidlink{0000-0002-1113-0041},\inst{7}
Georg Lamer,\inst{2}
Tathagata Saha\orcidlink{0000-0001-5647-3366},\inst{1,8}
David A.H.\ Buckley\orcidlink{0000-0002-7004-9956},\inst{9,10,11}
Malte Schramm,\inst{12}
Daniel E.\ Reichart,\inst{13}
Mara Salvato,\inst{14}
Pietro Baldini\inst{14}
}

\institute{
$^1$Nicolaus Copernicus Astronomical Center, Polish Academy of Sciences, ul.\ Bartycka 18, 00-716 Warsaw, Poland\\
$^2$Leibniz-Institut f\"ur Astrophysik Potsdam (AIP), An der Sternwarte 16, 14482 Potsdam, Germany \\
$^3$Institute of Astronomy, University of Cambridge, Madingley Road, Cambridge, CB3 0HA, United Kingdom \\
$^4$Center for Theoretical Physics, Polish Academy of Sciences, Al.\ Lotnik\'ow 32/46, 02-668, Warsaw, Poland\\
$^5$Astronomical Observatory, University of Warsaw, Al.\ Ujazdowskie 4, 00-478 Warsaw, Poland \\
$^6$Department of Physics, University of Johannesburg, PO Box 524, Auckland Park 2006, South Africa \\
$^7$Dr.\ Karl Remeis-Observatory and Erlangen Centre for Astroparticle Physics, Friedrich-Alexander Universit\"at Erlangen-N\"urnberg, Sternwartstr.\ 7, 96049 Bamberg, Germany \\
$^{8}$Inter-University Centre for Astronomy and Astrophysics, Post Bag 4, Ganeshkhind, Pune University Campus, Pune 411007, India\\
$^{9}$South African Astronomical Observatory, PO Box 9, Observatory, Cape Town 7935, South Africa\\
$^{10}$Department of Astronomy, University of Cape Town, Private Bag X3, Rondebosch 7701, South Africa\\
$^{11}$Department of Physics, University of the Free State, PO Box 339, Bloemfontein 9300, South Africa\\
$^{12}$Universit\"at Potsdam, Karl-Liebknecht-Str. 24/25, 14476 Potsdam, Germany \\
$^{13}$Department of Physics and Astronomy, University of North Carolina at Chapel Hill, Campus Box 3255, Chapel Hill, NC 27599-3255, USA\\
$^{14}$Max-Planck-Institut f\"ur Extraterrestrische Physik, Giessenbachstr.\ 1, 85748 Garching, Germany\\
             }

\date{Received December 31, 2025; accepted April 14, 2026 }

\abstract 
{Extreme sudden changes in the flow of accreting gas onto
  supermassive black holes manifest themselves via large-amplitude
  multiband continuum variability, as well as changes to broad Balmer
  emission profiles, driving changing-look active galactic nuclei (AGN). }
{X-ray flux monitoring with Spectrum Roentgen Gamma (SRG)/eROSITA
  revealed that in the Seyfert AGN HE~1237$-$2252 the soft X-ray flux
  dipped abruptly, by a factor of 17 within 18 months. We initiated a
  follow-up campaign that caught the luminosity recovery after the dip,
  and enabled us to study how the various accretion components,
  including the broad-line region (BLR)  and X-ray-emitting coronae, responded during this
  flux recovery.}
{Our campaign included multiband photometry, X-ray spectroscopy, and
  optical spectroscopy. We tracked as the accretion rate relative to
  Eddington increased by a factor of 7 in 3 years.}
{Based on broad H$\beta$ variability, HE~1237$-$2252 was subtype
  1.0--1.2 in 2002, transitioned to subtype 1.8 by the time of the
  luminosity dip, and then transitioned back to subtype 1.0 within 3
  months as luminosity recovered.  Both transitions saw broad H$\beta$
  integrated line flux change by factors of 4--6.  The broad Balmer
  profile is decomposed into a broad Gaussian consistent with
  virialized gas at 27$\pm$3 lt-dy, plus a double-peaked profile,
  consistent with a diskline structure at $\gtrsim$5 lt-dy. The
  diskline component's relative contribution to the total profile
  increases as continuum flux rises.}
{The lack of significant obscuration in the X-ray spectra, as well
    as the IR continuum dip, point to an intrinsic pause in the
    accretion rate as opposed to variable line-of-sight obscuration.
    Candidates for the underlying mechanisms include propagating cold
    and warm fronts in the accretion disk.  The increased prominence
    of the diskline BLR component's emission could be due to evolution
    in the physical extent of the X-ray corona, and in the fraction of
    $>$13.6~eV photons intercepted by the diskline, as the accretion rate
    increases.}
   \keywords{galaxies: active -- galaxies: Seyfert -- X-rays: galaxies -- individual objects: HE~1237$-$2252}

\authorrunning{A.\ Markowitz et al.}
\titlerunning{Two-component BLR in CLAGN \jotwelve}
   \maketitle

\nolinenumbers


\section{Introduction} 

Active galactic nuclei (AGN) are powered by accretion of matter onto a
supermassive black hole \citep{Soltan82} usually through an accretion
disk that feeds the black hole.  Accretion is stochastically variable
on a wide range of timescales.  Accretion episodes can be
intermittent, with the global supply of gas supplied by the host
galaxy turning on and off, likely on timescales of the order of $10^4-10^5$ years
\citep{Schawinski15,Shen21}.  In addition, during episodes of
persistent accretion, AGN emission varies stochastically on timescales
from hours to decades across the electromagnetic spectrum
\citep[e.g.,][]{Mushotzky93}.

In recent years, the community has been accumulating evidence for
events where AGN undergo major changes in their accretion rates on
timescales ranging from months to several years. Specifically, such
events are thought to be a driver behind observations of changes in a
given source's optical spectral type, changing-look AGN (CLAGN; 
e.g., \citet{Tohline76},
\citealt{Shappee14}, \citealt{Denney14}, and \citealt{LaMassa15}, to list
just a few early examples).

Broadly, Seyfert and quasar AGN are classified as type 1 or type 2
based on their optical line emission
properties \citep{Khachikian74}. Both types exhibit narrow (several
hundred km s$^{-1}$) emission lines emanating from the narrow-line
region (NLR), but only type 1 galaxies additionally exhibit Doppler-broadened
(several thousand km s$^{-1}$) emission lines emanating from the broad-line region (BLR).  Intermediate classification types (1.2, 1.5, 1.8,
1.9) can be assigned depending on the strength of the broad H$\beta$
component relative to the narrow H$\beta$ component
\citep{Osterbrock76,Osterbrock77,Osterbrock81,Cohen83,Runco16} or to [\ion{O}{iii}]~$\lambda$5007
\citep{Winkler92},
and  on the relative intensities of broad H$\beta$ and broad H$\alpha$. For example,
in type 1.8, broad H$\alpha$ is detected well, while broad H$\beta$ is detected but weakly; 
in type 1.9 broad H$\beta$ is not detected, only broad H$\alpha$,  
while in type 2 galaxies neither H$\alpha$ nor H$\beta$ display broad components.

For a given CLAGN, observations of optical spectral type changes are
often associated with observations of major concurrent changes in
continuum flux \citep[e.g.,][]{Denney14,McElroy16}.  Major changes in
accretion rate are a natural explanation: the ionizing luminosity
responds accordingly, driving changes in the illumination and
ionization in the BLR \citep[e.g.,][]{Korista00,Wu23}, and the optical
spectral type thus evolves.
However, the nature of these major changes in accretion rate is
unclear; they could be due to changes in the rate of global
accretion onto the
disk, for example cold chaotic accretion \citep{Gaspari13}. 
Alternately, they could be associated with locally operating instabilities in
the accretion flow, such as a radiation pressure-induced instability
\citep[e.g.,][]{Lightman74,Saxton15,Sniegowska23},
or magnetically driven instabilities \citep{Stern18,Scepi21}. CLAGN events can
also be triggered by stream-like accretion from tidal disruption
events (TDEs) occurring in AGN, \citep[e.g.,][]{Homan23}, including
cases where a TDE-like stream is suspected of directly impacting the
disk and corona \citep{Ricci20}.\footnote{In this paper, we do not
discuss optical type change attributed to transit of a dusty cloud or
wind across the line of sight (e.g.,
\citealt{Goodrich95,Wang09,Gaskell18,Zeltyn22}); such events have been
observed only very rarely.}

Another open question is whether the BLR is a static reservoir of gas,
passively getting illuminated by the disk and/or the corona, and then
reradiating line emission. Alternatively, the BLR may be a wind
outflowing from the disk, and it forms or dissipates when the
accretion rate relative to Eddington, {\mdotedd} $\equiv$ {\lboledd},
transitions across a certain critical value \citep{Elitzur09}. Recent
results for CL events support this notion, as CL quasars tend to
accrete near such a critical value of accretion
rate \citep{Green22,Panda24}.  Meanwhile, samples of (non-CL)
Seyferts and quasars suggest that type 2 galaxies are  more prominent
toward lower luminosities, possibly due to a decrease in the covering
fraction of BLR clouds toward lower luminosities \citep{Elitzur14}.
Alternately, the BLR in type 2 galaxies may become more obscured
toward lower luminosities due to an increase in
the covering fraction of the obscuring torus \citep{Ricci17}.  

Changing-look AGN events occur rarely on a per-object basis, so monitoring
a starting sample of AGN and quasars that is as large as possible is key for the
detection of new CL events.  Well over 150 CLAGN have been detected from
repeat optical spectroscopy, for example the Time Domain Spectroscopic Survey within the Sloan Digital Sky Survey (SDSS)
\citep{Runnoe16,Green22,Zeltyn24},
and with additional forthcoming detections 
expected from large spectroscopic surveys such as the Dark Energy Spectroscopic Instrument (DESI) survey.
Dozens more events have been
identified via photometric monitoring using large-area
(observed-frame) optical surveys such as the Zwicky Transient Facility, Catalina Real-time Transient Survey, and PanSTARRS
\citep{Yang18,MacLeod19,Ross20,Frederick21,LopezNavas22,Wang24}; 
mid-IR monitoring
\citep{Sheng20}; or combinations of wavelengths \citep{LopezNavas23,Yang25}.   
Here one looks for large rapid changes in flux 
in excess of the amplitudes of variability associated with persistent accretion
\citep[e.g.,][]{Graham20} or sudden strong deviations from the
standard variability behavior associated with non-CLAGN, for example
CLAGN straying from a damped random walk behavior
\citep{Suberlak21,SanchezSaez21}.   In all cases, though,
multiwavelength follow-ups while transitions are occurring
are key to best exploring how accretion flow components --- disk, X-ray
corona, BLR --- interact with each other during such major changes in
accretion and luminosity.

In this paper we report the detection of a CLAGN using the X-ray band,
specifically using the Extended ROentgen Survey with an Imaging
Telescope Array \citep[eROSITA;][]{Predehl21}, the soft X-ray
telescope  on board the \textit{Spectrum Roentgen/Gamma} (\textit{SRG})
spacecraft \citep{Sunyaev21}. From Decembeer  2019 to February 2022, eROSITA's all-sky surveys enabled X-ray
monitoring on the order of a million AGN and quasars, once every six months, thus
amplifying small numbers of rare transient AGN events, and providing
the first X-ray-based channel for the identification of major changes in
luminosity.
Our team monitored AGN for major changes in soft X-ray flux,
and in January 2022, we identified a Sy 1.0--1.2 (type based on archival
spectra) whose soft X-ray flux had dropped by a factor of roughly 17
in 1.5 yr.  We thus triggered target-of-opportunity observations
encompassing X-ray spectroscopy and photometry, space-based UV/optical
photometry, and ground-based optical photometry and spectroscopy,
spanning from late 2022 through early 2025.  As demonstrated below, by early
2022 the optical spectrum had transitioned into a subtype 1.8. Over
the next year, the X-ray continuum recovered and the optical/UV continuum
rose steadily; concurrently, the optical spectrum had transitioned
back into a type 1 by June--August 2022.  Then, during
2023--2024, the broad Balmer line profile evolved from being dominated by a single broad
Gaussian-like component into a Gaussian plus a double-peaked
structure.

The remainder of this paper is organized as follows. 
Section~2 describes the event detection and gives an overview of multiwavelength campaign.
Section~3 gives an overview of the multiwavelength continuum variability.
Sections~4, 5, and 6 describe, respectively, the X-ray spectral fits,
the broadband spectral energy distribution (SED) modeling, and the fits to optical emission-line spectra.
The results are discussed in Sect.~7,
and we provide our summary and conclusions in Sect.~8.
 
Throughout this paper, we assume the cosmological parameters of $H_0 = 70$
km s$^{-1}$ Mpc$^{-1}$, $\Omega_{\rm M} = 0.29$, and $\Omega_{\rm vac}                                                                                                    
= 0.71$.  The redshift of $\jotwelve$ thus corresponds to a luminosity
distance of 443~Mpc and a co-moving (proper) distance of 404~Mpc,
using Ned Wright's Cosmology Calculator
\citep{Wright06}.\footnote{\url{http://www.astro.ucla.edu/~wright/CosmoCalc.html}}

\section{Counterpart, follow-up observations, and data reduction}     \label{sec:section2}

\subsection{Detection of the event in eROSITA scans}

The position of the event was scanned in each of the five eROSITA
All-Sky Surveys \citep[eRASS;][]{Merloni24}, hereafter referred to as
eRASS1 to eRASS5.  Every six months (every eRASS), a given position
near the orbital equator is scanned six times at a four-hour cadence,
each with a roughly 40 second exposure. These exposures are added to
yield one flux point per eRASS.  After eRASS5 scanned the position of
the point source in question, we detected an extreme flux drop from
eRASS2 to eRASS5.  We identified the event with the eROSITA DR1
\citep{Merloni24} source 1eRASS J124028.2$-$230925.
The error of the DR1 position in each coordinate, as given in
catalogue column POS\_ERR, is $1{\farcs}2$.\footnote{Includes
  statistical and systematic uncertainties  
corresponding to the 1$\sigma$ confidence interval
given the R.A.\ and
  Dec.\ for this source.}  
We cross-matched with known AGN and quasar catalogs \citep{Salvato25} to
identify the counterpart.  The closest counterpart is the Seyfert HE
1237$-$2252, identified in the Hamburg/ESO survey by
\citet{Reimers96}, and located $1{\farcs}6$   from the X-ray position,\footnote{After
  $\jotwelve$, the next closest counterpart candidates listed in NED
  are all $\geq29\arcsec$ away.} at J2000 coordinates $\alpha = 12^{\rm h} 40^{\rm m}
28{\fs}32$, $\delta= -23^{\circ} 09\arcmin 26{\farcs}7$.  It is also
cross-listed as the IR source WISEA~J124028.32$-$230926.7 and
6dF~J1240283$-$230927; the event is thus referred to hereafter as
$\jotwelve$.\footnote{While the offset 
   between the X-ray position and
   position of this counterpart is slightly larger than the nominal
   X-ray position center, counterpart identification is not
   affected. In addition, as demonstrated below, X-ray fluxes determined from
   times outside the flux dip -- eRASS1 and eRASS2 in 2019--2020, an
   \textit{XMM-Newton} slew in 2013, and our \textit{Swift} and
   \textit{XMM-Newton} points from the end of our campaign -- are all
   in reasonable agreement with each other.}
Its redshift $z$ is 0.09643$\pm$0.00015 (6dF Galaxy
Survey; \citealt{Jones09}; see also \citealt{Reimers96}).

We conducted spectral fits for each individual eRASS scan,
as described in Sect.~\ref{sec:eR_XRT_NICER_spec}. We
determined that $\jotwelve$'s 0.5--2.0~keV flux had decreased by a
factor of $17.0^{+16.3}_{-7.2}$ over 18 months, from
$9.32^{+1.01}_{-1.26} \times10^{-13}$ \ecgs in eRASS2 (June 2020) to
$5.5^{+2.7}_{-2.4} \times10^{-14}$ \ecgs in eRASS5 (January 2022).
The start and stop times of the scans that included the position of
$\jotwelve$ are listed in Table~\ref{tab:Xobs}.  For brevity, all
details of eROSITA data reduction, including source
and background spectral extraction, and soft X-ray flux measurements
are detailed in Appendix~\ref{sec:appdx_er}.  Good exposure times
after correcting for vignetting effects were in the range 230--280~s,
as listed in Table~\ref{tab:Xobs}.

\subsection{Multiwavelength campaign overview}

Following the identification in January 2022 of the major change in soft X-ray flux in $\jotwelve$, 
we triggered multiple target-of-opportunity observations as follows:

We obtained a high signal-to-noise X-ray spectrum with the
\textit{XMM-Newton} \citep{Jansen01} European Photon Imaging Cameras
\citep[EPIC;][]{Strueder01} in January 2022 (hereafter XM1).
\textit{XMM-Newton}'s Optical Monitor \citep[OM;][]{Mason01} also
provided a concurrent set of optical/UV photometry points.

In February 2022, the eRASS scans were suspended, so we supplemented the
X-ray flux monitoring with \textit{Neil Gehrels Swift Observatory}
\citep[\textit{Swift};][]{Gehrels04} X-Ray Telescope
\citep[XRT;][]{Burrows05} monitoring, consisting of 15 pointings
between February 2022 and August 2024, hereafter referred to as Sw1--15.
These observations also provided optical and UV photometric monitoring
courtesy of \textit{Swift}'s UltraViolet/Optical Telescope (UVOT).
We also included X-ray fluxes from a 2013 \textit{XMM-Newton} slew, 
obtained via the \textit{XMM-Newton} Upper Limit
Server.\footnote{\url{http://xmmuls.esac.esa.int/upperlimitserver/}}

Meanwhile, we obtained B-, V-, R-, and I-band photometry at ground-based
facilities operated by the Las Cumbres Observatory global telescope
\citep[LCOGT;][]{Brown13} network, with nine observations obtained
between 1 March 2022 and 24 May 2023 (LCO1--9), as listed in
Table~\ref{tab:OUVobs}.  We also obtained 13 B-, V-, and/or R-band
photometric observations of $\jotwelve$ between 7 February 2022 and 7 May
2022 using the 0.4-meter PROMPT6 telescope at Cerro Tololo Inter-American
Observatory, operated as part of the Skynet Robotic Telescope
Network. They are also listed in Table~\ref{tab:OUVobs}.

The dates, ObsIDs, and exposures for all X-ray observations are given in
Table~\ref{tab:Xobs}.  The dates, filters, and exposures for all
optical/UV photometric observations are given in
Table~\ref{tab:OUVobs}.

To track the behavior of the Balmer emission line profiles, we
conducted optical spectroscopic monitoring, using the South African
Large Telescope (SALT) longslit Robert Stobie Spectrograph
\citep[RSS;][]{Burgh03, Kobulnicky03}, the FORS2 spectrograph
\citep{Appenzeller98} on the 8.2~m Very Large Telescope Array’s (VLT)
UT1 at Cerro Paranal, and the SpUpNIC spectrograph \citep{Crause19} at
the South African Astronomical Observatory (SAAO) 1.9~m telescope.  We
obtained a total of 19 new observations of $\jotwelve$ between
February 2022 and December 2024, hereafter referred to as spectra~\#3--21.
Spectra~\#1 and {\#}2 refer to archival spectra taken in 1993 at the
ESO 3.6~m telescope as part of the Hamburg/ESO (H/ESO) survey
\citep{Reimers96} and in 2002 at the UK Schmidt Telescope as part of
6dFGS \citep{Jones09}, respectively.  The dates, instruments, and
exposures for all spectra are listed in Table~\ref{tab:optspeclog}.

Throughout 2022, both X-ray and optical/UV fluxes rose quasi-steadily.
By December 2022, $F_{0.5-2.0}$ had risen back to $11.8^{+1.3}_{-1.1}
\times 10^{-13}$ \ecgs; the far-UV flux (UVW1, UVM2, UVW2 filters) had
concurrently increased by factors spanning 1.5--2.0 compared to
fluxes in January 2022.  We thus triggered a second \textit{XMM-Newton}
observation, which was executed in January 2023 (XM2), followed by
additional \textit{Swift} pointings through August 2024, as well as 28
visits with the Neutron Star Interior Composition Explorer
\citep[NICER;][]{Gendreau16} aboard the International Space Station,
between 15 February 2023 and 23 May 2023. We conducted a third
\textit{XMM-Newton} observation in January 2025 (XM3).

Finally, we supplemented these data with publicly available optical
and infrared photometry. We used data taken with the Asteroid
Terrestrial impact Last Alert System (ATLAS) during November 2017 --
December 2023 and with the \textit{Widefield Infrared Survey Explorer}
(\textit{WISE})/\textit{NEOWISE} during January 2012 -- June 2023,
respectively.

For brevity, further details of observations (e.g., the gratings used for
optical spectroscopy), as well as all details of data reduction,
calibration, and extraction of all X-ray spectra and photometric
fluxes can be found in the Appendices, and the resulting X-ray fluxes and
optical, UV, and IR magnitudes are given in tabular form in the
Appendices as follows: \ref{sec:appdx_er}: eROSITA;
\ref{sec:appdx_xmm_epic}: \textit{XMM-Newton} EPIC;
\ref{sec:appdx_sw_xrt}: \textit{Swift} XRT;
\ref{sec:appdx_nicer}: NICER;
\ref{sec:appdx_xmm_om}: \textit{XMM-Newton} OM;
\ref{sec:appdx_sw_uvot}:  \textit{Swift} UVOT;
\ref{sec:appdx_lcogt}: LCOGT;  
\ref{sec:appdx_prompt}: PROMPT-6;
\ref{sec:appdx_atlaswise}: ATLAS and \textit{WISE/NEOWISE}; and
\ref{sec:appdx_optspec}: SALT, VLT, \& SAAO 1.9m optical spectra.  

\section{Multiband continuum variability overview} \label{sec:VarOverview}

\begin{figure*}\sidecaption
\includegraphics[width=12cm]{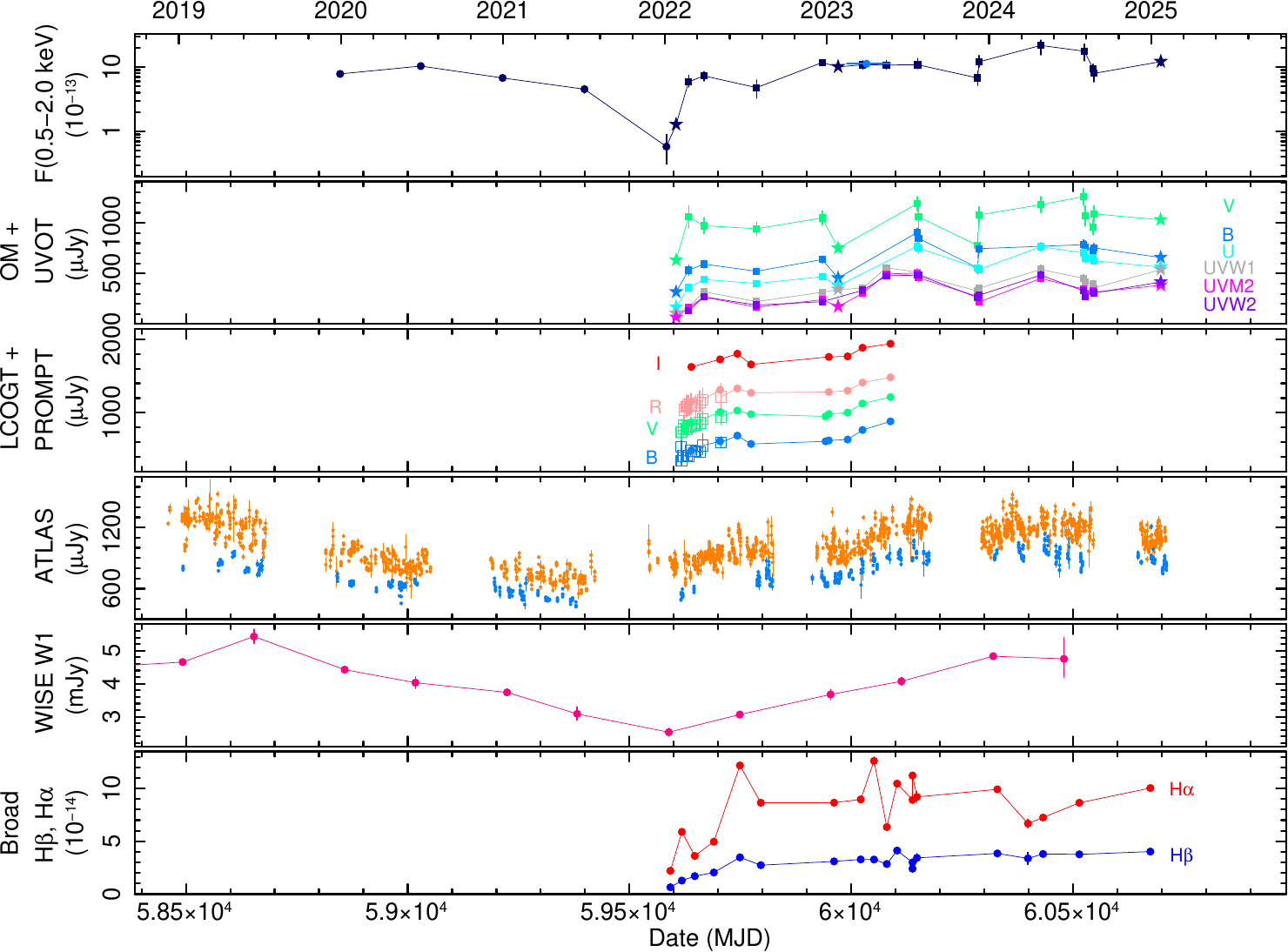}
\caption{  
Multiband continuum and broad Balmer line flux light curves.  In the
top panel, the black circles, stars, and squares denote X-ray fluxes from
eRASS, \textit{XMM-Newton}, and \textit{Swift}, respectively; the blue
point denotes the summed NICER data.  In the second panel, the stars and
squares denotes optical/UV flux densities from \textit{XMM-Newton} and
\textit{Swift}, respectively.  The third and fourth panels show
ground-based optical photometry from LCO+PROMPT and ATLAS,
respectively; orange and cyan  respectively denote  the ATLAS $o$ and
$c$ bands.  The fifth panel shows the \textit{NEOWISE} W1 band
photometry.  For the UV, optical, and IR fluxes plotted here, no host
galaxy subtraction has been done; differences in apertures and seeing
among the various datasets have resulted in different levels of host
contribution.  The last panel shows the broad Balmer line fluxes (see
$\S$\ref{sec:optspec}).  Across all panels, all data are plotted with
error bars, but some error bars are smaller than the data symbols.  }
\label{fig:ALL_LC_ZOOM}
\end{figure*}

The IR flux remained relatively steady from 2010 until 2019
(Fig.~\ref{fig:wise_lc_forappdx}), indicating a relatively stable
bolometric output during this period.  Meanwhile, the archival
0.5--2~keV flux from the 2013 \textit{XMM-Newton} slew was
$(1.53\pm0.44)\times10^{-12}$~erg~cm$^{-2}$~s$^{-1}$, with fluxes in
2019--2020 (eR1, eR2; discussed in Sect.~\ref{sec:eR_XRT_NICER_spec})
only a factor of 1.6 lower.\footnote{The XMM-ULS provided the 0.2--2 keV
  flux for this observation; we used the best-fitting model to XM3
  (Sect.~\ref{Xspec_xmm}) to infer its 0.5--2 keV flux.}

However, as shown in Fig.~\ref{fig:ALL_LC_ZOOM}
and Fig.~\ref{fig:wise_lc_forappdx}, from mid-2019 until
late 2021, the IR flux dropped by roughly 0.7 mag in both W1 and W2
bands. This change is similar to the average change in magnitudes
(${\Delta}$W1 = 0.65; ${\Delta}$W2 = 0.88) displayed by the ten
optically confirmed CLAGN with strong mid-IR variability identified by
\citet{Sheng17}.  Meanwhile, as illustrated in
Fig.~\ref{fig:ALL_LC_ZOOM}, the 0.5--2~keV X-ray flux decreased
steadily from 2020 through early 2022, when we triggered the
optical/UV photometric monitoring.

After early 2022, all bands increased steadily through late 2022:
$F_{0.5-2.0}$, far-UV (UVW1 and UVM2) flux, and IR (W1) flux 
saw increases by factors of roughly 26, 2.5--3.0, and 1.5 from
January 2022 through December 2022.
Then, starting in early 2023, the X-ray flux became rather stable, and varied only
minimally for the next two years; values of $F_{0.5-2.0}$
have varied by 37 percent  (fractional variability amplitude), and have
typically remained in the range 0.8--2.2 $\times10^{-12}$ \ecgs, consistent with
the fluxes measured in 2013 and 2019--2020.
Since mid-2023, the far-UV fluxes  have also  become overall stable, varying by
only 23--28$\%$.
Meanwhile, from 2020 through mid-2024, IR
flux recovered by roughly 0.4--0.5 mag in both bands.  Throughout
the flux drop and recovery, the W1$-$W2 color remained steady at 0.7--0.8
(Fig.~\ref{fig:wise_lc_forappdx}),
suggesting that AGN activity has remained persistent, following
\citet{Stern12} and \citet{Assef18}.
Importantly, the dip in the \textit{WISE/NEOWISE} light curve effectively
excludes that the variability in the UV/X-ray bands can be attributed to
obscuration by a cloud along the line of sight.

We performed cross-correlations on selected pairs of light curves,
using the interpolated correlation function \citep[ICF;][]{White94}
with bootstrap errors determined by 
flux randomization \citep{Peterson98}, due to the limited
number of data points.
All pairs of data points are generally well correlated at zero lag,
with zero-lag correlation coefficients of 
0.67--0.78 for $F_{0.5-2.0}$ to the IR bands;
0.67--0.79 for $F_{0.5-2.0}$ to the U, UVW1, UVM2, and UVW2 bands;
0.85--0.97 for comparisons of U, UVW1, UVM2, and UVW2 to each other;
and 0.64--0.88 for U, UVW1, UVM2, and UVW2 to the IR bands.
The data do not yield any evidence of lags or leads.
For instance, we find an upper limit
of $\pm$360 days for any lag from $F_{0.5-2.0}$ to \textit{WISE/NEOWISE} W1.
Similarly, the best-fitting values for the \textit{WISE/NEOWISE} bands
lagging the ATLAS bands spanned +150--240~d, but with uncertainties of
250--330~d, so we cannot comment on any dust reprocessing activity.

\section{X-ray spectral fits} \label{sec:Xspecfits}

Our spectral fitting strategy was to start by modeling the
\textit{XMM-Newton} EPIC spectra as they had the highest
S/N.  We then modeled the lower S/N eRASS
and \textit{Swift} XRT spectra.

All the X-ray spectral fits were done
using \textsc{Xspec} \citep{Arnaud96} version 12.13.0c.  All parameter
uncertainties are for one interesting parameter, and were derived via
Monte Carlo Markov chains (MCMC) using the \textsc{chain} routine
in \textsc{Xspec}.  We used the Goodman-Weare MCMC sampler \citep{Goodman10}, chains of length 10000, 20 walkers,
and a burn length of 5000.  Parameter errors are at the $90\%$
confidence level, and are taken from the 5th and 95th percentile values
of the parameter distribution.  In all models, we included a
\textsc{TBabs} component to account for Galactic absorption by
\ion{H}{i} and H$_2$ totaling $7.37 \times 10^{20}$~cm$^{-2}$
\citep{Willingale13}.  We assumed the abundances of \citet{Wilms00}.

\subsection{XMM-Newton EPIC spectral fits} \label{Xspec_xmm}

For each observation, we fit pn0 (pattern 0; 0.25--10 keV) + pn14
(patterns 1--4; 0.5--10 keV) + MOS1 + MOS2 (both 0.2--10 keV)
jointly. We applied instrumental constant components for
cross-calibration purposes, keeping the constant for pn0 fixed at
unity; constants for the other spectra were usually within a few percent of unity for the best-fitting models. All spectra were grouped to 20
counts per bin to ensure the use of $\chi^2$ statistics.

For all three observations, we found excellent fits consisting of the
following baseline model, which is very typical for nearby,
X-ray-unobscured Seyferts:

\begin{itemize}

\item A hard X-ray power law (\textsc{zpowerlw}) 
to model emission from the hot, optically thin corona.

\item A soft X-ray excess, which we modeled as warm Comptonization of
optical/UV thermal photons, following   \citet{Mehdipour11},
\citet{Porquet18}, and \citet{Petrucci18}, among others. We used \textsc{CompTT}   
\citep{Titarchuk94}, assumed the sphere geometry, and
fixed the seed photon temperature $T_{\rm seed}$ to 20 eV.

\item A narrow Fe K$\alpha$ line is detected in XM1 only: a simple preliminary
fit consisting of a narrow (width $\sigma$ fixed to 10 eV) Gaussian
component plus a power-law fit to 2--10 keV yields a line intensity
$I_{\rm Fe} = (8.7 \pm 5.6) \times10^{-7}$ ph cm$^{-2}$ s$^{-1}$
keV$^{-1}$ and equivalent width relative to a hard power law $EW = 220
\pm 142$ eV; a set of 500 Monte Carlo simulations using
\textsc{simftest} indicates that the line is detected at the 99.4$\%$
confidence level.

In our final model, we therefore modeled a narrow Fe~K$\alpha$ line as
well as a Compton reflection hump, produced by reflection from
distant, neutral material lying out of the line of
sight. Specifically, we used \textsc{UxClumpy} \citep{Buchner19},
which assumes a clumpy medium.  Given the lack of $>$10 keV data, we
left the following parameters frozen to arbitrary values: 
the cloud angular Gaussian
distribution $\sigma_{\rm TOR}$ set to $30^{\circ}$, cloud column
density $N_{\rm H}$ set to $1.0\times10^{24}$~cm$^{-2}$, and
Compton-thick inner ring covering fraction to 0.
In addition, we froze the system inclination at $15^{\circ}$, 
based on the results from fitting a diskline component
to the broad Balmer profiles (see Sect.~\ref{sec:optspec}).

For XM2--3, the narrow Fe K$\alpha$ line was not significantly
detected, with $I_{\rm Fe} < 3.4~(1.5) \times10^{-6}$ ph cm$^{-2}$
s$^{-1}$ and $EW < 178$ (80)~eV in XM2 (XM3).  In our final model, we
thus included an \textsc{UxClumpy} component, but with its
normalization frozen at the best-fitting value from XM1, under the
assumption that the hard X-ray power-law component has varied between
XM1 and XM2, but the distant material has not responded on timescales
of 1--3 years.

\end{itemize}

\begin{figure*}
 \includegraphics[width=1.98\columnwidth]{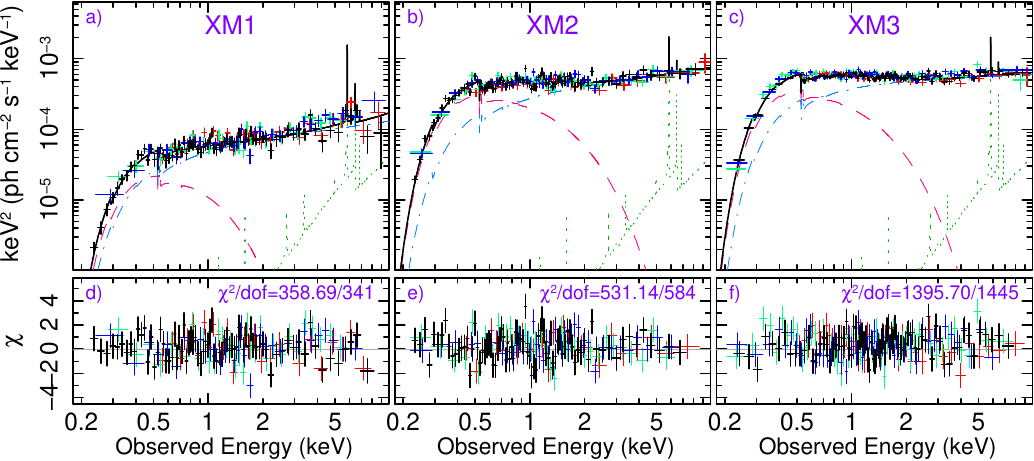}
  \caption{Spectral data and fits for XM1--3.
  Black, red, green, and blue denote pn0, pn14, MOS1, and
  MOS2, respectively.  The data have been rebinned by factors of 2, 3, and 6 in XM1, XM2, and XM3, respectively, for
  clarity. 
  Panels a--c display the best-fitting unfolded models and data.
  The \textsc{CompTT}, hard X-ray power-law, and \textsc{UxClumpy} components are   respectively denoted by
  red dashed, blue dash-dotted, and green dotted lines, and the 
  total model is denoted by a black solid line.  
  The corresponding $\chi$ residuals are plotted in panels d--f. } 
  \label{fig:XMMresids}
\end{figure*}   

We obtained excellent fits to all three observations, with
the best-fitting parameters listed in Table~\ref{tab:XMparms}. The spectral
data, best-fitting models, and data--model residuals are plotted in
Fig.~\ref{fig:XMMresids}.  From XM1 to XM2, the hard X-ray power law
and the soft X-ray excess increase in flux by factors of
$5.5^{+1.1}_{-0.7}$ and $14.1^{+10.3}_{-5.4}$, respectively. Then,
from XM2 to XM3, the soft X-ray excess increases by an additional
factor of $1.3\pm0.2$, while the hard X-ray power-law flux values are
consistent with each other.  Values for other spectral parameters,
namely $T_{\rm e}$ and $\tau$ in \textsc{CompTT} and $\Gamma_{\rm
HX}$, are consistent across the three observations.

As an alternate parameterization of the soft X-ray excess, we tested
the reflection off an ionized relativistic accretion disk.  We removed
the \textsc{CompTT} component and replaced it with \textsc{relxill}
\citep{Garcia14, Dauser14}, using version v.1.4.3, keeping the outer
radius fixed at $400~R_{\rm g}$, and the power-law cutoff energy fixed at
100~keV (though the fit was insensitive to thawing these
parameters). We kept the inner radius, black hole spin parameter, disk
inclination, disk ionization parameter, Fe abundance relative to the
solar value, and power-law emissivity index all free.  We also
included a narrow Gaussian component (width $\sigma$ fixed to 10~eV;
energy fixed to 6.4~keV) to model any narrow Fe~K$\alpha$ line
emission.  However, the best-fitting models have values of
$\chi^2/dof$ that are worse by 20.8, 30.0, and 473.5 for five more degrees of freedom ($dof$) 
for XM1, 2, and 3, respectively, compared to the best-fitting models
using \textsc{CompTT},\footnote{Furthermore, the Akaike information
criterion \citep[AIC;][]{Akaike73} with finite sample correction
by \citet{Sugiura78} yields that ${\Delta}$AIC going
from \textsc{CompTT} to \textsc{relxill} is always positive (+29.3,
+38.3, and +481.6, respectively.)} and with worse broadband data/model 
residuals. We thus did not consider this model further.

We also tested for the presence of line-of-sight obscuration. We tested
full-covering neutral obscuration modeled with \mbox{\textsc{zTBabs}},
partial-covering, neutral obscuration modeled with \textsc{TBpcf}, and
ionized obscuration modeled with \textsc{zxipcf}, with values of 
log($\xi$, erg cm s$^{-1}$) set to +1. For all three \textit{XMM-Newton}
observations, we found no improvement to our model in either
observation when adding various absorption components; AIC increases,
and upper limits on $N_{\rm H}$ were always less than a few times
$10^{21}$ cm$^{-2}$.

\renewcommand{\arraystretch}{1.18}
\begin{table*}
\caption[]{    Best-fitting models to \textit{XMM-Newton} EPIC spectra} 
        \centering
\label{tab:XMparms}
\begin{tabular}{lccc} \hline\hline
Component/               & XM1  & XM2 & XM3  \\ 
Parameter                & (MJD 59604.9) & (MJD 59970.2) & (MJD 60697.4) \\ \hline
\multicolumn{4}{c}{Hard X-ray power law}  \\
$\Gamma_{\rm HX}$          & $1.64^{+0.17}_{-0.09}$    & $1.65^{+0.16}_{-0.13}$  & $1.83^{+0.09}_{-0.13}$   \\
$F_{\rm PL,2-10}^{(a)}$ (\ecgs)       & $2.53^{+0.18}_{-0.33}\times10^{-13}$     & $1.40^{+0.06}_{-0.09}\times10^{-12}$  & $1.38^{+0.03}_{-0.07}\times10^{-12}$  \\  \hline
\multicolumn{4}{c}{\textsc{CompTT}}        \\
$T_{\rm seed}$ (eV)             & 20*  & 20* & 20* \\ 
$T_{\rm e}$ (keV)              &  $0.29^{+0.29}_{-0.15}$    &  $0.52^{+0.40}_{-0.18}$ & $0.54^{+0.36}_{-0.12}$ \\
$\tau$                   &   $12^{+9}_{-5}$   &   $9\pm3$  & $8\pm2$ \\
$F_{\rm CompTT,0.2-0.7}^{(b)}$ (\ecgs)    & $8.6^{+2.9}_{-3.2}\times10^{-14}$  &  $1.21^{+0.15}_{-0.21}\times10^{-12}$  & $1.59^{+0.21}_{-0.16}\times10^{-12}$ \\  \hline
\multicolumn{4}{c}{\textsc{UxClumpy}} \\      
Inclination, $i$        &  $15^{\circ}$*    &   $15^{\circ}$*  & $15^{\circ}$* \\
$\sigma_{\rm tor}^{(c)}$    &  $30^{\circ}$*    &   $30^{\circ}$*  &   $30^{\circ}$*  \\
CTK Cov.\ Frac.$^{(d)}$  & 0*  & 0* & 0* \\ 
log($N_{\rm H}$, cm$^{-2}$) &  24.0* & 24.0* & 24.0* \\
Normalization  (ph keV$^{-1}$ cm$^{-2}$~s$^{-1}$)  &   $6.3^{+4.5}_{-2.5} \times 10^{-4}$   &  $6.3 \times 10^{-4}$*  &  $6.3 \times 10^{-4}$*  \\ \hline
Observed $F_{2-10}$ (\ecgs)    & $2.93^{+0.17}_{-0.20} \times10^{-13}$  &  $1.47^{+0.06}_{-0.07}\times10^{-12}$ & $(1.44\pm0.04)\times10^{-12}$ \\
Observed $F_{0.5-2.0}$ (\ecgs) & $(1.28\pm0.04)\times10^{-13}$   &  $1.01^{+0.02}_{-0.03}\times10^{-12}$  &  $1.22^{+0.01}_{-0.08}\times10^{-12}$  \\
$L_{2-10}^{(e)}$ (erg s$^{-1}$) &  $6.0^{+0.4}_{-0.8}\times10^{42}$ & $3.2^{+0.1}_{-0.2}\times10^{43}$ & $3.3^{+0.1}_{-0.2}\times10^{43}$   \\
$\chi^2/dof$                   &  358.69/341  & 531.14/584 & 1395.70/1445 \\  \hline  \hline 
\end{tabular}
\tablefoot{  
An asterisk (*) denotes a fixed parameter.\\
\tablefoottext{a}{2--10~keV flux of the hard power-law component, corrected for Galactic absorption}\\
\tablefoottext{b}{0.2--0.7~keV flux of the \textsc{CompTT} component, corrected for Galactic absorption}\\
\tablefoottext{c}{Denotes the cloud angular Gaussian distribution in \textsc{UxClumpy}}\\
\tablefoottext{d}{Denotes that we fixed the covering fraction of the Compton-thick inner ring in \textsc{UxClumpy} to 0} \\
\tablefoottext{e}{Denotes the 2--10~keV luminosity corresponding to the hard X-ray power-law component.  }
}
\end{table*}

\renewcommand{\arraystretch}{1.18}
\begin{table*}
\caption[]{Best-fitting power-law models to eRASS and \textit{Swift} XRT spectra.} 
        \centering
\label{tab:eRASSandXRTparms}
\begin{tabular}{llcc} \hline\hline
Obs.\              &  C-stat/$dof$    & Power-Law      & Observed $F_{0.5-2.0}$  \\
(Date)             &                  & Photon Index   & ($10^{-13}$~\ecgs)$^{(a)}$ \\ \hline
eR1  (MJD 58847.5) & 12.93/10 & $2.31\pm0.29$   & $9.16^{+1.24}_{-0.96}$  \\
eR2  (MJD 59029.6) & 12.65/10 & $2.43\pm0.32$   & $9.32^{+1.01}_{-1.26}$ \\
eR3  (MJD 59213.9) &  0.73/4  & $2.04\pm0.51$   & $6.92^{+1.21}_{-1.21}$ \\
eR4  (MJD 59398.3) &  6.09/4  & $2.30\pm0.57$   & $3.08^{+0.99}_{-0.92}$ \\
eR5  (MJD 59583.3) &  2.97/2  & 2.0*            & $0.58^{+0.32}_{-0.27}$  \\ 
Sw1  (MJD 59633.3) &  0.75/3  & $1.95\pm0.46$   &  $5.9^{+1.3}_{-1.1}$ \\
Sw2  (MJD 59667.7) &  3.35/6  & $2.12\pm0.30$   &  $7.3\pm1.2$   \\ 
Sw3  (MJD 59786.3) &  2.55/2  & $1.83\pm0.79$   &  $4.8\pm1.5$ \\
Sw4  (MJD 59934.3) & 15.87/18 & $2.25\pm0.17$   &  $11.8^{+1.3}_{-1.1}$  \\
Sw5  (MJD 60025.5) &  2.71/4  & $2.66\pm0.48$   &  $7.2^{+1.5}_{-1.6}$  \\
Sw6  (MJD 60078.6) &  8.54/7  & $2.33\pm0.30$   & $10.9\pm1.6$  \\
Sw7--8 (MJD 60148--51) & 4.65/10 & $2.51^{+0.33}_{-0.30}$ & $10.9^{+2.8}_{-1.5}$ \\ 
Sw9  (MJD 60284.3) &  3.19/3  & $1.77\pm0.54$   &  $6.8^{+1.9}_{-1.6}$  \\
Sw10 (MJD 60288.1) &  1.87/3  & $2.04\pm0.52$   &  $12.1^{+3.1}_{-2.1}$  \\
Sw11 (MJD 60426.7) &  2.39/3  & $2.48^{+0.50}_{-0.48}$ &  $21.8^{+4.6}_{-6.3}$  \\ 
Sw12--13 (MJD 60523--7)& 3.70/4  & $2.39^{+0.78}_{-0.67}$ &  $17.7^{+5.1}_{-5.2}$  \\ 
Sw14 (MJD 60544.7) &  1.54/5  & $1.85\pm0.32$          &  $9.5^{+1.6}_{-1.5}$  \\
Sw15 (MJD 60547.1) &  9.01/2  & $2.48^{+0.60}_{-0.57}$ &  $8.0^{+2.3}_{-2.1}$  \\ \hline \hline
\end{tabular}
\tablefoot{
An asterisk (*) denotes a fixed parameter; specfically, the photon index in eR5 was completely unconstrained. \\
  \tablefoottext{a}{Observed (Galactic-absorbed) 0.5--2.0~keV flux.}\\
  }
\end{table*}

\subsection{Spectral modeling of eRASS, \textit{Swift} XRT, and NICER}   \label{sec:eR_XRT_NICER_spec}

For the \textit{Swift} XRT spectra, the number of 0.2--10 keV spectral
counts was always small, less than 330 in all cases.
We combined Sw7 -- Sw8 and Sw12 -- Sw13, as those
pairs of observations were separated by only three and four days,
respectively.

For eRASS and XRT data, we binned each spectrum to 15 counts per bin
when the number of spectral counts was above 100 (eR1, eR2, eR3, Sw1,
Sw2, Sw6, Sw7-8, Sw10, and Sw14) or to 10 counts per bin when the number
of counts was less than 100 (eR4, Sw3, Sw5, Sw9, Sw11, Sw12-13, Sw15).
We fit using the C-statistic. In all cases, a single power law
modified by the Galactic column provided an excellent fit.  The photon
index in eR5 was unconstrained, so we froze it to 2.0, the value
obtained from fitting a single power law to the 0.5--2.0~keV XM1
spectrum.  The best-fitting photon indices and fluxes are listed in
Table~\ref{tab:eRASSandXRTparms}.  Photon indices are usually poorly
constrained, and there is no evidence of spectral variability.

For the NICER data, we added all 28 observations to maximize
the S/N. We fit the 0.4--4.0 keV spectrum.  A simple power law
yielded a poor fit, so we applied the best-fitting model from XM2
(\textsc{CompTT} + hard power law + \textsc{UxClumpy}).  $T_{\rm e}$
and $\tau$ were highly degenerate, so we froze $\tau$ at 11, the
best-fitting value from XM2; similarly, we froze $\Gamma_{\rm HX}$ at
1.66.  Our best-fitting model has $\chi^2/dof$=47.22/30, $T_{\rm
e}=0.23^{+0.02}_{-0.01}$ keV, $F_{\rm CompTT,0.2-0.7} =
2.61^{+0.29}_{-0.27}\times10^{-12}$~\ecgs, a 0.4--4.0~keV Galactic
absorption-corrected power-law flux of
$(1.52\pm0.03) \times10^{-12}$~\ecgs, and an observed absorbed
0.5--2.0 keV flux of $(1.32\pm0.01)\times10^{-12}$~\ecgs.

\section{Broadband SED modeling}   \label{sec:OUVSED}

\begin{figure*}
\includegraphics[width=1.99\columnwidth]{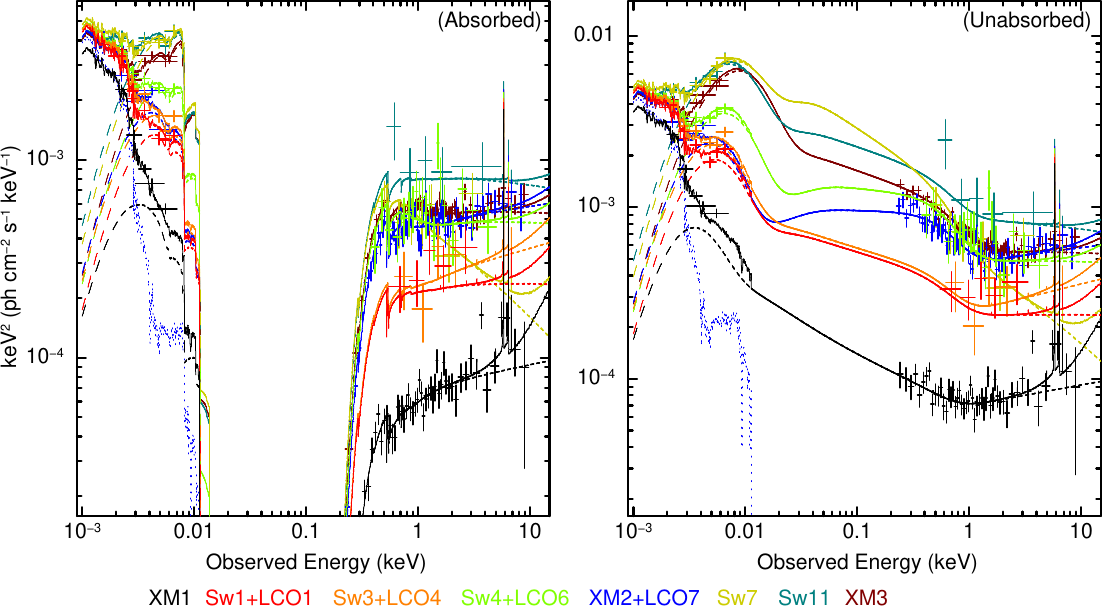} 
\caption{Optical/UV/X-ray SEDs and best-fitting models.  
 In the left panel the data points are not corrected for Galactic 
 reddening or obscuration, and best-fitting models are 
 reddened or obscured.  In the right panel, both data and models are
 unreddened and unobscured.  The dashed lines represent the best-fitting
 \textsc{fAGNSED} components. The dotted line indicates the host galaxy
  template (for clarity, only displayed for the XM2 fit). 
  The solid lines represent the total model; a torus component is included
  in the modeling but for clarity is displayed here only as part of the total model.  For the
  \textit{XMM-Newton} fits, only the EPIC pn0 data were used.
  XM1, XM2, and XM3 pn0 data were rebinned by factors of 3, 4, and 7 for
  plotting purposes only.  
    }
\label{fig:SED}
\end{figure*}

We constructed optical/UV/X-ray SEDs at several selected points in
time to track the broadband spectral evolution as the source
luminosity recovered.  We fit XM1, Sw1, Sw3, Sw4, XM2, Sw7, Sw11, and
XM3, using either OM+EPIC~pn or XRT+UVOT.  For Sw1, Sw3, Sw4, and XM2,
we augmented these SEDs with data points constructed
from quasi-simultaneous (to within 21 days) BVRI photometry obtained
from the LCOGT observations that were closest in time; the datasets
used are listed in Table~\ref{tab:agnsedfitresults}.  We did not
correct the optical and UV data points for Galactic reddening prior to fitting.  We
included a 5$\%$ systematic uncertainty to account for differences in
filter responses, as well as for the fact that most LCOGT and X-ray/UV
observations were not strictly simultaneous, and some mild variability
intrinsic to the AGN may have occurred.

The resulting SEDs are plotted in Fig.~\ref{fig:SED}; the spectral
variability, particularly in the optical/UV regime, is readily
apparent.
We fit a model consisting of \textsc{fAGNSED} \citep{Kubota18,Hagen23} plus an
Sb host galaxy template from the SWIRE template library
\citep{Polletta07} to account for host galaxy starlight, with both
components absorbed and reddened only by our Galaxy.  For Galactic
reddening, we set $E(B-V)$ to 0.057, based on the dust maps of
\citet{Schlegel98}.
\textsc{fAGNSED} posits three emission regions: an optically thick, 
geometrically thin outer disk that encircles  
a warm Comptonizing region, which in turn encircles a
geometrically thick, optically thin, hot Comptonizing region.  In all
fits we froze the co-moving distance to the source at 404 Mpc, the
normalization at unity, the black hole mass to $1.4\times10^8 \Msun$,
derived via the width of the broad H$\beta$ emission lines as
discussed in Sect.~\ref{sec:MBHestimate}, the dimensionless black hole
spin parameter $a*$ at zero, the electron temperature of the hot
corona, $k_{\rm B}T_{\rm e, hot}$, at 100~keV, the hot corona's scale
height at 10~$R_{\rm g}$, and the inclination angle of the warm corona
and outer disk at $15^{\rm \circ}$.  In addition, we
kept the  radial size of the warm corona, $R_{\rm warm}$, fixed to
twice that of the hot corona, $R_{\rm hot}$, in all fits.  The outer
radius of the disk was calculated via the self-gravity radius
from \citet{Laor89}, and reprocessing was included.  Compared
to \textsc{AGNSED}, \textsc{fAGNSED} includes a color--temperature
correction for the accretion disk emission, $f_{\rm col}$, which we
left as a free parameter.
For the XM1--3 fits, which had the best signal-to-noise ratio, the free
parameters were the accretion rate relative to Eddington,
$\dot{m}_{\rm Edd}$, the electron temperature of the warm corona,
$k_{\rm B}T_{\rm e, warm}$, the spectral indices of the warm
($\Gamma_{\rm warm}$) and hot ($\Gamma_{\rm hot}$) Comptonization
components, the outer radius of the hot Comptonizing component,
$R_{\rm hot}$, and the normalization of the host galaxy template.  For
the fits using \textit{Swift} data, to obtain reasonable constraints
on free parameters, we froze $R_{\rm hot}$ to 26~$R_{\rm g}$ and
$k_{\rm B}T_{\rm e,warm}$ to 0.16~keV, the averages of the
best-fitting values from XM1--3.  For all fits except Sw4+LCO6,
$\Gamma_{\rm warm}$ was very poorly constrained, so we froze it to
2.3.  Sw4 had the highest S/N  soft X-ray data of all XRT
observations, and thawing $\Gamma_{\rm warm}$ yielded a fit that was
an improvement at the 99.8$\%$ confidence level compared to leaving
$\Gamma_{\rm warm}$ frozen at 2.3 according to an $F$-test.  Finally,
we included a torus component modeled with \textsc{UxClumpy}, with all
parameters frozen to values used in the \textit{XMM-Newton}-only fits
(Table~\ref{tab:XMparms}).

The best-fitting model parameters are listed in
Table~\ref{tab:agnsedfitresults}.  For brevity, the best-fitting
values of the normalization of the host galaxy template are omitted
from that table; the best-fitting values are always in the range
(8--11)$\times10^{-16}$ erg cm$^{-2}$ s$^{-1}$ $\AA^{-1}$.  The
best-fitting values of $\dot{m}_{\rm Edd}$ increase steadily over 3
years, by a factor of 7 from XM1 to XM3.  
We list the estimates of the unabsorbed bolometric luminosity, $L_{\rm Bol}$,
at all epochs; $L_{\rm Bol}$ is as low as $6.4\times10^{43}$~erg~s$^{-1}$ during XM1.
We also list the 13.6~eV --
10~keV ionizing luminosity, $L_{\rm ion}$, calculated from integrating over the
best-fitting unabsorbed models, and useful for discussion in
Sect.~\ref{sec:Discussion} regarding Balmer line emission; they also increase
by a factor of roughly 7 from XM1 to XM3. 
The values of $L_{\rm Bol}/L_{2-10}$ for XM1, 2, and 3 are thus 11, 8, and
13, respectively, consistent with expectations for Seyfert AGN at this
luminosity \citep{Duras20}. The best-fitting
values of $\Gamma_{\rm hot}$ and $k_{\rm B}T_{\rm e,warm}$ are
consistent across XM1--3, precluding any evidence for evolution in
those parameters. In particular, photon counts above 3~keV 
are poor for Sw1, Sw3, Sw4, and Sw7, so those values of
$\Gamma_{\rm hot}$ may be subject to artefacts of modeling and should be interpreted with caution.
The best-fitting absorbed and unabsorbed models are plotted in
Fig.~\ref{fig:SED}.

\renewcommand{\arraystretch}{1.18}
\begin{table*}
\caption[]{    Best-fitting broadband SED models using \textsc{fAGNSED}.}    
        \centering
\label{tab:agnsedfitresults}
\begin{tabular}{lcccccccc} \hline\hline
Dataset                   &  log($\dot{m}_{\rm Edd}$) &   $f_{\rm col}$         &   $k_{\rm B}T_{\rm e,warm}$ & $\Gamma_{\rm warm}$     & $\Gamma_{\rm hot}$      & $R_{\rm hot}$         &   $L_{\rm ion}$$^{(a)}$  & $L_{\rm bol}$$^{(b)}$   \\
and date (MJD)             &                           &                         &   (keV)                      &                         &                         &                       &   (erg s$^{-1}$)                 \\ \hline

XM1 (59604)                &  $-2.44^{+0.01}_{-0.08}$  & $1.08^{+0.57}_{-0.04}$  & $0.14^{+0.74}_{-0.04}$       & $2.54\pm0.10$           & $1.88^{+0.17}_{-0.16}$  & $22.6^{+7.7}_{-0.4}$  & $3.2 \times10^{43}$  & $6.4 \times10^{43}$ \\   
$\chi^2/dof$ = 145.36/150  \\    \hline

Sw1 (59633)                & $-2.03^{+0.04}_{-0.02}$   & $1.53^{+0.34}_{-0.27}$  & 0.16*                        &  2.3*                   & $2.00^{+0.15}_{-0.13}$  & 26*                      & $1.1 \times10^{44}$  & $1.6 \times10^{44}$\\
+ LCO1 (59639)   \\ 
$\chi^2/dof$ = 15.52/11     \\ \hline

Sw3 (59786)                & $-1.94\pm0.03$        & $1.42^{+0.31}_{-0.20}$   & 0.16*                        &  2.3*                   & $1.84^{+0.31}_{-0.17}$  & 26*                      & $1.2 \times10^{44}$  & $2.0 \times10^{44}$\\
+ LCO4 (59773)   \\
$\chi^2/dof$ = 12.58/10  \\    \hline

Sw4 (59934)                &  $-1.75^{+0.03}_{-0.02}$  & $1.69^{+0.31}_{-0.17}$   & 0.16*                        & $2.09^{+0.12}_{-0.07}$   & $2.01^{+0.11}_{-0.12}$  & 26*                    & $2.4 \times10^{44}$   & $3.1 \times10^{44}$  \\
+ LCO6 (59949) \\
$\chi^2/dof$ = 21.62/25   \\  \hline

XM2 (59970)                & $-1.81^{+0.02}_{-0.01}$   & $1.50^{+0.26}_{-0.23}$   & $0.16^{+0.04}_{-0.03}$        &  $2.08^{+0.08}_{-0.05}$    & $1.91^{+0.11}_{-0.10}$      & $31.8^{+2.1}_{-1.2}$    & $1.9 \times10^{44}$ &  $2.7 \times10^{44}$   \\
+ LCO7 (59991) \\  
$\chi^2/dof$ = 261.06/248    \\ \hline

Sw7 (60148)                & $-1.50^{+0.05}_{-0.02}$    & $1.67^{+0.43}_{-0.24}$   & 0.16*               &  2.3*                    & $2.49^{+0.13}_{-0.10}$       & 26*           & $4.9 \times10^{44}$    & $5.5 \times10^{44}$  \\
$\chi^2/dof$=  9.97/9  \\  \hline

Sw11 (60426)               &  $-1.49^{+0.10}_{-0.03}$  & $1.57^{+0.85}_{-0.28}$  & 0.16*                        &  2.3*                    & $2.06^{+0.21}_{-0.19}$  & 26*             & $4.4 \times10^{44}$   & $6.1 \times10^{44}$ \\
$\chi^2/dof$=  5.83/6  \\ \hline

XM3 (60697)                &  $-1.60^{+0.06}_{-0.02}$  & $1.86^{+0.38}_{-0.23}$  & $0.18^{+0.03}_{-0.02}$       & $2.30^{+0.10}_{-0.04}$  & $2.00^{+0.06}_{-0.05}$  & $21.6^{+1.7}_{-2.7}$     & $3.6 \times10^{44}$  & $4.4 \times10^{44}$ \\   
$\chi^2/dof$ = 442.50/473  \\    \hline \hline

\end{tabular}
\tablefoot{ 
An asterisk (*) denotes a fixed parameter..\\
\tablefoottext{a}{13.6~eV -- 10 keV unabsorbed model luminosity. } \\
\tablefoottext{b}{Bolometric unabsorbed model luminosity.}
}
\end{table*}

\section{Fits to optical emission-line spectra}\label{sec:optspec}

\begin{figure*}
\includegraphics[width=1.99\columnwidth]{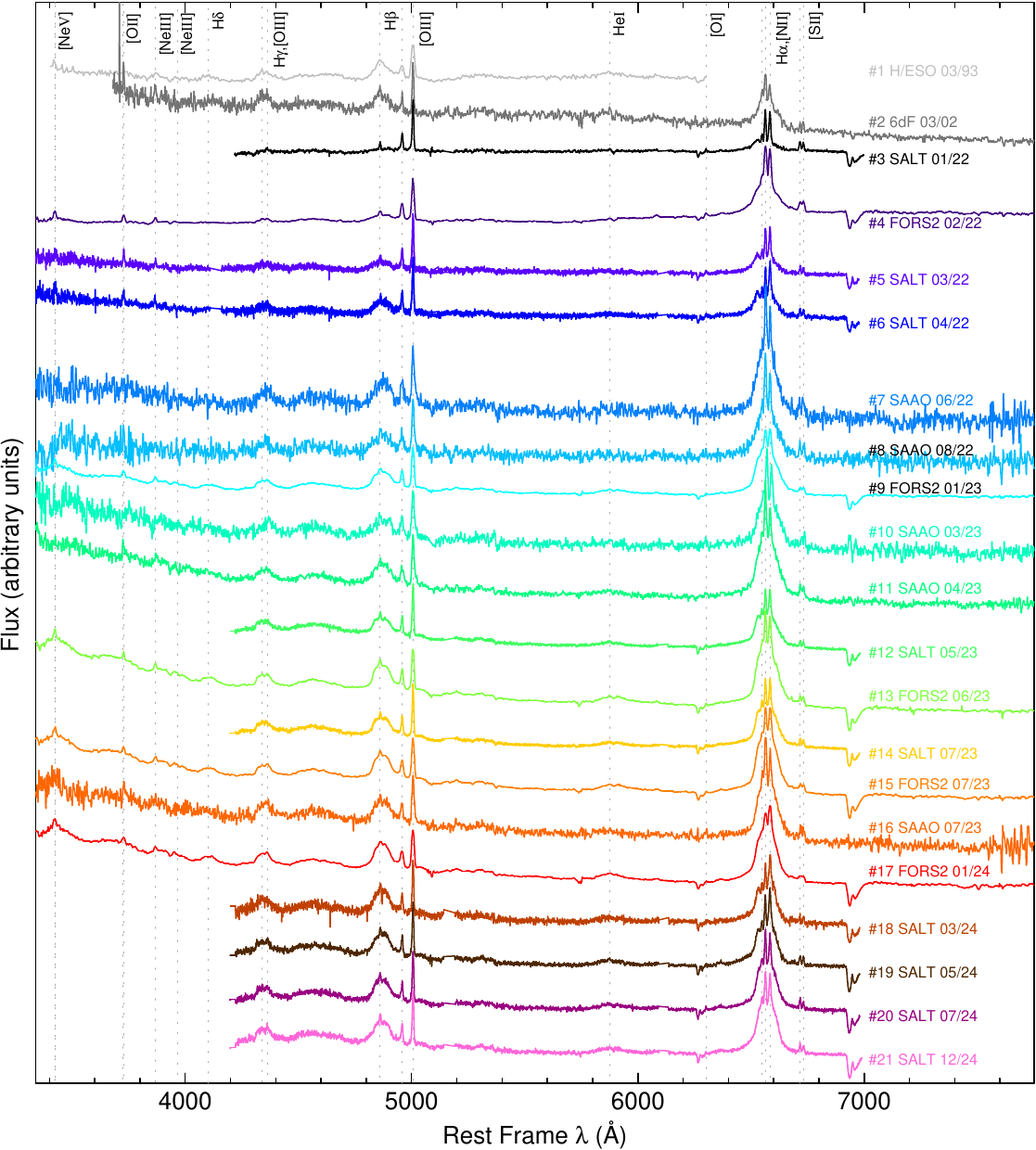}
\caption{All 21 optical spectra considered for spectral fitting, 
plotted with arbitrary shifts for visual clarity.   }
\label{fig:optspecwide}
\end{figure*}

\begin{figure*}
\includegraphics[width=0.99\columnwidth]{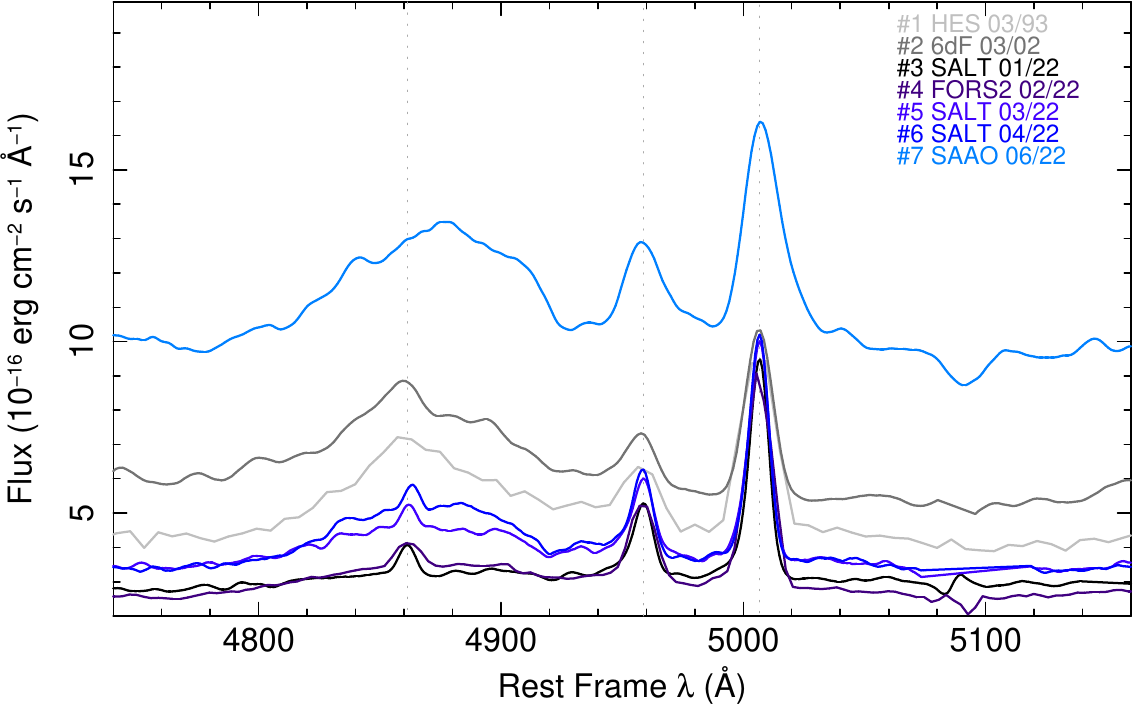}
\includegraphics[width=0.99\columnwidth]{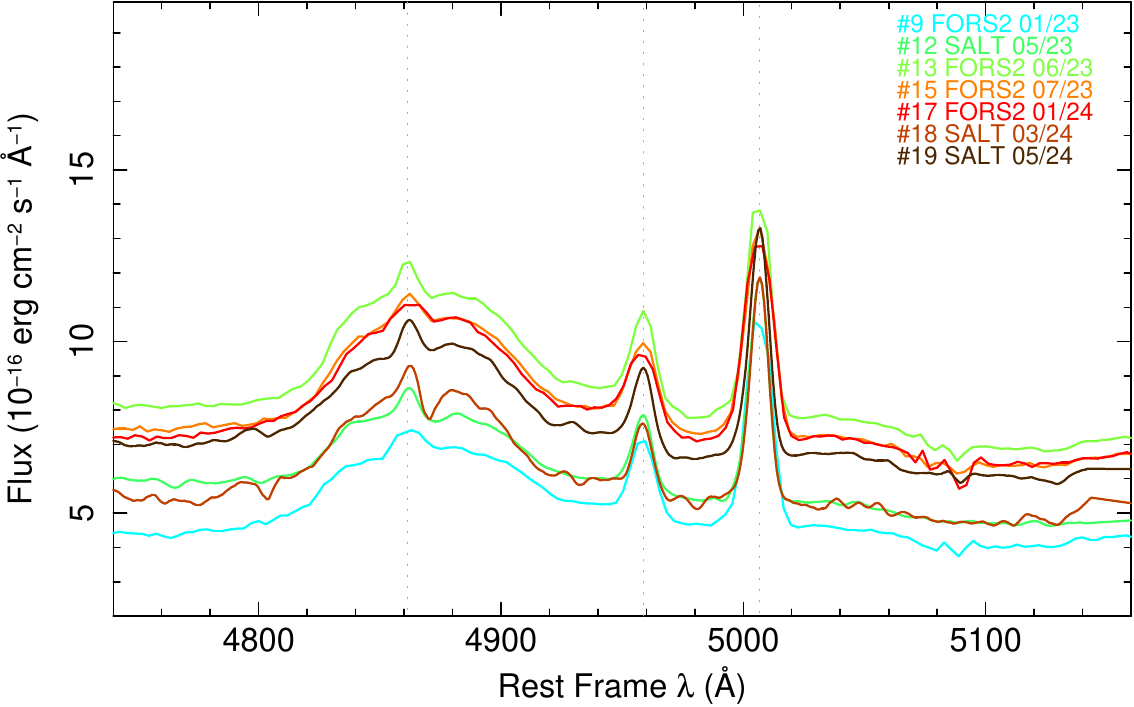}
\caption{H$\beta$ and [\ion{O}{iii}] regions of selected optical spectra.
The latter spectra (right panel) correspond to the resurgence of the
diskline component, as discussed below.  These spectra have been
gray-scaled so that the integrated [\ion{O}{iii}] fluxes match.  
Smoothing has been applied for visual clarity only, using a boxcar of
width 5~$\AA$ for all spectra, except for \#2 and \#7 where the width
was 10~$\AA$.}
\label{fig:optspeczoom}
\end{figure*}

The 21 rest-frame optical spectra are plotted in
Fig.~\ref{fig:optspecwide}, with a zoom-in on the H$\beta$ region for
selected spectra in Fig.~\ref{fig:optspeczoom}.  Visually, the broad
Balmer profiles are very weak during the low-continuum state in
January--February 2022 (\#3--\#4). During March--August 2022 (\#5--\#8), as
optical, UV, and X-ray continuum fluxes rise, the broad Balmer
profiles become stronger, but remain roughly Gaussian-like
(Fig.~\ref{fig:optspeczoom}, left panel).  However, by January 2023
(spectrum \#9), the broad Balmer profiles become much more boxy in
shape.

Other features in the spectra include \ion{Fe}{ii} emission near
4570~\AA, and narrow lines due to [\ion{O}{iii}], [\ion{N}{ii}],
[\ion{S}{ii}], [\ion{Ne}{v}], and the Balmer series.  There is also
some broad
\ion{He}{i} $\lambda$5876 emission, prominent in the H/ESO spectrum
(\#2), then very weak during spectra \#3--4, then with strength
gradually increasing throughout the rest of the campaign.

\subsection{Summary of spectral fit procedure}   \label{sec:performingoptfits}

Spectral modeling was done with the Python \textsc{lmfit}
package \citep{Newville14}.  We masked a region of atmospheric
telluric absorption, 6843--6900~$\AA$ (observed frame, corresponding
to 6242--6293~$\AA$ rest frame).  A second region of telluric
absorption impacted a narrow region near 7175~$\AA$ observed frame
(6544~$\AA$ rest frame), in the broad H$\alpha$ profile and very close
to [\ion{N}{ii}]~$\lambda$6548.
We modeled the H$\beta$ (4200-5400~$\AA$ rest frame) and H$\alpha$
regions (5600--6900~$\AA$ rest frame) separately.  For the AGN
continuum, we used a broken power-law model.  We used an Sb host
galaxy template from the SWIRE library \citep{Polletta07}.  We used
the \ion{Fe}{ii} templates of \citet{Kovacevic10} for H$\beta$ range
since it can model the F, S, G, P, and I Zw 1 groups of \ion{Fe}{ii}
separately, though in our fits, the P and Z group amplitudes fell to
zero.  We used the \citet{Bruhweiler08} template for the H$\alpha$
fits, given the limited wavelength coverage of the \citet{Kovacevic10}
templates. The \ion{Fe}{ii} velocity widths are typically of order 2500 km~s$^{-1}$.

Our fits to the H$\beta$ region contained Gaussians for narrow-line
emission from H$\beta$, [\ion{O}{iii}]~$\lambda\lambda$5007,4959, and
[\ion{O}{iii}]~$\lambda$4363, with all velocity widths $\sigma$ tied
to that of [\ion{O}{iii}]~$\lambda$5007 (typically
150--350~km~s$^{-1}$).  The [\ion{O}{iii}]~$\lambda\lambda$5007,4959
profiles each required an additional broad but weak blueshifted
Gaussian component, with velocity width typically 250--800~km~s$^{-1}$
and velocity offset typically 150--400~km~s$^{-1}$. Such features are
indicative of a NLR outflow in $\jotwelve$, similar to those inferred
for many nearby as well as high-redshift AGN \citep{Schmidt18,Leung19}.

In the H$\alpha$ region, we included narrow Gaussians for
[\ion{N}{ii}]~$\lambda\lambda$6583,6548,
[\ion{S}{ii}]~$\lambda\lambda$6731,6716, and
[\ion{O}{i}]~$\lambda$6300, with these lines' velocity widths tied.
We also included a broad Gaussian to model \ion{He}{i}~$\lambda$5876
emission; for all spectra except \#3--4, and \#8 (low S/N),
it was detected and modeled well with a Gaussian of width $\sigma$
typically 35--55 \AA, with no evidence of evolution in $\sigma$.  To
model the narrow atmospheric telluric feature at 7175~$\AA$ observed
frame (6544~$\AA$ rest frame), we obtained a good fit using a narrow
Pearson7 component, though this component was required only for
spectra \#3, 5, 6, 9, 12--19, and 21.

\begin{figure*}
\includegraphics[width=0.65\columnwidth]{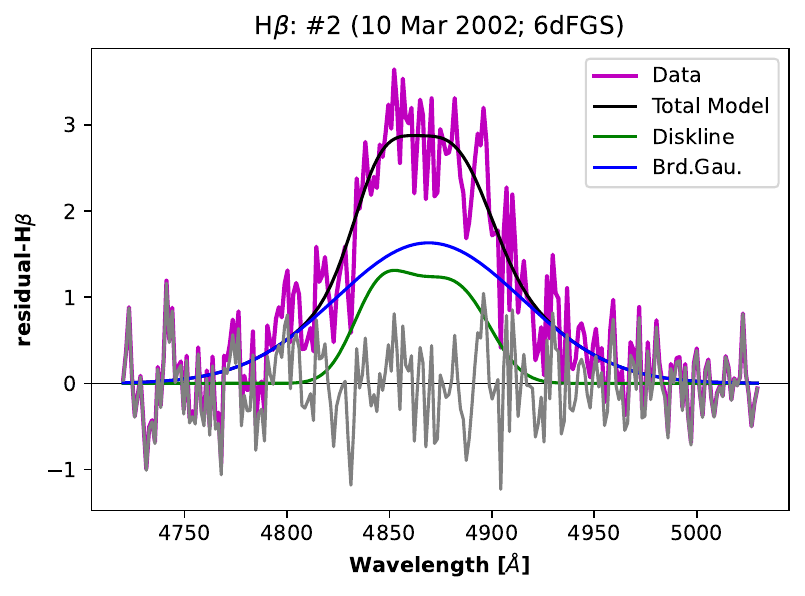}
\includegraphics[width=0.65\columnwidth]{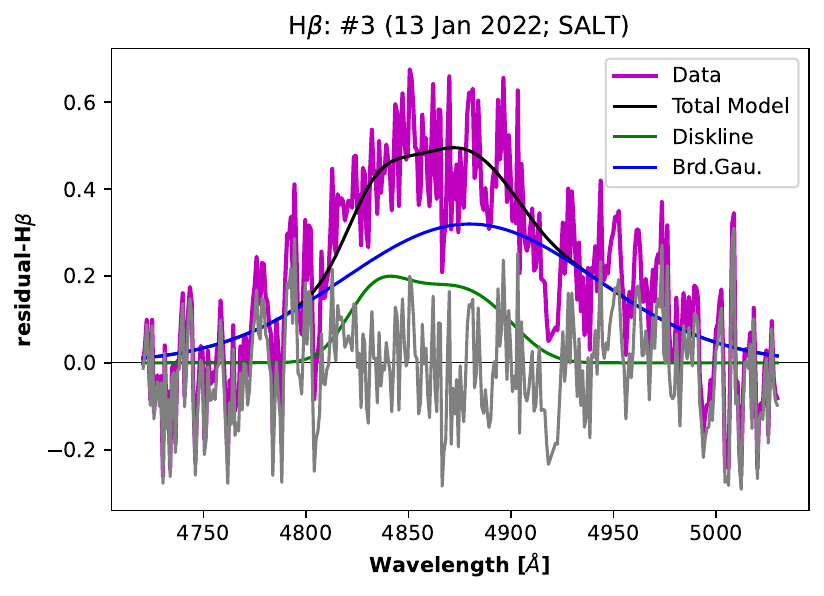}
\includegraphics[width=0.65\columnwidth]{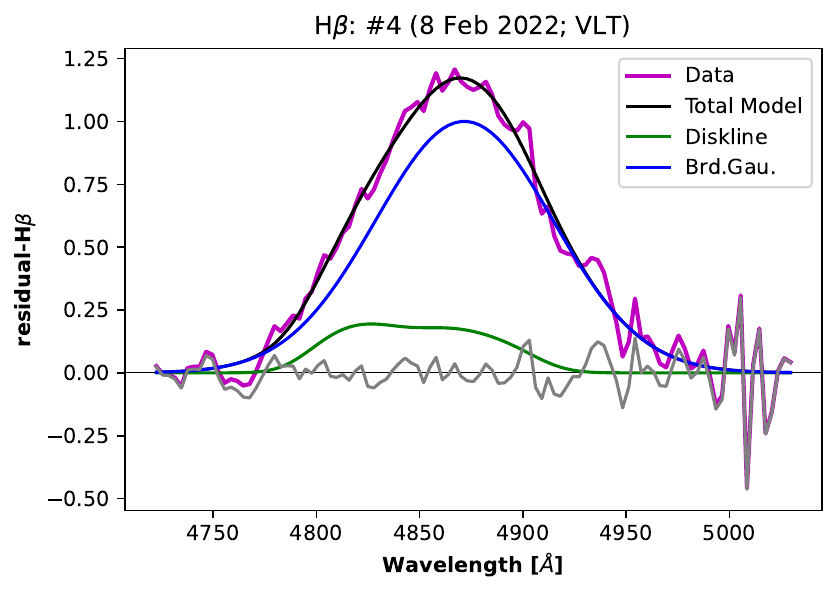}

\includegraphics[width=0.65\columnwidth]{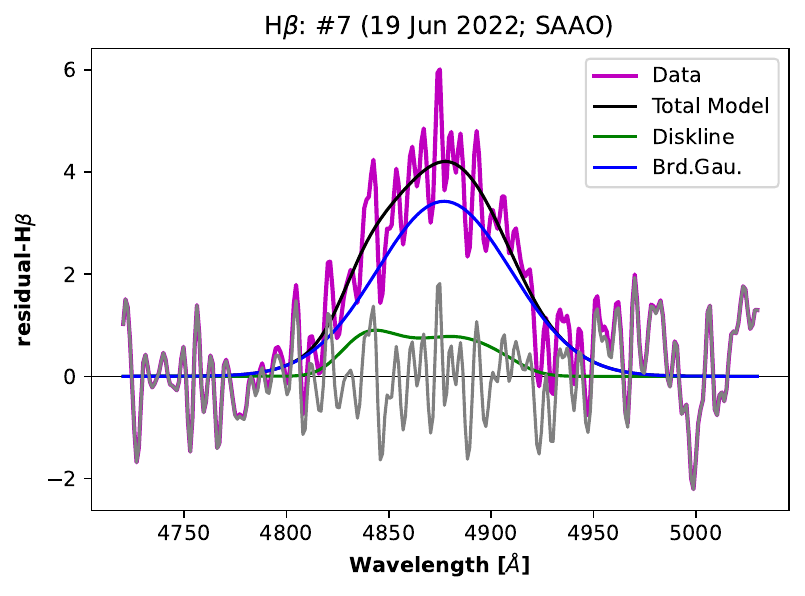}
\includegraphics[width=0.65\columnwidth]{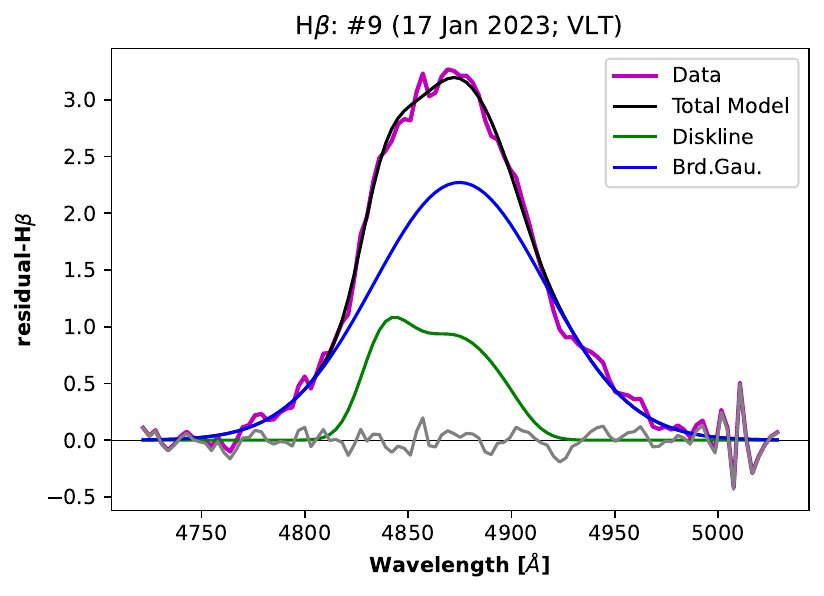}
\includegraphics[width=0.65\columnwidth]{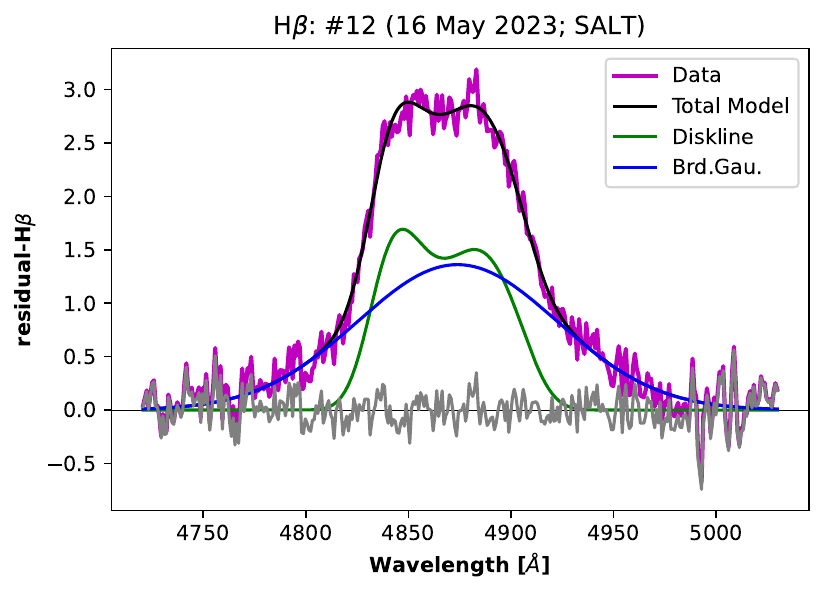}

\includegraphics[width=0.65\columnwidth]{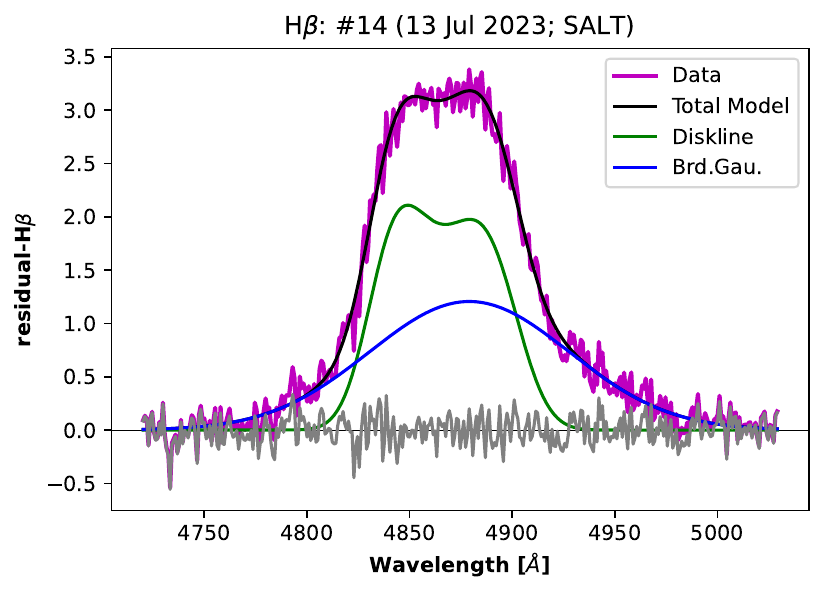}
\includegraphics[width=0.65\columnwidth]{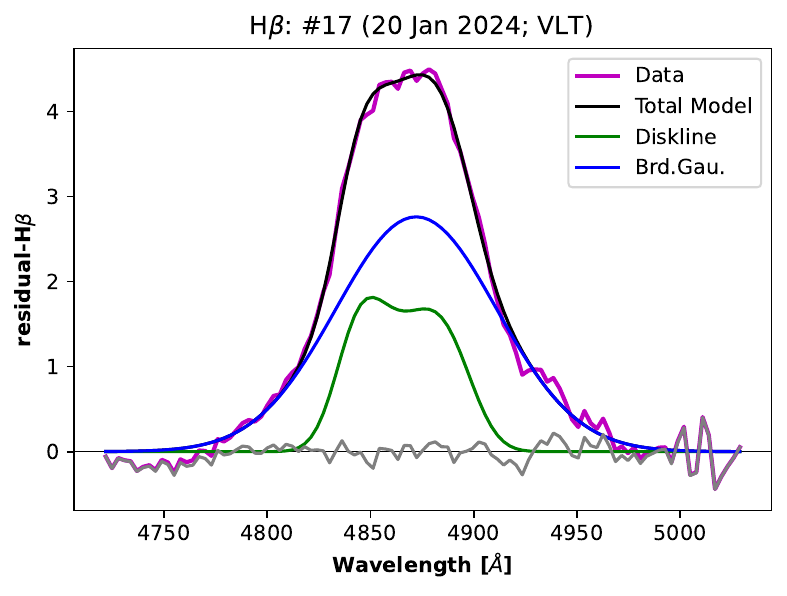}
\includegraphics[width=0.65\columnwidth]{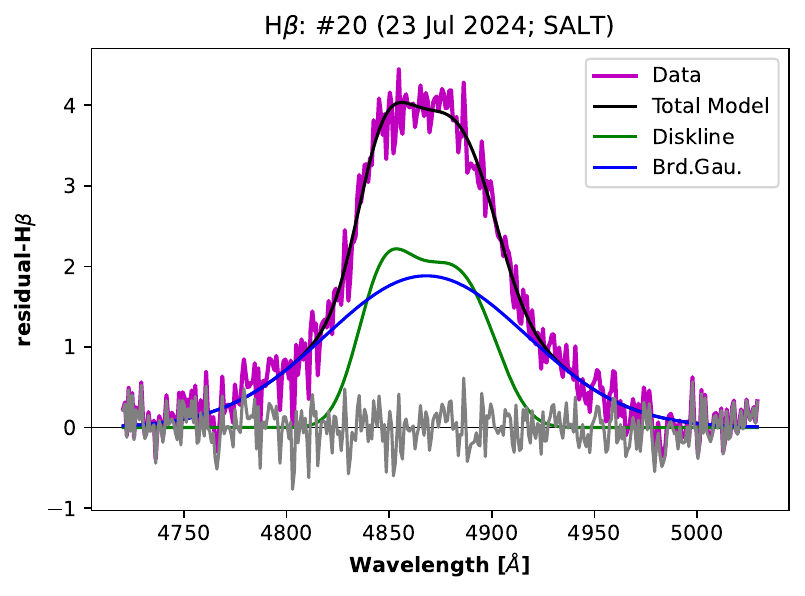}
\caption{Decomposition of diskline plus Gaussian components 
for fits to the H$\beta$ profiles of selected optical spectra. The data
(after subtraction of continuum components and [\ion{O}{iii}] lines)
are shown in purple; green, blue, and black denote the diskline
component, the broad Gaussian component, and the total model
respectively. The Y-axis units are $10^{-16}$~erg~cm$^{-2}$~s$^{-1}$~$\AA^{-1}$.} 
\label{fig:sampleDLfits}
\end{figure*}

For each of the broad H$\beta$ and H$\alpha$ Balmer profiles, a simple
broad Gaussian slightly redshifted relative to rest frame and width
$\sigma$ usually 30--50~$\AA$, yields decent fits for the spectra
taken in 2022 (\#3--\#8), but yields poor fits to all later
spectra. The later spectra are fit well phenomenologically by the sum
of a broad Gaussian plus two additional Gaussians, with equal widths
$\sigma$ and amplitudes on either side of the main broad peak.  These
two additional  Gaussians typically peak at
4845--4852 and 4883--4888~$\AA$ (H$\beta$) and
6535--6548 and 6583--6601~$\AA$ (H$\alpha$).
Consequently, we pursued a more physical modeling using the sum of a
broad Gaussian plus a relativistic diskline component \citep{Chen89}
to model emission across a range of radii in an assumed flat annular
region.  We fixed the diskline's outer radius $R_{\rm out}$ to 5000
$R_{\rm g}$ in all fits, but left inner radius $R_{\rm in}$ free.  We
incorporated a local broadening factor, $\sigma_{\rm 0}$, to account
for electron scattering in a photoionized atmosphere.  Some
parameters, when left thawed, were not constrained in some fits.  In
those cases, we froze $R_{\rm in}$ to 1000~$R_{\rm g}$, $\sigma_{\rm
0}$ to 10~$\AA$ (600~km~s$^{-1}$) for H$\beta$ or to 15~$\AA$
(700~km~s$^{-1}$) for H$\alpha$, and inclination angle
$i$\footnote{$i$ is defined such that 0$^{\rm \circ}$ indicates a
face-on disk.}  to 13$^{\rm \circ}$, the average of the best-fitting
values for the remaining fits.  We applied this component to all
H$\beta$ and H$\alpha$ profiles.  For all H$\alpha$ fits, the use of the
diskline plus broad Gaussian model is a better fit compared to either a
single broad Gaussian or a single diskline component (lower values of
AIC and $\chi^2$ in each case) for all spectra. For the H$\beta$ fits,
this statement holds true for all spectra except
\#4, \#8, and \#11, but we also adopt the diskline plus broad Gaussian model for
these fits, on physical grounds.
We note that diskline parameters are relatively
insensitive to the assumed value of $R_{\rm out}$. For instance,
assuming a fiducial value of $R_{\rm out}$=20000~$R_{\rm g}$, $i$
increases by an average of only 3--4$^{\rm \circ}$, and $R_{\rm in}$
increases by an average of only 
15$\%$.\footnote{Additionally, for this value of $R_{\rm out}$, the values of the
parameter $F_{\rm DL}$, discussed below, increase by an average of
only 5 percent, and the associated Pearson correlations are not
significantly impacted.}  
Sample model fits are shown in Fig.~\ref{fig:sampleDLfits}.
Best-fitting values using the diskline plus broad Gaussian model are
listed in Tables~\ref{tab:HBFITRESULTS}, \ref{tab:HAFITRESULTS}, and
\ref{tab:OTHEROPTFITRESULTS}. 
The total broad H$\beta$ line flux is plotted as function of time in
Fig.~\ref{fig:ALL_LC_ZOOM}; the line fluxes increase during 2022--2024,
qualitatively tracking the increase in continuum fluxes.  We do
not find evidence of evolution in the Balmer decrement of the broad
Gaussian component, diskline component, or total broad-line fluxes (average
values of 2.28, 2.76, and 2.73, respectively).

For the diskline component in the H$\beta$ and H$\alpha$ fits, the best-fitting values
of inclination angle $\theta$ (defined such that 0$^{\rm \circ}$
corresponds to a face-on orientation) are usually 10--15$^{\rm
\circ}$. 
The median value and standard deviation of $R_{\rm in}$ across all
fits where $R_{\rm in}$ was a free parameter is 1040$\pm$410~$R_{\rm
g}$.  We find no evidence of evolution in either $R_{\rm in}$ or
$\theta$ parameter with time.  For the broad Gaussian components,
energy centroids $E_{\rm cent}$ are redshifted by average values of
12$\pm$10~$\AA$ (770$\pm$620~km~s$^{-1}$) for H$\beta$ and
13$\pm$8~$\AA$ (620$\pm$370~km~s$^{-1}$) for H$\alpha$. Rest-frame
velocity widths $\sigma$ are typically 1850--3300 km s$^{-1}$ for
H$\beta$ and typically 1800--3900 km s$^{-1}$ for H$\alpha$, with no
evidence of time evolution of these parameters.

\begin{figure*}
\includegraphics[width=0.99\columnwidth]{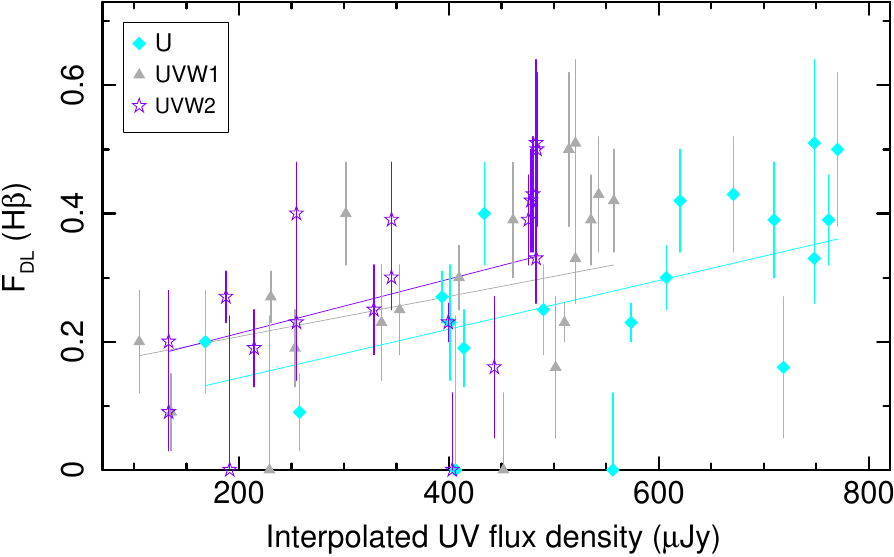}
\includegraphics[width=0.99\columnwidth]{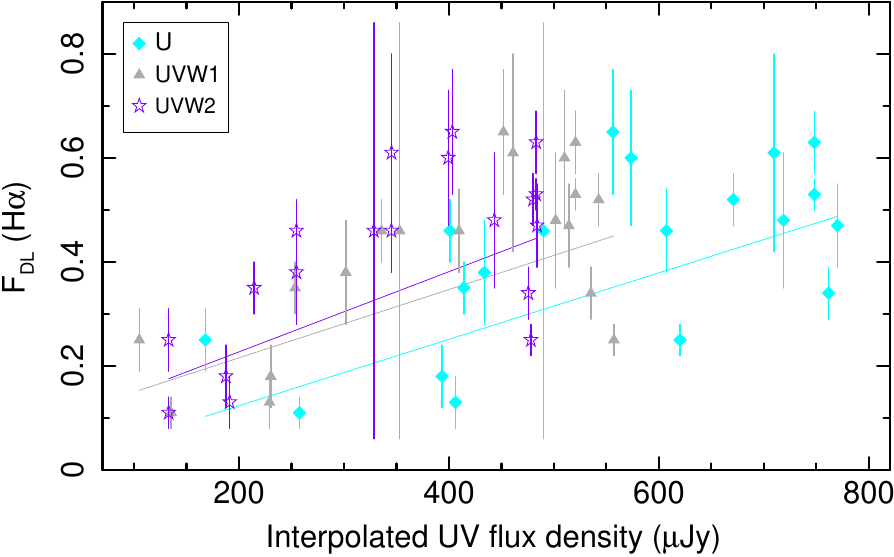}
\caption{$F_{\rm DL}$, the ratio of diskline component intensity to total broad-line intensity for the
H$\beta$ line (left panel) and H$\alpha$ line (right panel), 
as a function of UV flux density, interpolated to match the dates when the optical spectra
were taken. Cyan, gray, and purple denote U, UVW1, and UVW2 flux densities, respectively.
The solid lines indicate the best-fitting linear relations, meant to guide the eye.
Positive correlations occur in all bands for both H$\beta$ and H$\alpha$, indicating a link between
UV continuum emission and the broad Balmer profiles.}
\label{fig:fdl_vs_uvlum}
\end{figure*}

We define $F_{\rm DL}$ as the flux of the diskline component relative
to the total (broad Gaussian plus diskline) flux.  In
Fig.~\ref{fig:fdl_vs_uvlum}, we plot values of $F_{\rm DL}$ for both
H$\beta$ and H$\alpha$ as a function of U, UVW1, and UVW2 flux densities,
using values interpolated from the combined UVOT plus OM light curves. For
both H$\beta$ and H$\alpha$ there are moderate positive correlations
between $F_{\rm DL}$ and UV flux densities.  For H$\beta$, Pearson
correlation coefficients (null hypothesis probabilities) for U, UVW1,
and UVW2 are 0.614 (0.0052), 0.611 (0.0055), and 0.627 (0.0041),
respectively.  For H$\alpha$, these values are 0.601 (0.0065), 0.700
(0.0008), and 0.628 (0.0040), respectively.

\begin{figure*}
\includegraphics[width=0.99\columnwidth]{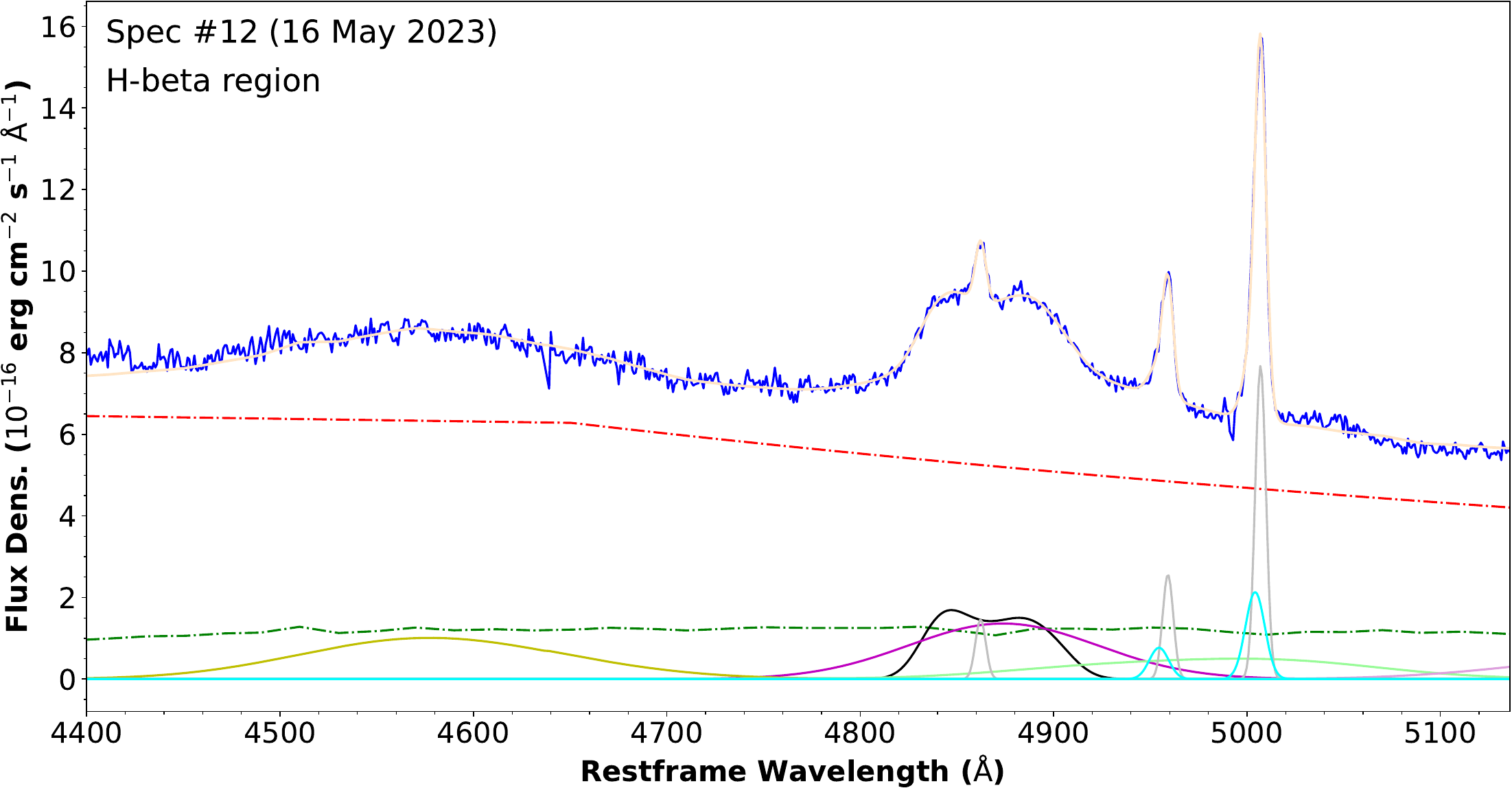}
\includegraphics[width=0.99\columnwidth]{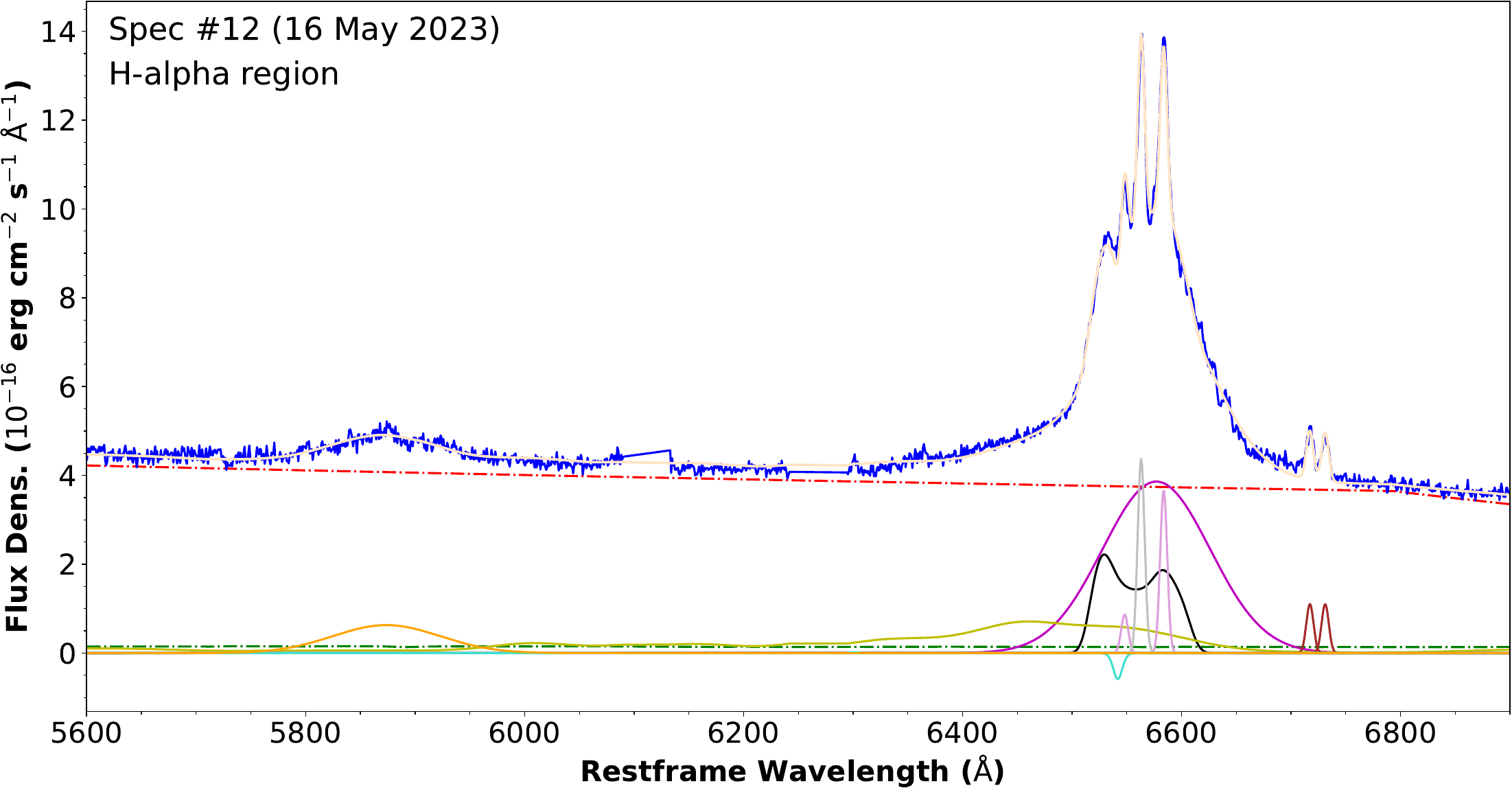}
\caption{Sample spectral decomposition in the H$\beta$ (left) and H$\alpha$ (right) 
regions for spectrum \#12.  In each panel, the blue solid line
denotes the data, the total model is shown in beige, the solid black
line denotes the Balmer diskline component, the solid magenta line
denotes the broad Balmer Gaussian component, the red dashed line
denotes the power-law component, and the dark green dashed line
denotes the host galaxy emission.  In the left panel (H$\beta$),
yellow, pale green, and pink respectively denote the F-, S-, and G-groups
of \ion{Fe}{ii} emission, while the narrow Balmer and
[\ion{O}{iii}] components are shown in gray, and the broad blueshifted
[\ion{O}{iii}] components are shown in cyan.  In the right panel
(H$\alpha$),
\ion{He}{i} emission is shown in orange, [\ion{S}{ii}] emission is
shown in brown, [\ion{N}{ii}] emission is shown in pink, and the telluric
atmospheric absorption feature is shown in cyan.}
\label{fig:optspecdecomp}
\end{figure*}

\begin{figure}
\includegraphics[width=0.99\columnwidth]{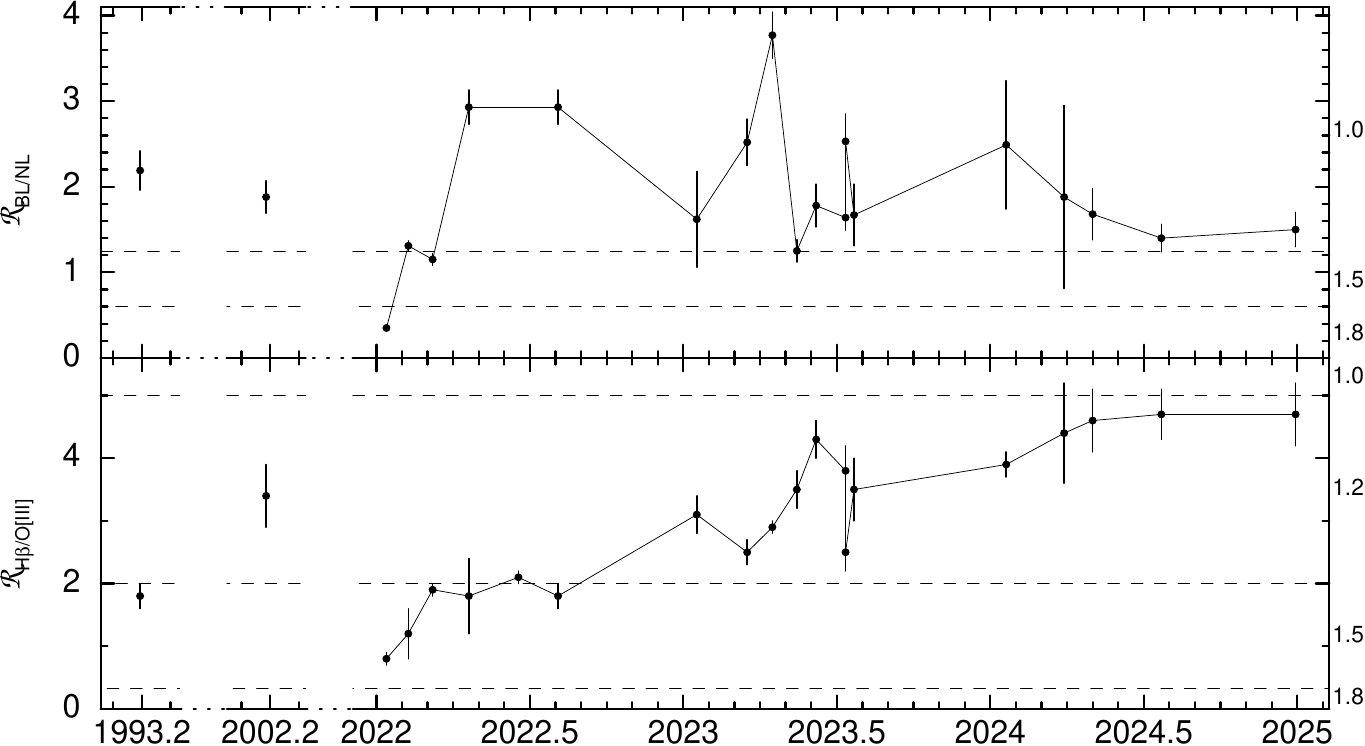}
\caption{$\mathcal{R}_{\rm BL/NL}$ (top panel) and $\mathcal{R}_{\rm H\beta/[\ion{O}{iii}]}$ (bottom panel)
as a function of time.
The horizontal dashed lines denote subtype classification boundaries following \citet[][top panel]{Runco16}
and \citet[][bottom panel]{Winkler92}.}
\label{fig:Rversustime}
\end{figure} 

\subsection{Subtype evolution}

To quantify the evolution in broad H$\beta$ flux, we consider the
ratio of the integrated broad H$\alpha$ flux (sum of diskline and
Gaussian; see Table~\ref{tab:HBFITRESULTS}) to the sum of the
[\ion{O}{iii}]~$\lambda$5007 line (sum of narrow and blueshifted
components; see Table~\ref{tab:OTHEROPTFITRESULTS}), hereafter
$\mathcal{R}_{\rm H\beta/[\ion{O}{iii}]}$, as defined
in \citet{Winkler92}.  We also consider the ratio of broad to narrow
H$\beta$ peak flux densities, hereafter $\mathcal{R}_{\rm BL/NL}$,
following \citet{Runco16}.  The values of $\mathcal{R}_{\rm
H\beta/[\ion{O}{iii}]}$ and $\mathcal{R}_{\rm BL/NL}$ are listed in
Table~\ref{tab:OTHEROPTFITRESULTS}, and plotted in
Fig.~\ref{fig:Rversustime}. From spectrum \#2 (March 2002) to \#3
(January 2002), $\mathcal{R}_{\rm H\beta/[\ion{O}{iii}]}$ falls by a
factor of 4.3$\pm$0.8. $\mathcal{R}_{\rm H\beta/[\ion{O}{iii}]}$ then
gradually increases by a factor of almost 6 for the rest of the
campaign.  The values of $\mathcal{R}_{\rm BL/NL}$ also reach a minimum
during spectrum \#3 (factors of 5--6 smaller than in \#1 and \#2), and
increase back to their approximate maximum values thereafter.

We can also use $\mathcal{R}_{\rm H\beta/[\ion{O}{iii}]}$ and
$\mathcal{R}_{\rm BL/NL}$ and to assign
approximate Seyfert subtype classification.
\citet{Winkler92} assigned subtypes 1.0, 1.2, 1.5, and 1.8 to objects with
$\mathcal{R}_{\rm H\beta/[\ion{O}{iii}]}$ $>5$, 
$2 < \mathcal{R}_{\rm H\beta/[\ion{O}{iii}]} < 5$,
$0.33 < \mathcal{R}_{\rm H\beta/[\ion{O}{iii}]} < 2$, and
$\mathcal{R}_{\rm H\beta/[\ion{O}{iii}]}$ $<0.33$, respectively.
Meanwhile, \citet{Runco16} assigned subtypes 1.0, 1.5, and 1.8 to
objects with $\mathcal{R}_{\rm BL/NL} > 1.25$, $0.6 < \mathcal{R}_{\rm BL/NL} < 1.25$, and
$\mathcal{R}_{\rm BL/NL} <0.6$, respectively.  
These classification boundaries, and the evolution  
with respect to them, are plotted in Fig.~\ref{fig:Rversustime}.
For example, under the \citet{Runco16} scheme,
$\jotwelve$ 
was type 1.0 in 1993 and 2002 (\#1--2), transitioned to type 1.8 by January 2022 (\#3),
and returned to type 1.0 by April 2022 (\#6  onward).

However, we recall some brief caveats regarding the use of
these quantities for subtype classification: In general, comparison from one
object to the next means comparing objects with potentially different NLR
properties and/or host galaxy star formation activity.  In
addition, the impact of the host galaxy continuum in the spectrum (and
how accurately it is modeled) can be strong when the broad H$\beta$
line is weak.  Finally, spectra taken with different apertures can
yield different measurements of [\ion{O}{iii}] flux and host galaxy
light, for example.

Nonetheless, in summary, we have tracked strong evolution in the
$\jotwelve$ broad H$\beta$ profile --- as quantified by large changes
in $R_{\rm H\beta/[\ion{O}{iii}]}$ and $R_{\rm BL/NL}$, and evolution
across subtype assignments --- both from 2002 to the low-continuum
state in January 2022, and during 2022--2024 as the continuum
rebrightened. This evolution is qualitatively consistent with that
observed in multiple other CLAGN, even on timescales as short as 
months, such as 1ES~1927+654 \citep{Trakhtenbrot19}, NGC~5548
\citep{Shapovalova09}, and Mkn~1018 \citep{Lu25}.

\subsection{Estimation of black hole mass}  \label{sec:MBHestimate}

We can obtain an estimate for the black hole mass, $M_{\rm BH}$, using
the width of the broad H$\beta$ line and the empirically derived
dependence of BLR H$\beta$ weighted emission radius, $R_{\rm BLR}$, on
optical \citep{Bentz09,Bentz13} and/or UV \citep{KilerciEser15}
luminosity.  To minimize the impact that the dip-then-recovery
transient event had on the illumination and ionization structure of
the BLR during 2022, we consider the average of widths of the broad
H$\beta$ Gaussian component during spectra \#12--19, which are likely
more representative of $\jotwelve$ in a persistent,
nontransient-event state. The H/ESO and 6dF archival spectrum were not
flux calibrated, so we do not consider them here.

We find an average flux density at 5100~$\AA$ (rest frame),
$F_{5100\AA}$ = $(5.1\pm0.3)\times10^{-16}$ \ecgsA.  For a luminosity
distance of 443 Mpc, the monochromatic optical luminosity
${\lambda}L_{5100\AA} = (6.1\pm0.4)\times10^{43}$ erg~s$^{-1}$.  We
apply the best-fitting ${\lambda}L_{5100\AA}-R_{\rm BLR}$ relation
of \citet{Bentz13}, log($R_{\rm BLR}$, lt-dy) = $(1.56\pm0.02)$ +
$(0.55\pm0.03)$log(${\lambda}L_{5100\AA}$/ ($10^{44}$~erg~s$^{-1}$)).
We obtain $R_{\rm BLR} =(7.1\pm0.8)\times10^{16}$~cm = $27\pm3$ lt-dy.
The average FWHM of the broad H$\beta$ Gaussian was
$5930\pm240$~km~s$^{-1}$.  Using the best-fitting relation between
virial factor $f$ and FWHM from \citet{MejiaRestrepo18}, we obtain
$f=0.8$. This yields $M_{\rm BH} = f R_{\rm BLR} v_{\rm FWHM}^2 G^{-1}
= (1.4\pm0.6)\times10^8 \Msun$, where we have increased the
uncertainty by 0.5~dex given the scatter in the relationship between
reverberation-mapped masses and single-epoch mass
estimation \citep[e.g.,][]{VP06}.

\section{Discussion}  \label{sec:Discussion}

\subsection{Review of main observational results}

\begin{itemize}

\item Thanks to eROSITA's all-sky X-ray surveys, we detected a dip in
the soft X-ray (0.5--2~keV) flux in $\jotwelve$ by a factor of
$17.0^{+16.3}_{-7.2}$ over 18 months, from eRASS2 to eRASS5.  

\item Our follow-up campaign tracked the subsequent recovery from the continuum
dip in the X-ray band, and tracked a concurrent increase in the UV
and optical bands.  Compared to the lowest fluxes measured, in January
2022 (XM1), the soft X-ray flux rose by a factor of roughly 9 by
2023--2024 (average of all data points from December 2022 -- April 2024).
Concurrently, the far-UV flux increased by a factor of roughly 5.

\item The IR continuum as traced with \textit{WISE} indicates a stable
  accretion rate through 2019, but also shows a flux dip starting in
  $\sim$2020 and a subsequent flux recovery during 2022--2024.  The
  variable component of IR continuum emission in Seyferts typically
  originates in warm circumnuclear dust that reprocesses disk and
  coronal photons. The IR variability behavior in $\jotwelve$ and the
  lack of a significant neutral X-ray-obscuring gas column along the
  line of sight both thus strongly favor a scenario where the
  continuum behavior is intrinsic to the emission disk and corona, as
  opposed to being caused by the transit of a dusty, compact obscuring
  cloud across the line of sight to the inner disk and corona.

\item A thermal Comptonization model (\textsc{fAGNSED}) can
  successfully describe all SEDs well. The value of $\dot{m}_{\rm
    Edd}$ increases by a factor of 7 over 3 years, from 0.0036 to
  0.025. There is no strong evidence for evolution in the warm corona
  temperature nor in the parameters of the hot corona.

\item Our optical spectral monitoring program uncovered significant
  evolution in the broad Balmer fluxes and profiles that tracked the
  continuum flux variability over timescales of months--years.
  $\mathcal{R}_{\rm H\beta/[\ion{O}{iii}]}$ ($\mathcal{R}_{\rm
    BL/NL}$) decreased by a factor of 2--4 (5--6) from 1993 and 2002
  to the low-flux state in January 2022.  As continuum flux increased
  during 2022--2024, $\mathcal{R}_{\rm H\beta/[\ion{O}{iii}]}$
  ($\mathcal{R}_{\rm BL/NL}$) increased by a factor of roughly 6 (5).
  Under the \citet{Runco16} subtype classification, $\jotwelve$ was
  subtype 1.0 in the archival spectra, 1.8 in January 2022, and back to
  1.0 by mid-2022.

\item The broad H$\beta$ and H$\alpha$ profiles can each be modeled as
  the sum of a broad (width $\sigma$ typically 30--50~$\AA$), slightly
  redshifted Gaussian plus a double-peaked diskline component.  The
  best-fitting diskline component has an inclination of 10--15$^{\rm
    \circ}$, and an inner radius typically of 600--1500 $R_{\rm g}$ (with
 the  outer radius assumed fixed at 5000 $R_{\rm g}$).  The diskline's
  fractional contribution to the total observed profile, $F_{\rm DL}$,
  is low during the continuum dip: 0.1--0.2 during spectra \#3--4 for
  both H$\beta$ and H$\alpha$.  However, as the overall continuum flux
  recovers, $F_{\rm DL}$ increases to roughly 0.4 for H$\beta$ and 0.6
  for H$\alpha$, and we find significant correlations between $F_{\rm
    DL}$ and UV continuum fluxes for both H$\beta$ and H$\alpha$.

\end{itemize}

\subsection{Mechanisms driving evolution in accretion rate }

$\jotwelve$ is yet another CLAGN where a change in accretion rate seems
the most likely driver; as mentioned above, a change in obscuration
would not cause mid-IR luminosity to dip and recover concurrently to the X-ray and optical/UV bands.
Using the IR flux as a proxy for optical/UV continuum flux prior to
2022, we can assign a time of 
2--3 years for the optical/UV luminosity drop.  The time for 
optical/UV luminosity to increase
can also be 2--3 years, but constraints are limited by the length of
our campaign.  Meanwhile, the X-ray dip and recovery timescales are
shorter, but difficult to define given the limited cadence of the
eRASS and WISE scans, and the difficulty in visually separating
short-term rapid variability from longer-term trends.  We define
$t_{\rm X,drop}$ $\sim$ 6~months and $t_{\rm X,incr.}$ $\sim$
3~months, with the caveat that these may be lower limits (values up to
18--24 months are not implausible).  Based on the X-ray and IR fluxes,
though, it seems that by roughly late 2024, both disk and coronal
luminosities had returned to the same flux levels as in
$\sim$2019--2020.

To investigate the likely mechanism(s) driving the accretion rate dip
and recovery, we can review the most relevant physical timescales for
an accretion disk around a $1.4\times10^8 \Msun$ SMBH.  For most
CLAGN and transient events observed to date, the associated timescales of
the event are usually not compatible with the viscous timescale for a
geometrically thin disk (scale height $H/R \ll 1$), on the order of 
millennia or longer,  nor with the light-crossing time.  The
Keplerian orbital timescale is $t_{\rm Kepl} = 2\pi [ r^3 / (GM_{\rm
    BH}) ]^{1/2}$.  The thermal timescale is $ t_{\rm th} =
1/(\alpha) [ r^3 / (GM_{\rm BH}) ]^{1/2}$, where
    $\alpha$ is the local effective viscosity parameter.  The viscous
  accretion timescale is $ t_{\rm visc} = t_{\rm th} (H/R)^{-2}$, and
  approaches $t_{\rm th}$ only as $H/R$ approaches 1.

We consider the timescales in the annulus from
$R \sim 30$~$R_{\rm g}$ to $\sim 100~R_{\rm g}$, 
the respective expected flux-weighted radii of far-UV and optical thermal continuum
emission in the disk, using Eq.~1 of \citet{Edelson19}. The Keplerian timescales
$t_{\rm Kepl}$ in this region range from 8 to 50~d, somewhat shorter than  
$t_{\rm X,drop}$ and $t_{\rm X,incr.}$. 
For the thermal timescales, however, arbitrarily chosen values of $\alpha \sim$
0.01--0.04 yield values of $t_{\rm th}$ of order 120--150~d, similar
to $t_{\rm X,drop}$ and $t_{\rm X,incr.}$.  These values of $\alpha$
are roughly consistent with the range ($\sim$0.01--0.1) expected
based on various simulations of accretion disks \citep{King2007},
for instance where disks are supported by magnetic pressure
\citep{Jiang2019} and where MRI-induced turbulence drives angular
momentum transport \citep{Simon2012,Ju2017}. However, exact values of
$\alpha$ can depend on magnetization strength and field geometry, for example
\citep[e.g., whether vertical components exist;][]{Hawley1995,Ju2017}.
For all choices of $\alpha$, within these broad ranges, values of
$t_{\rm th}$ (and $t_{\rm visc}$ for a geometrically thick disk) in
the region of the inner accretion disk where optical/UV continuum
emission peaks seem most relevant, as is the case with many
changing-look and flaring AGN \citep[e.g.,][]{Krishnan24}.  

Below we consider some candidate driving mechanisms in an attempt to
explain the  behavior in $\dot{m}_{\rm Edd}$ and the observed multiband
continuum variability:

\begin{itemize}

\item 1) The first possibility is a mechanism that could temporarily increase the local
  accretion timescale in the optical- and UV-emitting regions of the
  inner disk by temporarily reducing disk scale height $H/R$.  As one
  example, if the accretion disk scale height is supported by magnetic
  pressure \citep{Dexter19}, some mechanism could act to temporarily
  decrease that pressure.

Another possibility is that the drop and recovery in $\dot{m}_{\rm
  Edd}$ were caused by, respectively, a propagating cold front and
warm front within the disk. Such fronts were invoked by \citet{Ross18}
to explain a 20-year flux dip and recovery in a $z$=0.38 quasar.
The cold (warm) front would decrease (increase) $H/R$ locally as it
propagates.  The front propagation timescale can be written as $t_{\rm
  front}$~$\sim$~$t_{\rm th}$/($H/R$) \citep{Ross18}.

For $r$ = 30~$R_{\rm g}$, arbitrary values of $(H/R) \sim 0.05$ and
$\alpha \sim 0.05-0.10$ can yield values of $t_{\rm front}$ of 0.8~--~1.5 years,
although for these same values, $t_{\rm front}$ would be a factor of 6
longer near 100~$R_{\rm g}$. 
Ideally, there could be an optical/UV lag
that could indicate the front propagation speed and direction, but our
data are insufficient to detect or constrain such a lag.  As the cold
front causing the luminosity dip reaches the UV-producing region of
the inner disk, the reduction in seed photons from the disk reduces
the Comptonized X-ray emission in both the hot and warm coronae.
Another speculative possibility is that if heating   the hot and/or
warm coronae is facilitated by magnetic fields, then the cold front
might also disrupt these fields, reducing the luminosity output of the
coronae.  The low value of $t_{\rm X,incr.}$ could thus be influenced
by the time for the coronae to reconstitute itself.  However, our data
preclude any constraints on time lags between optical/UV and X-ray
bands (since the UV did not clearly capture the point of lowest UV
flux), so the exact sequence of events remains speculative.

\item 2) A second possibility is a temporary local (impacting at least the radii of peak
optical to far-UV emission) decrease, then increase, in viscosity
parameter $\alpha$, perhaps driven by changes in magnetic stress.

\item 3) There could be a strong stochastic variation in the local mass supply.
  Inwardly propagating variations in local mass accretion rate have
  been used to explain continuum variability in Seyfert AGN across a
  wide range of timescales \citep{Lyubarskii97,Arevalo06,Ingram11}.
  However, the change in X-ray flux over an 18-month timescale is
  greater here than for most (90$\%$) Seyfert 1 galaxies in the \textit{Rossi
    X-ray Timing Explorer} archive \citep[e.g.,][]{ME04} likely
  categorizing the variability event in $\jotwelve$ as an outlier.

\item 4) There could be an outflowing wind, such as a magnetically driven wind
\citep{Feng21}, which can remove some angular momentum from the disk.  It is
conjecturally possible that there had been such a wind in $\jotwelve$  whose
temporary suspension decreased the
removal of angular momentum, thus increasing the inflow
timescale. However, this situation is somewhat speculative and
contrived as we have no evidence (e.g., from X-ray spectroscopy) for
such an outflow.

\end{itemize}

As a final note, we consider flux ratios taken from the SEDs that use
\textit{XMM-Newton} data, hereafter XM2/XM1 and XM3/XM1, and listed in
Table~\ref{tab:SED_RATIO_TABLE}.  We note that the soft X-ray band
typically varies more than the hard band by a factor
    of 1.5--1.9.  In addition, we note that the normalization of the
    \text{CompTT} component increases by factors of 2.5--3.4 faster
    than the normalization of the hard X-ray power-law component
    (Table~\ref{tab:XMparms}).
These values support the notion that the warm corona can be driven by something
other than the hot corona. The accretion disk is a likely candidate,
given previously observed strong correlations between these two
components in persistently accreting AGN such as Mkn~509
\citep[e.g.,][]{Mehdipour11}.  However, it is also interesting to note
that in $\jotwelve$ the warm corona flux is also more variable than the UV
bands, qualitatively similar to behavior observed in the CLAGN
Mkn~1018 \citep{Saha25a}.  Unlike Mkn~1018, $\jotwelve$ likely did not
transition to an ADAF (see above). Nonetheless, the extreme variations
in warm corona output suggest significant changes in the energetics
and/or structure of $\jotwelve$'s warm corona, such as radial
expansion in the context of the \textsc{fAGNSED} model
\citep{Hagen24}. There could also be a large amount of heating at the
photosphere of the disk \citep{Palit24}; as one possibility, in
magnetically dominated disks, a rising magnetic dynamo could generate
such heating \citep{Gronkiewicz23}.

There are other potential mechanisms that seem disfavored for 
$\jotwelve$ given the multiband variability behavior observed: 

\begin{itemize}

\item A magnetic inversion in a magnetically arrested disk, following
  \citet{Scepi21}: An advection event travels inward, increasing
  UV/optical luminosity, and then temporarily destabilizing the X-ray
  corona, which is powered by magnetic fields via the Blandford-Znajek
  effect.  
  However, this scenario predicts a sequence of variability
events different to that observed here. It predicts a rise in
  UV/optical flux occurring well before the X-ray dip; no such rise
  is evident from the ATLAS or \textit{WISE/NEOWISE} light curves. The scenario also
  predicts that after the X-ray dip, disk luminosity gradually
  decreases, not increases as observed in $\jotwelve$.

\item Radiation-pressure \citep{Lightman74,Sniegowska20} or
  hydrogen-ionization \citep{Noda18} instabilities: Either of these
  two mechanisms can operate in the accretion disk and generate
  heating fronts that temporarily increase $H/R$.  However, such
  mechanisms are typically invoked to explain luminosity flares driven
  by the temporary increases in local accretion rate in black hole
  X-ray binaries or changing-look/flaring AGN
  \citep{Krishnan24,Saha25b}.  The predicted light curve profiles
  \citep[e.g.,][]{Sniegowska20,Sniegowska23} do not resemble the flux
  dip displayed here.

\item A switch to a radiatively inefficient (RIAF) or
  advection-dominated accretion flow \citep[ADAF;
    e.g.,][]{Narayan94,Narayan98} in the inner disk: The value of
  $\dot{m}_{\rm Edd}$ during the dip is indeed near the critical value
  below which such accretion structures can form in both stellar-mass
  and supermassive BHs \citep[e.g.,][and references therein; see also
    \citealt{Saha25a} for the CLAGN Mkn~1018]{Yang15}.  However, if a
  viscous geometrically thin disk switches to an RIAF/ADAF flow, then
  UV/optical emission should be suppressed by several orders of
  magnitude \citep[][with supporting observational evidence in quasars
    by \citealt{Hagen24}]{Narayan98, Abramowicz02, Abramowicz13}.  In
  $\jotwelve$, the optical/UV varies by less than the X-rays do, at
  least during the flux recovery.  In addition, it is not clear that
  such a transformation out to $\sim$100 $R_{\rm g}$ can occur on a
  timescale of only six months to a year for a $10^8~\Msun$ black
  hole.

\item An impact by a secondary body, such as the debris from a
  tidally disrupted star: Such impacts onto the inner regions of
  accretion disks of AGN have been invoked to explain the behavior in
  several CLAGN and transient AGN (e.g., \citet{Blanchard17} and
  \citet{Ricci20}).  The impact creates shocks that remove angular
  momentum in the inner disk interior to the impact point, rapidly
  depleting it \citep[e.g.,][]{Chan19}, and disrupting the processes
  that support the corona \citep{Ricci20}.  However, during such
  events, kinetic energy from the impacting debris leads to a large
  amount of energy dissipation and increased
  optical/UV radiation in the disk, although a
  substantial fraction of that radiation can be trapped in the disk
  and advected into the black hole \citep{Chan19}.

\end{itemize}

\subsection{The origins and evolution of the broad Balmer emission components in
$\jotwelve$}

Broad H$\beta$ in local Seyferts can display shifts between being
dominated by single- or double-peaked on timescales of years.  One
prominent example is that of NGC~5548 \citep[e.g.,][]{Shapovalova09},
whose broad H$\beta$ profile can display major changes on timescales
of 1--2 years. Its profile is sometimes dominated by a single
Gaussian, other times by a double-peaked diskline-like profile, and
the relative intensities of the double peaks can vary. Another example
is that of the CLAGN Mkn 1018: \citet{Lu25} report that Balmer
profiles are single-peaked toward   higher values of
\mdotEdd, when the source is subtype 1.0--1.2, and double-peaked toward
lower values of \mdotEdd, when the source is subtype 1.5--1.8,
qualitatively consistent with the observations of \citet{Elitzur14}
for samples of Seyferts.  In addition, distinct kinematic regions are
inferred from the broad Balmer profile shape and/or variability
characteristics of some Seyferts \citep[e.g.,][]{Shapovalova09,Hu20}.
These distinct BLR structures can sometimes be linked to other
individual accretion structures, such as the inner dusty torus or a
disk-like component \citep{Nagoshi24}.  Anistropic illumination of
ionizing radiation onto the BLR may also yield distinct BLR regions
with different dynamics and radial locations, for example \citep{Wang14}.

The broad H$\beta$ Gaussian component in $\jotwelve$ is consistent
with gas located at a radial distance of 27$\pm$3~lt-dy
(Sect.~\ref{sec:MBHestimate}); the symmetry of this component supports
the notion that the gas is virialized.  The best-fitting models
yielded mild redshifts for the peak wavelengths of both H$\beta$ and
H$\alpha$ of
600--800~km~s$^{-1}$ across all spectra (including the archival spectrum).
It is a strong possibility that nonvirial, infall motion can
contribute to the kinematics of the BLR, as inferred from
velocity-resolved reverberation studies of some Seyferts and quasars
\citep[e.g.,][]{Grier13,Pancoast14,Du16,Bao22}.

Meanwhile, $\jotwelve$'s broad Balmer profiles additionally contain a
double-peaked profile. Double-peaked line profiles are also not
uncommon to observe
\citep[e.g.,][]{Eracleous94,Eracleous03,Gaskell99}, particularly in
type 1.8--1.9 Seyferts \citep{Elitzur14}, and with variability in
profile shapes and/or peak heights occurring on timescales of years
\citep{Shapovalova04,Lewis10}. These components can often be modeled
as a diskline from a static, flattened, annulus-like geometry
\citep{Chen89,Eracleous95} as we have done for $\jotwelve$. As a
reminder, our best-fitting model yielded an inner radius of
600--1500~$R_{\rm g}$ = 5--12 lt-dy, with an outer radius of
5000~$R_{\rm g}$ (41~lt-dy) assumed. The inferred radial distances of
the diskline and broad Gaussian are commensurate; however, it is
unclear whether the two components are physically distinct or connected.

A flattened, annulus-like geometrical structure is also possible in a
failed radiatively accelerated dusty outflow \citep[FRADO;][]{CH11},
in which dusty clouds are launched from the disk, and are driven by
radiation pressure acting on the dust.  The BLR geometry is thus
determined by the aggregate of clouds and their trajectories
\citep{Naddaf21}.  It is quite intriguing to note that a radius of
5--12 lt-dy is consistent with the annulus in the disk where dusty
outflows can be generated \citep{CH11}: Using Eq.~1 of
\citet{Edelson19}, and assuming log($\dot{m}_{\rm Edd}$) = $-$1.6,
$T=1000$~K occurs at a radius of 9 lt-dy. \citet{Naddaf22}
demonstrated that emission from FRADOs can yield double- or
single-peaked broad Balmer emission profiles depending on the system
parameters. For instance, 
for a given black hole mass of $10^8\Msun$,
    diskline-like distributions of clouds having relatively low scale heights
    can be preferentially generated toward values
    of log($\dot{m}_{\rm Edd}$) $\sim$ --1 to --2, depending on the
    metallicity, and yielding double-peaked Balmer profiles.  
    This range in $\dot{m}_{\rm Edd}$ overlaps with the
    values we infer from SED fits, supporting the notion that the
    diskline component is present during all timescales and accretion
    rates sampled during our campaign.

\subsubsection{Evaluating whether the BLR components intrinsically vanished during the flux dip}  

Our multiband monitoring tracked $\jotwelve$ as log(\mdotEdd) climbed
from $-$2.4 during the flux dip  to $-$1.6.  This range overlaps
well with the distributions of log(\mdotEdd) recorded for several
samples of changing-look AGN and quasars, which generally peak around
values of log(\mdotEdd) roughly $-$2.2 to $-$1.5
\citep[e.g.,][]{Green22,Panda24,Zeltyn24}.

As mentioned in the Introduction, there is evidence to support the
disappearance of the BLR when the accretion rate drops below a
critical value, either in a statistical sense \citep{Elitzur09} or in
individual changing-look quasars \citep{Green22,Panda24}.  As
suggested by \citet{Elitzur09}, if the BLR is a wind fed by the
accretion disk, then accretion below the critical value means the
supply of gas from the disk into the BLR is insufficient to sustain
that wind.   \citet{Elitzur09} estimated that the critical value of bolometric
luminosity $L_{\rm Bol,crit}$ is $5\times 10^{39}$ ($M_{\rm BH}$ /
$10^7 \Msun$)$^{2/3}$ erg~s$^{-1}$.  In the case of $\jotwelve$, given
$M_{\rm BH} = 1.4\times10^8 \Msun$, the resulting critical value for
$L_{\rm Bol}$ is $3\times10^{40}$ erg s$^{-1}$.  However, from the SED fit to 
XM1, we estimate
$L_{\rm Bol}$ to be $6.4\times10^{43}$ erg s$^{-1}$ during the
flux dip, safely above the expected critical value.  The observed
spectral type changes, and the inferred BLR activity in $\jotwelve$ thus do
not seem connected to the notion of the presence or lack of the BLR
having a dependence on {\mdotEdd}, i.e., we did not observe that
$\jotwelve$ became a type~2 during its low-flux state.

\subsubsection{Explaining the evolution in diskline component emission } 

We now discuss the observed evolution in fractional
    contribution of the diskline component's flux to the total broad
    profile flux, parameterized via $F_{\rm DL}$.
In the context of the FRADO model, we can consider whether the increase in
$F_{\rm DL}$ could be due to an increase in covering factor as dusty gas
rises upward.  The FRADO simulations of \citep{Naddaf22} indicate that for a
$\sim$$10^8~\Msun$ BH with an accretion rate relative to Eddington of log($\dot{m}_{\rm Edd}$) $\sim$
$-1.5$, BLR clouds may reach heights of order 50~$R_{\rm g}$
($5.5\times10^9$~km).  
In the simplest most optimistic scenario, gas can uplift vertically
from the disk at a velocity $\lesssim$100~km~s$^{-1}$ for an accretion
rate of log($\dot{m}_{\rm Edd}$)~=~$-$1.5 \citep{Naddaf21}.  Over a
timescale of 1--1.5 years, gas can travel a distance of order
$3\times10^9$~km, so such a scenario is optimistically plausible for $\jotwelve$.
However, simulations of FRADOs so far have not considered the effects
of sudden changes in luminosity. 
It is thus not clear if a two-year decline in
bolometric luminosity is sufficient to allow clouds to fall back
toward the disk faster than they are generated, 
thus enabling a low scale height for clouds by the time of the low-X-ray flux state.

An alternate explanation for the evolution in $F_{\rm DL}$ is that the
BLR components remain dynamically static while the $>$13.6~eV 
illumination intercepted from the coronae evolves. 
We note that the coronae are not the only source of ionizing photons,
but we consider here the potential impact of coronal evolution on the inner BLR.
We consider that the double-peaked BLR component
originates in a vertically  and radially static diskline-like geometry with a low scale
height just above the surface of the accretion disk.  As the source
luminosity increases, the physical extent of warm
    and/or hot coronae, each a source of $>$13.6~eV ionizing continuum
    emission, could increase.  For instance, the scale height of the
hard X-ray corona in the lamppost geometry could increase, as could
the scale height or radial extent \citep{Hagen24} of a
vertically extended warm corona as modeled in
\citet{Partington24}; the major changes in warm corona
    luminosity observed could qualitatively support this notion.
Alternately, the scale height of the inner disk itself could increase
as radiation pressure increases.  Such structural changes are feasible
on timescales of $\sim$1--2 years, which is longer than dynamical
timescales for the innermost accretion disk.  Finally, the disk may
contain a warp or its surface may not be completely smooth.  As a
consequence of any of these possibilities, the fraction of total
ionizing photons emitted by the hard corona and particularly the warm
corona intercepted by the diskline component can increase as total
luminosity and the scale heights of the innermost disk and/or coronae
increase.  Meanwhile, there can be a vertically extended BLR component
that yields the broad Gaussian, but it intercepts ionizing photons in
both low- and high-flux states with a static covering fraction as seen
from the coronae.

A final possible explanation for the evolution in $F_{\rm DL}$ is an
obscuring wind launched from the inner disk, interior to the diskline
component of the BLR, and which filters the $>$13.6~eV continuum
emission traveling from the coronae to the diskline
\citep{Dehghanian19} only during the low-flux state. Such a wind could
be launched and driven by magnetocentrifugal processes and
radiation driving (e.g., \citealt{Proga00,Progaetal00,Everett05}). 
However, such a scenario may be contrived given that the
geometry of the wind would have to be such that it does not impact the line of
sight from the coronae to the region of the BLR producing the broad
Gaussian (e.g., the wind has a lower scale height than the gas
producing the broad Gaussian emission component) and not intersecting
the line of sight from the coronae to us, so as to avoid X-ray
obscuration.  Such a wind would have to be prominent during the
low-flux state only, and then later dissipate, or have its number
density drop below $\sim$$10^{11-12}$~cm$^{-3}$ such that it no
longer absorbs $>$13.6~eV photons \citep{Dehghanian19}. However, it is
not clear what mechanism could cause such a wind to be launched and/or
be sufficiently dense only in the low-flux state, so this scenario does not seem
likely.

\section{Conclusions}   \label{sec:Conclusions} 

This paper contributes to eROSITA's window into time-domain
astrophysics studies of a variety of accretion channels powering
extragalactic nuclear transient
events \citep[e.g.,][]{Malyali21,Grotova25,Baldini25}, including X-ray
detections of changing-look AGN, namely those undergoing major changes
in $\dot{m}_{\rm Edd}$ \citep{Homan23,Krishnan24,Saha25b} and those
undergoing changes in line-of-sight obscuration
\citep{Markowitz24}. It also highlights the utility
of multiwavelength photometric and spectroscopic follow-up programs
in constraining the behavior of the X-ray corona, accretion disk, and
BLR during such changing-look events.

eROSITA detected a drop in soft X-ray (0.5--2~keV) flux in the
$z=0.096$ Seyfert $\jotwelve$ by a factor of 17 over 18 months, from
June 2020 to January 2022; this dip was accompanied by a $\sim$2-year
decrease in IR flux as tracked with \textit{NEOWISE}, with the
recorded minima of both bands occurring in early 2022.  Our three-year
follow-up campaign tracked how the X-ray flux recovered within only
3~months, while the optical, UV, and IR bands all recovered more
slowly, over $\sim$3 years; $\dot{m}_{\rm Edd}$ increased by a factor
of 7 during this time.  One possible explanation for the drop and
recovery in $\dot{m}_{\rm Edd}$ is the propagation of cool and warm
fronts across the optical/UV continuum emitting regions of the
accretion disk, as described in
\citet{Ross18}.  However, our campaign's optical/UV sampling was too
sparse to thoroughly test this hypothesis via detection of interband
lags or leads. If such fronts are relevant for explaining the behavior of
other CLAGN, then we encourage future optical/UV monitoring campaigns
to adopt a high cadence (e.g., a few days) to better test and
constrain this class of models.

$\jotwelve$ also displays strong evolution in both the strength and
profile shape of its broad Balmer emission lines.  The broad H$\beta$ line
dropped by a factor of roughly 4--6 from March 2002 (Sy~1.0, using the
\citealt{Runco16} classification) to January 2022 (Sy~1.8), during the continuum dip.
The broad H$\beta$ flux subsequently recovered by the same factor through
late 2024, regaining subtype~1.0 status within a few months after the
continuum dip.

We deconvolved the broad H$\beta$ and H$\alpha$ profiles into two
kinematic components.  A broad Gaussian, with velocity widths $\sigma$
typically 1800--3900~km~s$^{-1}$ (30--50~$\AA$), dominated the spectra
taken during the first six months after the continuum dip (January 2022
-- June 2022); it is consistent with virialized gas at a radial
distance of 27$\pm$3~lt-dy.  The second component is double-peaked,
and becomes more prominent in the total broad Balmer profiles starting
one year after the continuum dip.  It can be modeled as a diskline
(flat annulus) with an inner radius of $\sim$1000~$R_{\rm g}$, or
5~lt-dy.  The fractional contribution of the diskline to the total
profile, $F_{\rm DL}$, increases as UV flux and $\dot{m}_{\rm Edd}$
increase during our campaign.  One possible explanation is that the
diskline component has a low scale height, but as $\dot{m}_{\rm Edd}$
increases, there is an increase in the scale height and/or radial
extent of the hot and/or in particular the warm X-ray-emitting
coronae; the extreme variations in warm corona luminosity that we
observe could qualitatively support this notion.  This increase allows
a larger fraction of $>$13.6~eV photons to be intercepted by the
diskline component.  One can speculate if the increase in scale height
may be connected to the putative warm propagating front that may drive
the increase in $\dot{m}_{\rm Edd}$.  In the context of such
propagating fronts, this change in scale height could provide a
physical mechanism linking the change in accretion rate with the
evolution in BLR illumination.

It is curious to note that these inferences about the geometry and
location of the BLR components in $\jotwelve$ differ from
what \citet{Nagoshi24} concluded for the two-component BLR in the
changing-state quasar SDSS J125809.31+351943.0. That object's broad
H$\beta$ profile was also decomposed into a single- and double-peaked
component. However, its single-peaked BLR component is responsive to
continuum variations, and inferred to be located at distances
commensurate with the dusty torus; meanwhile, the stable diskline
component is inferred to be associated with the outer accretion disk.
This divergence in implied BLR geometries likely reflects a diversity
in individual kinematic components across changing-look
Seyferts and quasars. We hope that future high-cadence, multiwavelength
observations of changing-look Seyferts and quasars can uncover, in
addition to evidence for structural evolution in the inner disk,
additional multicomponent BLRs, and take advantage of strong and
rapid flux variability to study their individual components.

\begin{acknowledgements}
AM and TS acknowledge partial or full support from Narodowe Centrum Nauki
(NCN) grants 2016/23/B/ST9/03123 and 2018/31/G/ST9/03224.  AM also
acknowledges partial support from NCN grant 2019/35/B/ST9/03944.  DH
acknowledges support from DLR grant FKZ 50 OR 2003. MK is supported by
DFG grant KR 3338/4-1. SH is partly supported by the German Science
Foundation (DFG grant numbers WI 1860/14-1 and 434448349).
DAHB acknowledges support from the National Research Foundation.

AM thanks Dr.\ Gergely Hajdu for assistance in planning LCOGT observations, and
Prof.\ W{\l}odzimierz Klu\'zniak and Dr.\ Debora Lan\v{c}ov\'a
for useful discussions on accretion disk properties.

This work is based on data from eROSITA, the soft X-ray instrument
aboard SRG, a joint Russian-German science mission supported by the
Russian Space Agency (Roskosmos), in the interests of the Russian
Academy of Sciences represented by its Space Research Institute (IKI),
and the Deutsches Zentrum für Luft- und Raumfahrt (DLR). The SRG
spacecraft was built by Lavochkin Association (NPOL) and its
subcontractors, and is operated by NPOL with support from the Max
Planck Institute for Extraterrestrial Physics (MPE). The development
and construction of the eROSITA X-ray instrument was led by MPE, with
contributions from the Dr. Karl Remeis Observatory Bamberg \& ECAP (FAU
Erlangen-Nuernberg), the University of Hamburg Observatory, the
Leibniz Institute for Astrophysics Potsdam (AIP), and the Institute
for Astronomy and Astrophysics of the University of Tübingen, with the
support of DLR and the Max Planck Society. The Argelander Institute
for Astronomy of the University of Bonn and the Ludwig Maximilians
Universität Munich also participated in the science preparation for
eROSITA.

The eROSITA data shown here were processed using the eSASS software
system developed by the German eROSITA consortium.

Based on observations obtained with XMM-Newton, an ESA science mission
with instruments and contributions directly funded by ESA Member
States and NASA.  The authors thank the \textit{XMM-Newton} director
for approving the DDT observations, and the \textit{XMM-Newton}
operations team for executing them.  This research has made use of
data and/or software provided by the High Energy Astrophysics Science
Archive Research Center (HEASARC), which is a service of the
Astrophysics Science Division at NASA/GSFC.

This research has made use of the XRT Data Analysis Software (XRTDAS)
developed under the responsibility of the ASI Science Data Center
(ASDC), Italy.

Part of this work is based on archival data, software or online services
provided by the Space Science Data Center - ASI.

This work was supported in part by NASA through the NICER mission and
the Astrophysics Explorers Program. NICER data used in this work were
gathered under a Guest Observer (GO) approved programme.

Some of the observations reported in this paper were obtained with the
Southern African Large Telescope (SALT) under programme 2021-2-LSP-001
for transients (PI: Buckley), conducted within the eROSITA M.O.U.\ as
part of the eROSITA-SALT Transient collaboration, as well as under
programmes 2021-2-MLT-003, 2023-1-MLT-001, and 2024-1-MLT-002
(PI: Markowitz).  Polish
participation in SALT is funded by grant No.\ MEiN nr 2021/WK/01.

This paper uses observations made using the South African Astronomical
Observatory (SAAO).

The ATLAS science products have been made possible through the
contributions of the University of Hawaii Institute for Astronomy, the
Queen’s University Belfast, the Space Telescope Science Institute, the
South African Astronomical Observatory, and The Millennium Institute
of Astrophysics (MAS), Chile.

This work makes use of observations from the Las Cumbres Observatory
global telescope network.

This publication makes use of data products from the Wide-field
Infrared Survey Explorer, which is a joint project of the University
of California, Los Angeles, and the Jet Propulsion
Laboratory/California Institute of Technology, funded by the National
Aeronautics and Space Administration.  This publication also makes use
of data products from NEOWISE, which is a project of the Jet
Propulsion Laboratory/California Institute of Technology, funded by
the Planetary Science Division of the National Aeronautics and Space
Administration.

This research has made use of the NASA/IPAC Extragalactic Database
(NED), which is funded by the National Aeronautics and Space
Administration and operated by the California Institute of Technology.

The Skynet Robotic Telescope Network is supported by the National
Science Foundation, the Department of Defense, the North Carolina
Space Grant Consortium, and the Mount Cuba Astronomical Foundation.

This work has made use of data from the European Space Agency (ESA) mission
{\it Gaia} (\url{https://www.cosmos.esa.int/gaia}), processed by the {\it Gaia}
Data Processing and Analysis Consortium (DPAC,
\url{https://www.cosmos.esa.int/web/gaia/dpac/consortium}). Funding for the DPAC
has been provided by national institutions, in particular the institutions
participating in the {\it Gaia} Multilateral Agreement.

This research has made use of ISIS functions (ISISscripts) provided by
ECAP/Remeis observatory and MIT
(http://www.sternwarte.uni-erlangen.de/isis/).

\end{acknowledgements}

\bibliographystyle{aa}
\bibliography{mybib}

\begin{appendix}

\section{Summary of X-ray, optical/UV, and optical spectroscopic observations}

Below, we list observation logs summarizing the X-ray observations (Table~\ref{tab:Xobs}),
the optical/UV space- and ground-based photometric observations (Table~\ref{tab:OUVobs}), and
the optical spectroscopic observations (Table~\ref{tab:optspeclog}).

\begin{table*}[!h]    
\caption[]{X-ray observation log of \jotwelve}
        \centering
\label{tab:Xobs}
        \begin{tabular}{lccccc} \hline\hline
Telescope     & ObsID            & Date                  & Date    & Exposure &  Abbr. \\
              &                  &                       & (MJD)   & (ks)     &        \\ \hline
\textit{XMM} EPIC &  (slew)   & 1 Jan.\ 2013 & 56293.9 & 0.008 & \\
eROSITA/eRASS1 &              & 29--31 Dec.\ 2019 & 58847.5 & 0.14 & eR1 \\
eROSITA/eRASS2 &              & 29--30 Jun.\ 2020 & 59029.6 & 0.11 & eR2 \\
eROSITA/eRASS3 &              & 30--31 Dec.\ 2020 & 59213.9 & 0.11 & eR3 \\
eROSITA/eRASS4 &              & 2--3 Jul.\ 2021   & 59398.3 & 0.12 & eR4 \\
eROSITA/eRASS5 &              & 3--4 Jan.\ 2022   & 59583.3 & 0.12 & eR5 \\
\textit{XMM} EPIC & 0862770901 & 25--26 Jan.\ 2022 & 59604.9 & 32.4, 36.9, 36.9 & XM1 \\
\textit{Swift} XRT & 00015025001 & 23 Feb.\ 2022 & 59633.3 & 2.7 & Sw1 \\
\textit{Swift} XRT & 00015025002 & 29 March 2022 & 59667.7 & 2.9 & Sw2 \\
\textit{Swift} XRT & 00015025003 & 26 Jul.\ 2022 & 59786.3 & 1.7 & Sw3 \\
\textit{Swift} XRT & 00015440001 & 21 Dec.\ 2022 & 59934.3 & 4.4 & Sw4 \\
\textit{XMM} EPIC &  0903991901 & 26 Jan.\ 2023 & 59970.2 & 8.3, 11.7, 11.7 & XM2 \\
NICER             & (see Appdx.~\ref{sec:appdx_nicer}) & 15 Feb.\ 2023 -- 23 May 2023 & 59990--60087 & 48.5 & N\
ICER \\
\textit{Swift} XRT & 00015025004 & 22 Mar.\ 2023 & 60025.5 & 1.9 & Sw5 \\
\textit{Swift} XRT & 00016018001 & 14 May   2023 & 60078.6 & 2.8 & Sw6 \\
\textit{Swift} XRT & 00016142001 & 23 Jul.\ 2023 & 60148.4 & 2.1 & Sw7 \\
\textit{Swift} XRT & 00016142002 & 26 Jul.\ 2023 & 60151.6 & 0.9 & Sw8 \\
\textit{Swift} XRT & 00016409001 &  6 Dec.\ 2023 & 60284.3 & 1.6 & Sw9 \\
\textit{Swift} XRT & 00016409002 & 10 Dec.\ 2023 & 60288.1 & 1.4 & Sw10 \\
\textit{Swift} XRT & 00097539001 & 26 Apr.\ 2024 & 60426.7 & 0.5 & Sw11 \\
\textit{Swift} XRT & 00097539002 &  1 Aug.\ 2024 & 60523.3 & 0.6 & Sw12 \\
\textit{Swift} XRT & 00097539003 &  5 Aug.\ 2024 & 60527.0 & 0.2 & Sw13 \\
\textit{Swift} XRT & 00097539004 & 22 Aug.\ 2024 & 60544.7 & 2.1 & Sw14 \\
\textit{Swift} XRT & 00097539005 & 25 Aug.\ 2024 & 60547.1 & 0.9 & Sw15 \\
\textit{XMM} EPIC  & 0903992001  & 22 Jan.\ 2025 & 60697.4 & 28.1, 29.9, 30.0  & XM3 \\
\hline \end{tabular}
\tablefoot{All MJD dates refer to the midpoint of the observation.
Exposure refers to good time after
screening. For eRASS, exposure times have been corrected for vignetting.
For \textit{XMM-Newton}, the three exposure values refer to
pn, MOS1, and MOS2, respectively.
For NICER, there were 28 observations from 15 February 2023 (MJD 59990) to 23 May 2023 (MJD 60087);
we combined all observations for spectral analysis, and here
we list the summed exposure. The NICER exposure-weighted MJD midpoint was 60034.4.}
\end{table*}

\begin{table*}      
\caption[]{Optical/UV photometric observations of \jotwelve}
        \centering
\label{tab:OUVobs}
        \begin{tabular}{lcccc} \hline\hline
Observation \& & Date & Date     & Filters & Exposure   \\
Instrument     &      & (MJD)    &        & (s)        \\ \hline
\textit{XMM-Newton} OM (XM1) &  25--26 Jan.\ 2022 & 59604.9 &  V, B, U, UVW1, UVM2 & 4400, 4400, 4400, 8800, 26400 \\
PROMPT-6 (P1)      &  6 Feb.\ 2022 & 59616.4 &       B & 600 \\
PROMPT-6 (P2)      &  7 Feb.\ 2022 & 59617.4 &    V, B & 600, 600 \\
PROMPT-6 (P3)      &  9 Feb.\ 2022 & 59619.4 &       B & 600 \\
PROMPT-6 (P4)      & 11 Feb.\ 2022 & 59621.4 &    V, B & 600, 600 \\
PROMPT-6 (P5)      & 12 Feb.\ 2022 & 59622.4 &    V    & 600 \\
PROMPT-6 (P6)      & 13 Feb.\ 2022 & 59623.4 & R       & 600 \\
PROMPT-6 (P7)      & 19 Feb.\ 2022 & 59629.1 & R, V, B & 600, 600, 600 \\
\textit{Swift} UVOT (Sw1) & 23 Feb.\ 2022 & 59633.3 & V, B, U, UVW1, UVM2, UVW2 & 138, 138, 138, 415, 1085, 691\\
PROMPT-6 (P8)      & 23 Feb.\ 2022 & 59633.1 & R, V, B & 600, 600, 600 \\
PROMPT-6 (P9)      & 28 Feb.\ 2022 & 59638.3 & R, V, B & 600, 600, 600 \\
LCOGT/LSC (LCO1)   &  1 Mar.\ 2022 & 59639.3 & I, R, V, B &  180, 240, 240, 360 \\
PROMPT-6 (P10)     & 11 Mar.\ 2022 & 59649.1 & R, V, B & 600, 600, 600 \\
PROMPT-6 (P11)     & 20 Mar.\ 2022 & 59658.1 & R, V, B & 600, 600, 600 \\
PROMPT-6 (P12)     & 27 Mar.\ 2022 & 59665.3 & R, V, B & 600, 600, 600 \\
\textit{Swift} UVOT (Sw2) & 29 Mar.\ 2022 & 59667.7 & V, B, U, UVW1, UVM2, UVW2 & 148, 148, 148, 446, 1210, 746\\
PROMPT-6 (P13)     &  7 May 2022   & 59706.1 & R, V, B & 600, 600,600 \\
LCOGT/LSC (LCO2)   &  5 May   2022 & 59704.2 & I, R, V, B &  180, 240, 240, 360 \\
LCOGT/LSC (LCO3)   & 13 Jun.\ 2022 & 59743.0 & I, R, V, B &  180, 240, 240, 360 \\
LCOGT/CPT (LCO4)   & 13 Jul.\ 2022 & 59773.8 & I, R, V, B &  180, 240, 240, 360 \\
\textit{Swift} UVOT (Sw3) & 26 Jul.\ 2022 & 59786.3 & V, B, U, UVW1, UVM2, UVW2 & 139, 139, 139, 276, 417, 553\\
\textit{Swift} UVOT (Sw4) & 21 Dec.\ 2022 & 59934.3 & V, B, U, UVW1, UVM2, UVW2 & 368, 368, 368, 739, 933, 1480 \\
LCOGT/CPT (LCO5)   &  29 Dec.\ 2022 & 59942.0 &  V, B  &  60, 80 \\
LCOGT/LSC (LCO6)   &  5 Jan.\ 2023 & 59949.3 & I, R, V, B &  180, 240, 240, 360 \\
\textit{XMM-Newton} OM (XM2) &      26 Jan.\ 2023 &  59970.2  & V, B, U, UVW1, UVM2 & 2200, 2200, 2200, 2200, 2200 \\
LCOGT/COJ (LCO7)   & 16 Feb.\ 2023 & 59991.7 & I, R, V, B &  180, 240, 240, 360 \\
LCOGT/LSC (LCO8)   & 22 Mar.\ 2023 & 60025.1 & I, R, V, B &  180, 240, 240, 360 \\
\textit{Swift} UVOT (Sw5) & 22 Mar.\ 2023 & 60025.5 &          UVW1, UVM2, UVW2 & 567, 637, 637 \\
\textit{Swift} UVOT (Sw6) & 14 May   2023 & 60078.6 &          UVW1, UVM2, UVW2 & 832, 951, 951 \\
LCOGT/LSC (LCO9)   & 24 May   2023 & 60088.2 & I, R, V, B &  180, 240, 240, 360 \\
\textit{Swift} UVOT (Sw7) & 23 Jul.\ 2023 & 60148.4 & V, B, U, UVW1, UVM2, UVW2 & 164, 169, 205, 411, 440, 655 \\
\textit{Swift} UVOT (Sw8) & 26 Jul.\ 2023 & 60151.6 & V, B, U, UVW1, UVM2, UVW2 & 76, 76, 76, 153, 182, 307 \\
\textit{Swift} UVOT (Sw9) &  6 Dec.\ 2023 & 60284.3 & V, B, U, UVW1, UVM2, UVW2 & 125, 125, 125, 251, 96, 162 \\
\textit{Swift} UVOT (Sw10) & 10 Dec.\ 2023 & 60288.1 & V, B, U, UVW1, UVM2, UVW2 & 113, 113, 113, 226, 322, 454 \\
\textit{Swift} UVOT (Sw11) & 26 Apr.\ 2024 & 60426.7 & V, U, UVW1, UVM2, UVW2 & 86, 86, 86, 86, 86\\

\textit{Swift} UVOT (Sw12) &  1 Aug.\ 2024 & 60523.3 & V, B, U, UVW1, UVM2, UVW2 & 101, 82, 101, 101, 101, 101 \\
\textit{Swift} UVOT (Sw13) &  5 Aug.\ 2024 & 60527.0 & V, B, U, UVW1, UVM2, UVW2 & 57,  48, 57, 57, 51, 57 \\
\textit{Swift} UVOT (Sw14) & 22 Aug.\ 2024 & 60544.7 & V, B, U, UVW1, UVM2, UVW2 & 190, 154, 190, 190, 190, 190 \\
\textit{Swift} UVOT (Sw15) & 25 Aug.\ 2024 & 60547.1 & V, B, U, UVW1, UVM2, UVW2 & 155, 128, 155, 155, 155, 155 \\
\textit{XMM-Newton} OM (XM3) & 22 Jan.\ 2025 &  60697.3  & V, B, U, UVW1, UVM2, UVW2 & 4400, 4400, 4400, 13200, 8800, 13200 \\
\hline \end{tabular}
\tablefoot{MJD date refers to the midpoint of the observation (time-average of all exposures).
Exposure times are summed over all exposures for a given filter.
LCOGT/LSC, CPT, and COJ indicate the
1-meter telescopes at Cerro Tololo Inter-American Observatory,
South African Astronomical Observatory,
and Siding Spring Observatory, respectively,
all operated by the Las Cumbres Observatory global telescope network.}
\end{table*}

\begin{table*}[h]  
\caption[]{Optical spectroscopic observations of $\jotwelve$}
        \centering
\label{tab:optspeclog}
        \begin{tabular}{lllcc} \hline\hline
\# & Telescope     & Date  &  MJD     & Total Exposure  \\
   & \& Instrument &       &          &  (s)             \\ \hline
 1 & H/ESO 3.6m EFOSC1 & 13 Mar.\ 1993 & 49059 & 300 \\
 2 & UK Schmidt 6dF   & 10 Mar.\ 2002 & 52343 & 1200 \\
 3 & SALT RSS  & 13 Jan.\ 2022 & 59592 & 600 \\
 4 & VLT FORS2 &  8 Feb.\ 2022 & 59618 & 450 (300I), 750 (300V), 750 (1400V)\\
 5 & SALT RSS  &  9 Mar.\ 2022 & 59647 & 525 (H$\beta$), 600 (H$\alpha$)\\
 6 & SALT RSS  & 21 Apr.\ 2022 & 59690 & 630 (H$\beta$), 630 (H$\alpha$)\\
 7 & SAAO 1.9m & 19 Jun.\ 2022 & 59749 &  1200 \\
 8 & SAAO 1.9m &  5 Aug.\ 2022 & 59796 &  2400 \\
 9 & VLT FORS2 & 17 Jan.\ 2023  & 59961 & 450 (300I), 450 (300V), 1200 (1400V) \\
10 & SAAO 1.9m & 18 Mar.\ 2023  & 60021 & 2400 \\
11 & SAAO 1.9m & 17 Apr.\ 2023  & 60051 & 2400 \\
12 & SALT RSS  & 16 May   2023  & 60080 & 680 \\
13 & VLT FORS2 &  8 Jun.\ 2023  & 60103 & 1020 (300I), 1020 (300V) \\
14 & SALT RSS  & 13 Jul.\ 2023  & 60138 & 680 \\
15 & VLT FORS2 & 13 Jul.\ 2023  & 60138 & 1020 (300I), 1020 (300V) \\
16 & SAAO 1.9m & 23 Jul.\ 2023  & 60148 & 2400 \\
17 & VLT FORS2 & 20 Jan.\ 2024  & 60329 & 1020 (300I), 1020 (300V) \\
18 & SALT RSS  & 29 Mar.\ 2024  & 60398 & 680 \\
19 & SALT RSS  &  2 May   2024  & 60432 & 700 \\
20 & SALT RSS  & 23 Jul.\ 2024  & 60514 & 700 \\
21 & SALT RSS  & 30 Dec.\ 2024  & 60674 & 700 \\
 \hline  \end{tabular}
\tablefoot{We refer to \citet{Reimers96} and \citet{Jones09} for details on
spectra \#1 and \#2, and to Appendix~\ref{sec:appdx_optspec} for details on instrumental setups for spectra \#3--21.}
\end{table*}

\section{Additional details on observations and data reduction}

All \textit{XMM-Newton}, eROSITA/eRASS, and \textit{Swift} reductions
were performed using HEASOFT v.~6.32.1 software.

\subsection{eROSITA/eRASS reduction}  \label{sec:appdx_er}

All eROSITA/eRASS data were extracted using event processing version
c020 and eSASS version 21121\_0\_4 \citep{Brunner22}.  We combined
data from all seven telescope modules (TMs).  We extracted counts for
$\jotwelve$ using a circular extraction region with the radius scaled
to the 0.2--2.3~keV maximum likelihood (ML) count rate from the eRASS
source catalog; that is, the extraction regions have larger radii
corresponding to higher count rates.  ML count rates take into account
the time when the source was in the field of view, and with
corrections for vignetting effects applied.  Similarly, background
regions were extracted using annuli whose inner and outer radii depend
on ML count rate, and with excising point sources detected in the
background extraction region, again using circular regions whose radii
depended on ML count rate.  Extraction radii, vignetting-corrected
exposure times, and ML count rates are listed in
Table~\ref{tab:erassreduction}.

\begin{table*}[h]
\caption[]{eRASS extraction regions, exposure times, and ML count rates}
        \centering
\label{tab:erassreduction}
        \begin{tabular}{ccccc} \hline\hline
eRASS    & Source             & Background                &  Exposure & ML Count Rate  \\
         & Radius ($\arcsec$) & Annulus Radii ($\arcsec$) &       (s) & (ct s$^{-1}$; 0.2--2.3~keV)     \\ \hline
1        & 92.3   & 194.3, 1147.9 &   240 &  $1.31 \pm 0.11$  \\  
2        & 95.0   & 199.1, 1180.9 &   113 &  $1.45 \pm 0.12$  \\      
3        & 84.2   & 179.7, 1047.8 &   114 &  $0.95 \pm 0.10$    \\  
4        & 74.1   & 161.1,  922.7 &   122 &  $0.60 \pm 0.08$ \\  
5        & 37.9   &  91.0,  474.4 &   119 &  $0.057 \pm 0.024$\\  \hline
\end{tabular}
\tablefoot{Exposure refers to good time after screening and after correction for vignetting effects.
ML count rate \citep{Merloni24}
denotes the maximum likelihood count rate taking into
account the time when the source was in the field of view, and with
corrections for vignetting effects applied.}
\end{table*}

\subsection{\textit{XMM-Newton} EPIC reduction}   \label{sec:appdx_xmm_epic}

We observed $\jotwelve$ with \textit{XMM-Newton} on 25--26 January 2022
(XM1; revolution 4052) for a net duration of 55~ks, on 26 January 2023
(XM2; revolution 4236) for 15~ks, and on 22 January 2025 (XM3;
revolution 4601) for 41 ks, using both the EPIC pn and MOS cameras
each time.  For XM1, all EPIC cameras were operated in Full Frame
mode.  For XM2, when the source was brighter, the pn was in Small
Window mode, while both MOS cameras were in Large Window mode.  For
XM3, all three cameras were in Large Window mode.  The medium optical
blocking filter was used in all cases.  We reduced the data using XMM
Science Analysis Software (XMMSAS) version 21.0, following standard
extraction procedures for point sources.  Source spectra were
extracted from circles 40$\arcsec$ in radius; background spectra were
extracted from source-free regions with the same size located a few
arcminutes away and on the same CCD chip.  We screened data against
strong, time-localized background flares due to proton flux by
filtering on the 5--13 keV background rate.  For XM1, six flares
totaling 17~ks were excised; for XM2, three short flares totaling 2~ks
were excised; for XM3, the first $\sim$10 ks were excised.  For the
pn, we selected data from pattern 0 and pattern 1--4 separately (pn0
and pn14 throughout this paper) due to their slightly different energy
resolutions and due to potential calibration issues between patterns 0
and 1--4 in Small Window mode.  We checked for pileup using the XMMSAS
task \textsc{epatplot} but found no evidence for any pileup.  Good
time exposures after screening are listed in Table~\ref{tab:Xobs}.
The numbers of spectral counts for XM1 are 3659, 1213, 1446, and 1520
for 0.25--10~keV pn0, 0.5--10~keV pn14, 0.2--10~keV MOS1, and
0.2--10~keV MOS2, respectively; for XM2, the corresponding spectral
counts are 6534, 1922, 2894, and 2951, respectively; for XM3, values
are 26251, 7382, 9672, and 9132, respectively.

\subsection{\textit{Swift} XRT reduction}  \label{sec:appdx_sw_xrt}

\textit{Swift} performed 15 observations of $\jotwelve$, as listed in
Table~\ref{tab:Xobs}.  We examined XRT photon counting (PC) mode data.
We reprocessed raw event files using \textsc{xrtpipeline} version
0.13.7 and the latest XRT calibration files.  We extracted source
spectra using circular regions of radius 20~pixels (47$\arcsec$);
background spectra were extracted from annular regions of inner radius
40~pixels (94$\arcsec$) and outer radius 80~pixels (188$\arcsec$), and
confirmed to be free of background sources.  We generated ancillary
response files using \textsc{xrtmkarf}, and we selected the PC mode
response files from the calibration database.  Good exposure times
after screening are listed in Table~\ref{tab:Xobs}.

\subsection{NICER}  \label{sec:appdx_nicer}

We used NICERDAS version 11a software, and followed standard
procedures to screen data, produce cleaned event files, and extract
spectra.  We discarded data from detectors 14 and 34, which are prone
to excessive noise.  We rejected time intervals when the detector
undershoot rate\footnote{Detector undershoots are reset events that
occur when incoming photons trigger a cascade of accumulate charge.}
exceeded $150$ ct~s$^{-1}$ per module.  We also screened out time
intervals when the detector overshoot rate (caused when high-energy
particles deposit excess charge) exceeded 1.5~ct~s$^{-1}$.  Given the
source faintness, we discarded data taken during the ISS' passage
through the  South Atlantic Anomaly (SAA) boundary, which is defined to
be more conservative and cover a slightly larger area than for the
standard NICERSAA boundary. We used the 3C50 background estimation
method, screening out times where the background rate in the hard band
(13--15~keV) exceeded 0.5~ct~s$^{-1}$ in the hard band.  As noted in
Sect.~\ref{sec:eR_XRT_NICER_spec}, we summed all 28 observations to
maximize the signal-to-noise ratio.  Given the danger of underestimated optical
loading impacting the softest energies, and given the faintness of the
source, we discarded data below 0.4~keV and above 4~keV.

\subsection{XMM-Newton OM reduction} \label{sec:appdx_xmm_om}
  
\begin{table*}    
\caption[]{\textit{XMM-Newton} OM photometric Vega magnitudes and flux density measurements }
        \centering
\label{tab:OM_mags}
        \begin{tabular}{lccccccc} \hline\hline
Obs.        & Date                     &         V          &       B          &          U       &       UVW1       &         UVM2     &     UVW2   \\
            & (MJD)                    &                    &                  &                  &                  &                  &            \\ \hline
\multirow{2}{*}{XM1} & \multirow{2}{*}{59604.9} &   $16.937\pm0.018$ & $17.832\pm0.015$ & $17.404\pm0.017$ & $17.476\pm0.019$ & $17.699\pm0.032$ & -- \\
                                              &                          &   $631\pm11$       & $317\pm4$        & $168\pm3$        & $105\pm2$        & $67\pm2$         & -- \\    \hline
\multirow{2}{*}{XM2} & \multirow{2}{*}{59970.2} &   $16.752\pm0.022$ & $17.618\pm0.015$ & $16.526\pm0.013$ & $16.193\pm0.018$ & $16.666\pm0.045$ & -- \\
                                              &                          &   $750\pm15$       & $456\pm6$        & $378\pm5$        &  $343\pm6$       & $173\pm7$        & -- \\    \hline
\multirow{2}{*}{XM3} & \multirow{2}{*}{60697.3} &   $16.421\pm0.014$ & $17.172\pm0.008$ & $16.095\pm0.007$ & $15.684\pm0.004$ & $15.754\pm0.011$ & $15.681\pm0.016$ \\
                                              &                          &   $1017\pm 13$ & $689\pm5$ & $563\pm4$ & $548\pm2$ & $400\pm4$ & $405\pm6$ \\ \hline \hline
\end{tabular}\tablefoot{The top value is the Vega magnitude. The bottom value is the flux density in $\mu$Jy.}
\end{table*}

During XM1, the Optical Monitor (OM) aboard \textit{XMM-Newton}
observed $\jotwelve$ with one exposure each at V, B, and U bands,
two exposures at UVW1, and six exposures for M2, 
with each exposure lasting 4.4~ks.
During XM2, one exposure was obtained at each of V, B, U, UVW1, and UVM2 bands,
with each exposure lasting 2.2 ks.
All exposures were taken in image mode + fast mode.

We reduced the data using the XMM\_SAS routines \textsc{omichain} and
\textsc{omfchain} for the image and fast modes, respectively.  These
routines apply flat-fielding, source detection, and aperture
photometry for each exposure, and they combine all exposure images
into a mosaiced image, and perform source detection and aperture
photometry on the mosaiced image.  The source extraction radius was 12
pixels = 5$\farcs$7.  These routines also correct fluxes for detector
dead time.  We verified that the source was well detected within each
exposure, that there were no obvious imaging artifacts in any
exposure. The resulting flux densities are listed in Table~\ref{tab:OM_mags}.
To generate spectral files, we used the SAS tool \textsc{om2pha}.

In order to combine these flux densities with those from
\textit{Swift} UVOT purely for the purpose of creating a combined
light curve, it is neccesary to account for the slight offsets in
effective wavelengths between UVOT and OM.  We estimated local
spectral slopes from the SED via simple linear interpolations, and we
estimated the flux densities at the effective wavelengths of the
corresponding UVOT filter.

We also examined the flux densities of two nearby stars, Gaia DR3
3501932993890436224 and 3501932169256712448, located $0\farcm$7 to the
north and $1\farcm2$ to the east, respectively.  For both stars, flux
densities are consistent with being constant to within their
uncertainties at all bands, indicating a lack of significant
systematic uncertainty.\footnote{Systematic corrections based on the
  stellar light curves were thus not applied to the target.}  Again,
we estimated local spectral slopes from the SED, and estimated flux
densities at the effective wavelengths of the corresponding UVOT
filter (discussed in $\S$\ref{sec:appdx_sw_uvot}).

\subsection{\textit{Swift} UVOT reduction} \label{sec:appdx_sw_uvot}

We performed aperture photometry using the ftool \textsc{uvotsource}.
We extracted source spectra using circular regions of radius
6$\arcsec$.  
Background spectra
were extracted using annular regions, free of nearby point sources,
with inner and outer radii 17$\farcs$5 and 25$\arcsec$, respectively.
We obtained flux densities for $\jotwelve$ as well as for the same two
stars considered in the OM observations.
We confirmed that in both stars, flux densities are consistent with being
constant in all filters both for the UVOT-only light curves and the combined
OM+UVOT light curves, indicating that the variability measured in
$\jotwelve$ is real and intrinsic to the AGN.
Magnitudes and flux densities for $\jotwelve$ are listed in Table~\ref{tab:UVOT_mags}.
Finally, we used the Ftool \textsc{uvotsource} to generate source and
background spectral files, using the same source and background
extraction regions as for the aperture photometry.

\begin{sidewaystable*}   
\caption[]{UVOT photometric Vega magnitudes and flux density measurements }
        \centering
\label{tab:UVOT_mags}
        \begin{tabular}{lccccccc} \hline\hline
Observation \&                              & Date                     &         V          &       B          &          U       &       UVW1       &         UVM2     &     UVW2   \\
Instrument                                  & (MJD)                    &                    &                  &                  &                  &                  &            \\ \hline
\multirow{2}{*}{\textit{Swift} UVOT (Sw1)}  & \multirow{2}{*}{59633.3} & $16.34 \pm 0.12$   & $17.20 \pm 0.12$   & $16.50 \pm 0.11$   & $16.79 \pm 0.12$   & $16.69 \pm 0.10$  & $16.86 \pm 0.11$  \\
                                            &                          & $1060\pm110$ &   $534\pm50$ &   $362\pm36$ &   $171\pm19$ &   $163\pm11$ &   $133\pm12$ \\ \hline
\multirow{2}{*}{\textit{Swift} UVOT (Sw2)}  & \multirow{2}{*}{59667.7} & $16.43 \pm 0.09$   & $17.09 \pm 0.09$   & $16.29 \pm 0.08$   & $16.11 \pm 0.08$   & $16.15 \pm 0.07$  & $16.09 \pm 0.07$  \\
                                            &                          & $973\pm78$   &   $593\pm38$ &   $441\pm31$ &   $320\pm24$ &   $268\pm12$ &   $270\pm15$ \\ \hline
\multirow{2}{*}{\textit{Swift} UVOT (Sw3)}  & \multirow{2}{*}{59786.3} & $16.47 \pm 0.08$   & $17.22 \pm 0.08$   & $16.39 \pm 0.07$   & $16.50 \pm 0.08$   & $16.65 \pm 0.09$  & $16.48 \pm 0.07$  \\
                                            &                          & $943\pm67$   &   $523\pm32$ &   $402\pm25$ &   $223\pm18$ &   $168\pm10$ &   $189\pm11$ \\ \hline
\multirow{2}{*}{\textit{Swift} UVOT (Sw4)}  & \multirow{2}{*}{59934.3} & $16.35 \pm 0.08$   & $17.01 \pm 0.07$   & $16.22 \pm 0.07$   & $16.13 \pm 0.08$   & $16.26 \pm 0.08$  & $16.30 \pm 0.07$  \\
                                            &                          & $1050\pm70$  &   $639\pm34$ &   $468\pm27$ &   $315\pm23$ &   $241\pm14$ &   $222\pm13$ \\ \hline
\multirow{2}{*}{\textit{Swift} UVOT (Sw5)}  & \multirow{2}{*}{60025.5} & -- & -- & -- & $16.00 \pm 0.10$  & $16.01 \pm 0.11$   & $15.86 \pm 0.09$  \\
                                            &                          &    --      &       --    &      --     &   $354\pm34$ &   $304\pm25$ &   $334\pm26$ \\ \hline
\multirow{2}{*}{\textit{Swift} UVOT (Sw6)}  & \multirow{2}{*}{60078.6} & -- & -- & -- & $15.51 \pm 0.07$  & $15.48 \pm 0.07$   & $15.47 \pm 0.06$  \\
                                            &                          &    --      &       --    &      --     &   $558\pm35$ &   $496\pm23$ &   $478\pm25$ \\ \hline
\multirow{2}{*}{\textit{Swift} UVOT (Sw7)}  & \multirow{2}{*}{60148.4} & $16.22 \pm 0.08$   & $16.63 \pm 0.07$   & $15.68 \pm 0.06$   & $15.60 \pm 0.07$   & $15.46 \pm 0.07$  & $15.46 \pm 0.06$ \\
                                            &                          &  $1190\pm70$  &   $906\pm42$ &   $771\pm41$ &   $514\pm35$ &   $505\pm22$ &   $484\pm23$ \\ \hline
\multirow{2}{*}{\textit{Swift} UVOT (Sw8)}  & \multirow{2}{*}{60151.6} & $16.34 \pm 0.13$   & $16.70 \pm 0.10$   & $15.71 \pm 0.08$   & $15.66 \pm 0.10$   & $15.56 \pm 0.11$  & $15.46 \pm 0.08$  \\
                                            &                          &  $1060\pm120$ &   $847\pm66$ &   $753\pm57$ &   $485\pm45$ &   $458\pm36$ &   $485\pm31$ \\ \hline
\multirow{2}{*}{\textit{Swift} UVOT (Sw9)}  & \multirow{2}{*}{60284.3} & $16.68 \pm 0.18$   & $17.16 \pm 0.14$   & $16.04 \pm 0.10$   & $16.09 \pm 0.11$   & $16.12 \pm 0.12$  & $16.12 \pm 0.09$  \\
                                            &                          &  $778\pm125$  &   $556\pm61$ &   $554\pm49$ &   $326\pm34$ &   $274\pm26$ &   $263\pm21$ \\ \hline
\multirow{2}{*}{\textit{Swift} UVOT (Sw10)} & \multirow{2}{*}{60288.1} & $16.32 \pm 0.09$   & $16.84 \pm 0.08$   & $16.06 \pm 0.07$   & $16.00 \pm 0.08$   & $16.37 \pm 0.09$  & $16.02 \pm 0.07$  \\
                                            &                          &  $1080\pm80$  &   $746\pm41$ &   $541\pm32$ &   $355\pm27$ &   $219\pm13$ &   $287\pm15$ \\ \hline
\multirow{2}{*}{\textit{Swift} UVOT (Sw11)} & \multirow{2}{*}{60426.7} & $16.22 \pm 0.08$   &     --             & $15.69 \pm 0.06$   & $15.54 \pm 0.09$   & $15.58 \pm 0.11$  & $15.46 \pm 0.09$  \\
                                            &                          &   $1180\pm80$  &        --   &   $765\pm44$ &   $540\pm46$ &   $451\pm35$ &   $484\pm38$ \\ \hline
\multirow{2}{*}{\textit{Swift} UVOT (Sw12)} & \multirow{2}{*}{60523.3} & $16.15 \pm 0.08$   & $16.78 \pm 0.08$       & $15.78 \pm 0.06$  & $15.74 \pm 0.09$   & $15.86 \pm 0.11$  & $15.87 \pm 0.10$  \\
                                            &                          &  $1260\pm80$  &   $785\pm48$ &   $704\pm40$ &   $453\pm40$ &   $347\pm29$ &   $331\pm28$ \\ \hline
\multirow{2}{*}{\textit{Swift} UVOT (Sw13)} & \multirow{2}{*}{60527.0} & $16.33 \pm 0.12$   & $16.81 \pm 0.10$       & $15.86 \pm 0.08$  & $15.83 \pm 0.12$   & $16.00 \pm 0.15$  & $16.07 \pm 0.13$  \\
                                            &                          &  $1070\pm100$ &   $763\pm60$ &   $654\pm46$ &   $414\pm45$ &   $307\pm37$ &   $274\pm31$ \\ \hline
\multirow{2}{*}{\textit{Swift} UVOT (Sw14)} & \multirow{2}{*}{60544.5} & $16.45 \pm 0.09$   & $16.87 \pm 0.08$       & $15.86 \pm 0.06$  & $15.86 \pm 0.08$   & $15.92 \pm 0.10$  & $15.89 \pm 0.08$  \\
                                            &                          & $957\pm75$   &   $722\pm40$ &   $654\pm33$ &   $403\pm31$ &   $329\pm22$ &   $325\pm23$ \\ \hline
\multirow{2}{*}{\textit{Swift} UVOT (Sw15)} & \multirow{2}{*}{60547.1} & $16.31 \pm 0.09$   & $16.83 \pm 0.08$       & $15.91 \pm 0.06$  & $15.99 \pm 0.09$   & $15.97 \pm 0.10$  & $15.95 \pm 0.09$  \\
                                            &                          & $1090\pm80$  &   $753\pm44$ &   $626\pm35$ &   $358\pm30$ &   $315\pm24$ &   $309\pm24$  \\ \hline  \hline
\end{tabular}\tablefoot{The top value is the Vega magnitude. The bottom value is the flux density in $\mu$Jy.}
\end{sidewaystable*}

\subsection{LCOGT photometry}  \label{sec:appdx_lcogt}

The nine ground-based photometric observations of $\jotwelve$ were
obtained using Sinistro 1-meter telescopes operated by the Las Cumbres
Observatory global telescope network from March 2022 to May 2023
(LCO1--9). LCO1--4 and 6--9 each 
used Johnson-Cousins/Bessell I-, R-, V-, and B-band
filters, with two exposures of 90~s for I, two exposures of 120~s for
V, two exposures of 120~s for R, and two exposures of 180~s for B;
LCO5 used the V and B filters with exposures of 60~s and 80~s, respectively.
LCOGT data are automatically processed using the BANZAI 
pipeline,\footnote{\url{https://lco.global/documentation/data/BANZAIpipeline/}}
which performs standard calibration steps including masking bad
pixels, biasing, dark subtraction, and flat fielding.

We performed aperture photometry by stacking the pairs of exposures,
and extracting source-centered circles of radius 6$\arcsec$ and
background annuli with inner and outer radii of 19$\arcsec$ and
26$\arcsec$, chosen to avoid nearby faint stars, and extracted
magnitudes. 
We also extracted magnitudes for five reference stars located   
roughly 4--5$\arcmin$ away,
GAIA DR3 3501934849316325248, 
GAIA DR3 3501928973801041408, 
GAIA DR3 3501928905081564288, 
GAIA DR3 3501929076880260736, 
and GAIA DR3 3501932920876028544 (UCAC4 335-068176),  
calibrating against synthetic photometric magnitudes generated from \textit{Gaia}
DR3 \citep{GaiaCollab16,GaiaCollab23} mean spectra in the performance
verification catalog \citep{GaiaCollab22}.   
We converted magnitudes to flux densities using standard Vega
magnitudes for Johnson-Cousins/Bessell filters.\footnote{e.g.,
\url{https://www.astronomy.ohio-state.edu/martini.10/usefuldata.html}}
The resulting magnitudes and flux densities are listed in Table~\ref{tab:LCO_mags}.

\begin{table*}   
\caption[]{LCOGT photometric Vega magnitudes and flux density measurements }
        \centering
\label{tab:LCO_mags}
        \begin{tabular}{lccccc} \hline\hline
Observation                   & Date                     &         I              &       R            &          V       &      B    \\
                              & (MJD)                    &                        &                    &                  &           \\ \hline
\multirow{2}{*}{LCO1}         & \multirow{2}{*}{59639.3} &     $15.430 \pm 0.003$ & $16.058 \pm 0.007$ & $16.561 \pm 0.003$ & $17.298 \pm 0.003$ \\
                              &                          &        $1624.8\pm4.5$  &  $1155.5^{+7.5}_{-7.4}$   &  $862.8\pm2.4$ &   $489.0^{+1.4}_{-1.3}$ \\   \hline 
\multirow{2}{*}{LCO2}         & \multirow{2}{*}{59704.2} &     $15.364 \pm 0.003$ & $15.919 \pm 0.002$ & $16.388 \pm 0.003$ & $17.051 \pm 0.004$ \\
                              &                          &        $1726.6\pm4.8$  &  $1313.3\pm2.4$    &  $1011.8\pm2.8$ &   $613.9\pm2.3$ \\  \hline
\multirow{2}{*}{LCO3}         & \multirow{2}{*}{59743.0} &     $15.317 \pm 0.004$ & $15.905 \pm 0.003$ & $16.368 \pm 0.005$ & $16.926 \pm 0.005$ \\
                              &                          &$1803.0^{+6.7}_{-6.6}$  &  $1330.4\pm3.7$    &  $1030.6^{+4.8}_{-4.7}$ &   $688.8\pm3.2$ \\ \hline
\multirow{2}{*}{LCO4}         & \multirow{2}{*}{59773.8} &     $15.408 \pm 0.007$ & $15.953 \pm 0.006$ & $16.422 \pm 0.010$ & $17.119 \pm 0.015$ \\
                              &                          &       $1658.0\pm10.7$  &  $1272.8^{+7.1}_{-7.0}$ &  $980.6^{+9.1}_{-9.0}$ &   $576.6^{+8.0}_{-7.9}$ \\  \hline
\multirow{2}{*}{LCO5} & \multirow{2}{*}{59942.0} &               &              &  $16.456\pm0.016$ & $17.057\pm0.022$ \\ 
                              &                          &                      &               &  $950.4^{+14.0}_{-13.9}$ & $610.5^{+12.5}_{-12.2}$ \\   \hline 
\multirow{2}{*}{LCO6}         & \multirow{2}{*}{59949.3} &     $15.343 \pm 0.003$ & $15.945 \pm 0.003$ & $16.419 \pm 0.004$ & $17.035 \pm 0.005$   \\
                              &                          &        $1760.3\pm4.9$  &  $1282.3\pm3.5$    &  $983.3\pm3.6$ &   $623.0\pm2.9$ \\ 
\hline 
\multirow{2}{*}{LCO7}         & \multirow{2}{*}{59991.7} &     $15.338 \pm 0.003$ & $15.930 \pm 0.002$ & $16.398 \pm 0.002$ & $17.013 \pm 0.003$   \\
                              &                          &        $1768.5\pm4.9$  &  $1300.1\pm2.4$    &  $1002.5\pm1.8$ &   $635.8\pm1.8$ \\   \hline
\multirow{2}{*}{LCO8}         & \multirow{2}{*}{60025.1} &     $15.269 \pm 0.002$ & $15.841 \pm 0.002$ & $16.272 \pm 0.002$ & $16.812 \pm 0.002$ \\
                              &                          &        $1884.5\pm3.5$  &  $1411.2\pm2.6$    &  $1125.9\pm2.1$ &   $765.1\pm1.4$ \\   \hline
\multirow{2}{*}{LCO9}         & \multirow{2}{*}{60088.2} &     $15.237 \pm 0.002$ & $15.788 \pm 0.001$ & $16.191 \pm 0.002$ & $16.656 \pm 0.002$   \\  
                              &                          &        $1940.9\pm3.6$  &  $1481.8\pm1.4$    & $1213.1\pm2.2$  &  $883.3\pm1.6$ \\ 
\hline \hline
\end{tabular}\tablefoot{The top value is the Vega magnitude. The bottom values are the flux densities in $\mu$Jy. 
Values are not corrected for Galactic extinction.}
\end{table*}

\subsection{PROMPT-6 optical photometry}  \label{sec:appdx_prompt}

All observations were obtained with the PROMPT 6 telescope at Cerro
Tololo Inter-American Observatory (CTIO), operated as part of Skynet
\citep[for details about the network, see][]{Martin19}, and using
Johnson–Cousins B, V, and R filters. For each filter, we observed the
target with five 120~s exposures with a field of view of
13$\farcs$5$\times$13$\farcs$5 arranged in a 3$\times$3 dither pattern
with a 40$\arcsec$ dither step, providing improved sampling of
flat-field and detector inhomogeneities. The five exposures in each
filter were subsequently aligned and combined and the time assigned to
each stacked image corresponds to the median MJD of its individual
components. Image reduction followed standard procedures, including
bias subtraction, dark correction, and flat-fielding using nightly
calibration frames. The dither pattern also enabled effective removal
of cosmic rays and bad pixels.  Observed (not corrected for Galactic
reddening) Vega magnitudes are listed in Table~\ref{tab:PROMPTmags}.

\begin{table}
\label{tab:PROMPTmags}
\caption{PROMPT-6 photometric Vega magnitudes}
\centering
\begin{tabular}{lccc}
\hline\hline
Date (MJD)                     &           R            &         V             &    B   \\  \hline
\multirow{2}{*}{59616.4 (P1)}  &   \multirow{2}{*}{--}  &  \multirow{2}{*}{--}  &  $17.20 \pm 0.17$  \\ 
                               &                        &                       &  $535^{+91}_{-78}$ \\  \hline

\multirow{2}{*}{59617.4 (P2)}  &   \multirow{2}{*}{--}  &  $16.73 \pm 0.16$           &  $17.67 \pm 0.20$   \\
                               &                        & $738^{+117}_{-101}$      &  $347^{+70}_{-58}$ \\  \hline

\multirow{2}{*}{59619.4 (P3)}  &   \multirow{2}{*}{--}  &  \multirow{2}{*}{--}  & $17.48 \pm 0.17$    \\
                               &                        &                       & $414^{+70}_{-60}$\\  \hline

\multirow{2}{*}{59621.4 (P4)}  &   \multirow{2}{*}{--}  &  $16.66 \pm 0.17$     & $17.48 \pm 0.18$  \\
                               &                        &  $788^{+134}_{-114}$     & $414^{+75}_{-63}$ \\  \hline

\multirow{2}{*}{59622.4 (P5)}  &   \multirow{2}{*}{--}  &  $16.60 \pm 0.15$           &  \multirow{2}{*}{--}    \\
                               &                        &  $832^{+123}_{-107}$     &   \\  \hline

\multirow{2}{*}{59623.4 (P6)}  &  $16.17 \pm 0.15$          &   \multirow{2}{*}{--}  &   \multirow{2}{*}{--}   \\
                               & $1042^{+154}_{-134}$      &                        &  \\  \hline

\multirow{2}{*}{59629.1 (P7)}  &  $16.13 \pm 0.15$          & $16.67 \pm 0.14$       & $17.47 \pm 0.14$ \\
                               & $1081^{+160}_{-140}$      &$780^{+107}_{-94}$       & $417^{+57}_{-50}$\\  \hline

\multirow{2}{*}{59633.1 (P8)}  &  $16.11 \pm 0.13$          & $16.62 \pm 0.12$       & $17.49 \pm 0.14$    \\
                               & $1101^{+140}_{-124}$      &$817^{+95}_{-86}$        & $410^{+56}_{-50}$\\  \hline

\multirow{2}{*}{59638.3 (P9)}  &  $16.10 \pm 0.14$          & $16.61 \pm 0.14$       &   $17.31 \pm 0.12$   \\
                               & $1112^{+153}_{-134}$      & $825^{+114}_{-100}$     & $484^{+57}_{-51}$\\  \hline

\multirow{2}{*}{59649.1 (P10)} &  $16.12 \pm 0.14$          & $16.60 \pm 0.14$       &  $17.33 \pm 0.12$  \\
                               & $1091^{+150}_{-132}$      &$832^{+115}_{-101}$	     & $475^{+55}_{-50}$\\  \hline

\multirow{2}{*}{59658.1 (P11)} &  $16.07 \pm 0.14$          & $16.56 \pm 0.14$       & $17.35 \pm 0.13$    \\
                               & $1143^{+157}_{-138}$      &$864^{+119}_{-104}$	     & $466^{+59}_{-53}$\\  \hline

\multirow{2}{*}{59665.3 (P12)} &  $16.04 \pm 0.14$          & $16.50 \pm 0.14$       &  $17.16 \pm 0.28$ \\
                               & $1175^{+162}_{-142}$      &$913^{+126}_{-110}$	& $555^{+163}_{-126}$\\  \hline

\multirow{2}{*}{59706.1 (P13)} &  $16.00 \pm 0.15$          & $16.46 \pm 0.13$             & $17.08 \pm 0.13$   \\
                               & $1219^{+181}_{-157}$      &$947^{+120}_{-107}$	& $598^{+76}_{-67}$ \\  
\hline
\end{tabular} 
\tablefoot{The top value is the Vega magnitude. The bottom value is the flux density in $\mu$Jy. These values are not corrected for Galactic extinction.}  
\end{table}

\subsection{ATLAS and \textit{WISE/NEOWISE}}  \label{sec:appdx_atlaswise}

To access publicly available photometry from the Asteroid Terrestrial
impact Last Alert System \citep[ATLAS;][]{Tonry18,Smith20}, we used
ATLAS' online forced-photometry pipeline to generate o-band (orange;
560-820~nm) and c-band (cyan; 420--650 nm) light curves. We used
data from 22 November 2017 (MJD 58097) to 24 December 2023 (MJD 60302), and
filtered against flux density data points that were obvious outliers
or had error bars larger than 25$\%$. They are plotted in
Fig.~\ref{fig:ALL_LC_ZOOM}. 

\begin{figure}  
\includegraphics[width=0.99\columnwidth]{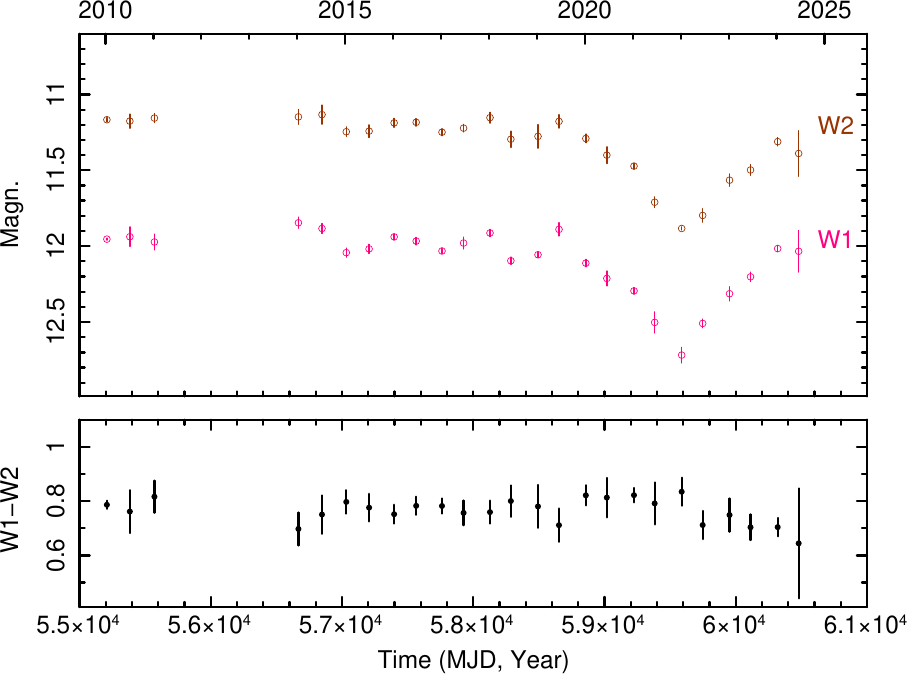}
\caption{\textit{WISE}/\textit{NEOWISE} W1 and W2 band light curves, binned to 180 days (top panel). 
W1$-$W2 is plotted in the lower panel. }
\label{fig:wise_lc_forappdx}
\end{figure}

We also obtained public IR photometric monitoring courtesy of the
\textit{Widefield Infrared Survey Explorer}
(\textit{WISE})/\textit{NEOWISE} mission \citep{Wright10, Mainzer14}.
We downloaded all single-epoch W1 and W2 band photometry points from
the AllWISE Multiepoch Photometry Table and the NEOWISE-R Single
Exposure Source Table catalogs hosted at the NASA/IPAC Infrared
Science Archive (IRSA\footnote{\url{https://irsa.ipac.caltech.edu}}),
collecting data taken from 12 January 2010 (MJD 55208) until 18 June
2023 (MJD 60113).  We rebinned the light curves to one flux point
every six months; the resulting W1 and W2 light curves, as well as
W1$-$W2, are plotted in Fig.~\ref{fig:wise_lc_forappdx}.

\subsection{Optical spectroscopy}  \label{sec:appdx_optspec}

The VLT-FORS2 spectra were taken with a number of different
instrumental setups and exposure times. All spectra presented in this
work are composite spectra, formed by combining multiple observations
taken in a single run. Spectra \#4 and \#9 are a combination of spectra
taken with grism 300V (4450--8650~$\AA$), 300I and filter OG590
(6000--11000~$\AA$), and grism 1400V (4560--5860~$\AA$; H$\beta$). The
other FORS2 observations only made use of grisms 300V and
300I+OG590. We give an overview of the spectral observations in
Table~\ref{tab:fors2obs}. The slit width for all observations was
1$\farcs$3 and the seeing over the observing nights ranged from 0$\farcs$3 to
2$\farcs$2. With regard to the level of seeing, we note that the
weather on all observing nights was calm, which means the upper limits
of the seeing range likely represent only brief spikes during the
run. For the reduction of the VLT data we made use of the
\texttt{esoreflex} pipeline \citep{esoreflex_2013}, which conducts all
standard CCD processing, as well as the spectral extraction, wavelength
calibration, and the flux calibration. Arc and standard star spectra
were taken on the night. FORS2 spectra \#4 and \#9 were taken as
part of ESO programs 105.20UT and 109.23MH (PI: M.\ Krumpe),
respectively, and spectra \#13, 15, and 17 as part of program
109.22XH (PI: D.\ Homan).

\begin{table}
\label{tab:fors2obs}
\caption{VLT-FORS2 instrumental setups}
\centering
\begin{tabular}{lcccc} 
\hline\hline
    & Date         & Setup  & Exposure & Seeing   \\
    &              &        & (s)      &          \\
\hline
4   & 8 Feb.\ 2022  & 300I   & 450      & 0$\farcs$4--1$\farcs$9 \\
    &              & 300V   & 750      &          \\
    &              & 1400V  & 750      &          \\
9   & 17 Jan.\ 2023 & 300I   & 450      & 0$\farcs$3--2$\farcs$2 \\
    &              & 300V   & 450      &          \\
    &              & 1400V  & 1200     &          \\
13  & 8 Jun.\ 2023  & 300I   & 1020     & 0$\farcs$5--1$\farcs$6 \\
    &              & 300V   & 1020     &          \\
15  & 13 Jul.\ 2023 & 300I   & 1020     & 0$\farcs$3--0$\farcs$8 \\
    &              & 300V   & 1020     &          \\
17  & 20 Jan.\ 2024 & 300I   & 1020     & 0$\farcs$5--2$\farcs$2 \\
    &              & 300V   & 1020     &          \\
\hline
\end{tabular}
\end{table}

Observations at SALT RSS used the pg0900 grating (to achieve spectral
resolution approximately 800--1200), and a slit width of 1$\farcs$5.
All observations used a grating angle of 16.25$^{\rm \circ}$ (camera
angle 32.5$^{\rm \circ}$) to cover both the H$\alpha$ and H$\beta$
regions.  Observations \#5 and \#6 additionally used a grating angle
of 13.25$^{\rm \circ}$ (camera angle 26.5$^{\rm \circ}$) to sample
from the H$\beta$ region to 3200~$\AA$.  All observations used
normal mode, the faint gain, and 2$\times$2 binning.  Exposure times,
listed in Table~\ref{tab:optspeclog}, were in the range 525--700~s.

Observations at SAAO used the low-resolution grating on the SpUpNIC
spectrograph, with a slit width of 2$\farcs$7.  Exposures were
$2\times 1200$~s for all observations except for spectrum \#7 (19
June 2022), which had a $1\times1200$~s exposure.

All CCD data were reduced using standard bias corrections
and flat-fielding. Wavelength calibration used arc-lamp spectra
taken the same night. Spectrophotometric calibrations were applied 
using standard stars (although such calibrations were not possible for spectra \#1 and 2,
the archival Hamburg/ESO and 6dFGS spectra).

We cross-calibrated all optical spectra using the framework of
\citet{vangroningen92} and \citet{fausnaugh17}, which accounts for
differing apertures, wavelength resolutions, slit widths, weather
conditions, air masses, and absolute wavelength calibration issues by
matching [\ion{O}{iii}] integrated fluxes and profile widths.  Further
details can be found in \citet{Saha25b}.  We used spectrum \#3 (SALT)
as the reference spectrum, and applied flux scaling (gray-shifted)
factors.  Because the two archival spectra, \#1 and 2, were not
flux-calibrated, we must treat all continuum and absolute line fluxes
as approximate, although parameters based on relative line fluxes
derived from lines that are very close in wavelength (e.g., $F_{\rm
  DL}$, $R_{\rm H\beta/[\ion{O}{iii}]}$, $R_{\rm BL/NL}$) are still
valid, albeit with some potential additional systematic uncertainty
compared to the flux-calibrated spectra.  We note from
Fig.~\ref{fig:optspeczoom} that spectrum \#7, taken with SAAO SpUpNIC
on 19 June 2022, seems to have a continuum level that is roughly
40$\%$ higher compared to other spectra taken in mid-2022, including
another SpUpNIC spectrum. As plotted in Fig.~\ref{fig:ALL_LC_ZOOM},
LCOGT photometry taken close in time (LCO3) do indicate a relatively
higher optical continuum at this time.  Another possibility is strong
host galaxy contamination: an absorption feature near 5090~$\AA$
(rest-frame), associated with host-galaxy absorption, is somewhat
strong here.  Finally, we applied de-reddening corrections to all
spectra, using $E(B-V)=0.057$ \citep{Schlegel98}, $R$=3.1, and the
Galactic extinction function of \citep{Fitzpatrick99}.
Best-fitting parameters from our optical spectral fits
(Sect.~\ref{sec:performingoptfits}) are listed in
Tables~\ref{tab:HBFITRESULTS}, \ref{tab:HAFITRESULTS}, and
\ref{tab:OTHEROPTFITRESULTS}.

\setcounter{table}{9}
\begin{table}[!hp]
\label{tab:SED_RATIO_TABLE}
\caption{Ratios of SED model flux densities}
\centering
\begin{tabular}{lccccc}
\hline\hline
        &   U & W1 & M2 &  0.5 keV  &  4 keV \\ \hline
XM2/XM1 &  2.9 & 3.2 & 3.8 & 9.7 & 6.5   \\
XM3/XM1 &  4.7 & 5.6 & 8.5 & 12.4 & 6.6    \\
\hline
\end{tabular}
\end{table}

\setcounter{table}{6}

\renewcommand{\arraystretch}{1.20}

\begin{sidewaystable*}
\caption[]{Best-fitting model parameters for the broad H$\beta$ line}
\centering
\label{tab:HBFITRESULTS}
\begin{tabular}{l|lll|lllll|ll} \hline\hline
                & \multicolumn{3}{c}{Gaussian}             &  \multicolumn{5}{|c|}{Diskline}                                                                                                     &  Total broad                     \\
 Date           & $\lambda_{\rm cent}$ (\AA) & $\sigma$ (\AA) & Flux$^{a}$ &   $\lambda_{\rm o}$ (\AA)  & $\sigma_{\rm 0}$ (\AA) & $R_{\rm in}$ ($R_{\rm g}$) &  Incl.\ $i$ ($^{\rm \circ}$)& Flux$^{a}$     &  line flux$^{a}$ &   $F_{\rm DL}$$^{b}$    \\   \hline
\#1 H/ESO 03/93   & 4878$\pm$3      & 44$\pm$3     & 201$\pm$32$^{c}$ & 4861*         & 10*         & 1440$\pm$520 & 13*      &  63$\pm$23$^{c}$  &  264$\pm$16$^{c}$  &  0.24$\pm$0.09    \\      
\#2 6dF 03/02   & 4869$\pm$2      & 45$\pm$5     & 185$\pm$44$^{c}$ & 4861*         & 10*         &  940$\pm$200 & 13*      &  88$\pm$24$^{c}$  &  273$\pm$15$^{c}$  &  0.32$\pm$0.12    \\      
\#3 SALT 01/22  & 4877$\pm$4      & 60$\pm$6     &  51$\pm$11 & 4853$\pm$3    & 10$\pm$5    &  500$\pm$180 & 13*      &  13$\pm$5   &   64$\pm$4   &  0.20$\pm$0.08    \\      
\#4 VLT 02/22   & 4870$\pm$2      & 43$\pm$11    & 115$\pm$9  & 4845$\pm$4    & 10*         & 1000*        & 22$\pm$4 &  12$\pm$8   &  127$\pm$4   &  0.09$\pm$0.06    \\      
\#5 SALT 03/22  & 4889$\pm$48     & 61$\pm$4     & 124$\pm$14 & 4855$\pm$2    & 10*         &  640$\pm$110 & 13*      &  46$\pm$5   &  170$\pm$8   &  0.27$\pm$0.04    \\      
\#6 SALT 04/22  & 4889$\pm$50     & 39$\pm$3     & 121$\pm$16 & 4855$\pm$2    & 10*         &  830$\pm$100 & 13*      &  83$\pm$12  &  204$\pm$5   &  0.40$\pm$0.08    \\      
\#7 SAAO 06/22  & 4877$\pm$3      & 33$\pm$2     & 280$\pm$46 & 4861*         & 10*         &  660$\pm$110 & 13*      &  67$\pm$55  &  347$\pm$16  &  0.19$\pm$0.16    \\      
\#8 SAAO 08/22  & 4874$\pm$2      & 37$\pm$2     & 274$\pm$40 & 4861*         & 10*         & 1000*        & 13*      &  $<$73      &  274$\pm$18  &  $<$0.24          \\      
\#9 VLT 01/23   & 4875$\pm$1      & 42$\pm$1     & 237$\pm$21 & 4855$\pm$3    & 10$\pm$2    &  540$\pm$140 & 11$\pm$1 &  72$\pm$26  &  309$\pm$27  &  0.23$\pm$0.09    \\      
\#10 SAAO 03/23 & 4877$\pm$3      & 56$\pm$5     & 247$\pm$36 & 4873$\pm$2    & 10*         &  920$\pm$19  & 13*      &  81$\pm$20  &  328$\pm$16  &  0.25$\pm$0.07    \\      
\#11 SAAO 04/23 & 4868$\pm$1      & 33$\pm$1     & 327$\pm$27 & 4861*         & 10*         & 1000*        & 13*      & $<$42       &  327$\pm$6   &  $<$0.12          \\      
\#12 SALT 05/23 & 4874$\pm$1      & 49$\pm$2     & 167$\pm$16 & 4863$\pm$1    & 10$\pm$1    &  830$\pm$120 & 14$\pm$1 & 118$\pm$19  &  285$\pm$13  &  0.42$\pm$0.08    \\      
\#13 VLT 06/23  & 4876$\pm$1      & 44$\pm$2     & 234$\pm$23 & 4861$\pm$1    & 10$\pm$1    &  840$\pm$130 & 13$\pm$1 & 178$\pm$32  &  412$\pm$22  &  0.43$\pm$0.09    \\      
\#14 SALT 07/23 & 4879$\pm$2      & 49$\pm$2     & 147$\pm$13 & 4862$\pm$1    & 21$\pm$1    & 1160$\pm$230 & 16$\pm$1 & 152$\pm$36  &  299$\pm$12  &  0.51$\pm$0.13    \\      
\#15 VLT 07/23  & 4867$\pm$4      & 25$\pm$1     & 161$\pm$9  & 4864$\pm$1    &  6$\pm$1    & 1220$\pm$230 & 18$\pm$2 &  79$\pm$16  &  240$\pm$15  &  0.33$\pm$0.07    \\      
\#16 SAAO 07/23 & 4889$\pm$7      & 46$\pm$3     & 170$\pm$36 & 4860$\pm$1    & 11$\pm$2    &  830$\pm$80  & 13*      & 173$\pm$22  &  343$\pm$40  &  0.50$\pm$0.12    \\      
\#17 VLT 01/24  & 4872$\pm$1      & 39$\pm$2     & 270$\pm$32 & 4861$\pm$1    & 14$\pm$2    & 1000*        & 13$\pm$1 & 115$\pm$13  &  385$\pm$21  &  0.30$\pm$0.05    \\      
\#18 SALT 03/24 & 4869$\pm$1      & 35$\pm$1     & 282$\pm$24 & 4861$\pm$1    &  3$\pm$2    & 1000*        & 15$\pm$1 &  56$\pm$36  &  338$\pm$58  &  0.16$\pm$0.11    \\      
\#19 SALT 05/24 & 4864$\pm$2      & 40$\pm$3     & 231$\pm$34 & 4867$\pm$1    & 14$\pm$1    & 1000*        & 13$\pm$1 & 149$\pm$15  &  380$\pm$25  &  0.39$\pm$0.07    \\      
\#20 SALT 07/24 & 4868$\pm$1      & 49$\pm$2     & 229$\pm$24 & 4863$\pm$1    & 11$\pm$1    &  800$\pm$160 & 12$\pm$1 & 147$\pm$35  &  376$\pm$18  &  0.39$\pm$0.09    \\      
\#21 SALT 12/24 & 4866$\pm$1      & 45$\pm$1     & 311$\pm$20 & 4866$\pm$1    & 10$\pm$2    & 1000*        & 13$\pm$1 &  91$\pm$8   &  402$\pm$22  &  0.23$\pm$0.03    \\      \hline \hline        
\end{tabular}\tablefoot{
An asterisk (*) indicates a fixed parameter.\\
$^{a}$: Line flux units are $10^{-16}$ erg cm$^{-2}$ s$^{-1}$.\\
$^{b}$: $F_{\rm DL}$ is the fraction contribution of the flux of the diskline component to the total broad-line flux.
For the diskline component, $\lambda_{\rm o}$ is the rest-frame wavelength of emission, $\sigma_{\rm 0}$ quantifies the local broadening due to electron scattering
in a photoionized atmosphere, $R_{\rm in}$ is the inner radius,
and $i$ is the disk inclination, defined such that 0$^{\rm \circ}$ indicates a face-on disk.  The outer radius $R_{\rm out}$ was always held fixed at 5000 $R_{\rm g}$.   \\
$^{c}$: Absolute flux calibration was not available for spectra \#1 and \#2; values of absolute line flux should be considered as estimates only. \\
}
\end{sidewaystable*}

\begin{sidewaystable*}
\caption[]{Best-fitting model parameters for the broad H$\alpha$ line}
\centering
\label{tab:HAFITRESULTS}
\begin{tabular}{l|lll|lllll|ll} \hline\hline
                & \multicolumn{3}{c}{Gaussian}             &  \multicolumn{5}{|c|}{Diskline}                                                                                                     &  Total broad                     \\
 Date           & $\lambda_{\rm cent}$ (\AA) & $\sigma$ (\AA) & Flux$^{a}$ &   $\lambda_{\rm o}$ (\AA)  & $\sigma_{\rm 0}$ (\AA) & $R_{\rm in}$ ($R_{\rm g}$) &  Incl.\ $i$ ($^{\rm \circ}$)& Flux$^{a}$     &  line flux$^{a}$ &   $F_{\rm DL}$$^{b}$    \\   \hline
\#2 6dF 03/02   & 6572$\pm$3      & 41$\pm$2     & 351$\pm$43$^{c}$ & 6560$\pm$2    & 15*         &  320$\pm$190          &  8$\pm$3 &  83$\pm$76$^{c}$  &  434$\pm$13$^{c}$  &  0.19$\pm$0.17    \\      
\#3 SALT 01/22  & 6557$\pm$1      & 55$\pm$2     & 166$\pm$11$^{c}$ & 6548$\pm$1    & 10$\pm$7    & 1110$\pm$210          & 15$\pm$1 &  54$\pm$13$^{c}$  &  220$\pm$4$^{c}$   &  0.25$\pm$0.06    \\      
\#4 VLT 02/22   & 6569$\pm$6      & 51$\pm$1     & 525$\pm$14 & 6544$\pm$2    &  8$\pm$2    &  600$\pm$120          & 10$\pm$1 &  63$\pm$17  &  589$\pm$9   &  0.11$\pm$0.03    \\      
\#5 SALT 03/22  & 6560$\pm$1      & 47$\pm$1     & 297$\pm$16 & 6552$\pm$1    &  6$\pm$2    & 1420$\pm$390          & 16$\pm$2 &  64$\pm$19  &  361$\pm$5   &  0.18$\pm$0.06    \\      
\#6 SALT 04/22  & 6560$\pm$1      & 52$\pm$3     & 310$\pm$51 & 6555$\pm$1    & 12$\pm$2    &  760$\pm$100          & 12$\pm$1 & 186$\pm$39  &  496$\pm$7   &  0.38$\pm$0.10    \\      
\#7 SAAO 06/22  & 6575$\pm$2      & 53$\pm$2     & 793$\pm$21 & 6559$\pm$2    & 15*         & 1000*                 & 11$\pm$1 & 425$\pm$62  & 1218$\pm$22  &  0.35$\pm$0.05    \\      
\#8 SAAO 08/22  & 6576$\pm$1      & 50$\pm$1     & 756$\pm$48 & 6554$\pm$4    & 15*         & 1000*                 & 12$\pm$1 & 108$\pm$41  &  864$\pm$18  &  0.13$\pm$0.05    \\      
\#9 VLT 01/23   & 6578$\pm$1      & 71$\pm$1     & 467$\pm$18 & 6563$\pm$1    & 11$\pm$1    &  840$\pm$40           & 16$\pm$1 & 399$\pm$20  &  866$\pm$20  &  0.46$\pm$0.06    \\      
\#10 SAAO 03/23 & 6589$\pm$5      & 51$\pm$5     & 483$\pm$256& 6566$\pm$5    & 20$\pm$4    &  800$\pm$390          & 12$\pm$3 & 413$\pm$269 &  896$\pm$15  &  0.46$\pm$0.40    \\      
\#11 SAAO 04/23 & 6594$\pm$1      & 99$\pm$4     & 436$\pm$37 & 6566$\pm$1    & 22$\pm$1    &  790$\pm$120          & 12$\pm$1 & 825$\pm$139 & 1261$\pm$37  &  0.65$\pm$0.12    \\      
\#12 SALT 05/23 & 6577$\pm$6      & 49$\pm$2     & 475$\pm$10 & 6554$\pm$5    &  6$\pm$1    & 1130$\pm$80           & 15$\pm$1 & 159$\pm$15  &  634$\pm$13  &  0.25$\pm$0.03    \\      
\#13 VLT 06/23  & 6587$\pm$2      & 87$\pm$3     & 504$\pm$35 & 6566$\pm$5    &  8$\pm$1    & 1100$\pm$90           & 18$\pm$1 & 542$\pm$49  & 1046$\pm$22  &  0.52$\pm$0.05    \\      
\#14 SALT 07/23 & 6563$\pm$1      &100$\pm$22    & 327$\pm$14 & 6564$\pm$4    & 18$\pm$1    & 1000$\pm$80           & 14$\pm$1 & 564$\pm$36  &  892$\pm$13  &  0.63$\pm$0.07    \\      
\#15 VLT 07/23  & 6584$\pm$1      & 85$\pm$2     & 533$\pm$23 & 6566$\pm$4    & 11$\pm$1    & 1250$\pm$100          & 18$\pm$1 & 590$\pm$22  & 1123$\pm$16  &  0.53$\pm$0.03    \\      
\#16 SAAO 07/23 & 6590$\pm$2      & 53$\pm$1     & 486$\pm$37 & 6564$\pm$8    & 11$\pm$1    & 1250$\pm$80           & 16$\pm$1 & 434$\pm$69  &  920$\pm$53  &  0.47$\pm$0.08    \\      
\#17 VLT 01/24  & 6569$\pm$2      & 84$\pm$2     & 534$\pm$27 & 6566$\pm$1    & 10$\pm$1    & 1640$\pm$240          & 20$\pm$1 & 459$\pm$75  &  992$\pm$15  &  0.46$\pm$0.08    \\      
\#18 SALT 03/24 & 6576$\pm$2      & 50$\pm$2     & 346$\pm$34 & 6562$\pm$8    &  9$\pm$1    & 2000$^{+0\dagger}_{500}$  & 18$\pm$2 & 322$\pm$125 &  668$\pm$40  &  0.48$\pm$0.13  \\      
\#19 SALT 05/24 & 6579$\pm$1      & 43$\pm$6     & 482$\pm$23 & 6557$\pm$6    &  8$\pm$2    & 1390$\pm$180          & 16$\pm$1 & 243$\pm$33  &  725$\pm$19  &  0.34$\pm$0.05    \\      
\#20 SALT 07/24 & 6585$\pm$1      & 72$\pm$2     & 329$\pm$20 & 6566$\pm$5    & 17$\pm$1    & 1630$\pm$290          & 16$\pm$1 & 536$\pm$169 &  865$\pm$17  &  0.61$\pm$0.19    \\      
\#21 SALT 12/24 & 6567$\pm$3      & 74$\pm$2     & 403$\pm$20 & 6569$\pm$5    & 18$\pm$1    & 2000$^{+0\dagger}_{430}$  & 17$\pm$2 & 601$\pm$285 & 1004$\pm$18  &  0.60$\pm$0.13  \\  \hline \hline        
\end{tabular}\tablefoot{
An asterisk (*) indicates a fixed parameter.\\
A dagger ($\dagger$) indicates parameter uncertainty pegging at a limit. \\  
$^{a}$: Line flux units are $10^{-16}$ erg cm$^{-2}$ s$^{-1}$.\\
$^{b}$: $F_{\rm DL}$ is the fraction contribution of the flux of the diskline component to the total broad-line flux.
For the diskline component, $\lambda_{\rm o}$ is the rest-frame wavelength of emission, $\sigma_{\rm 0}$ quantifies the local broadening due to electron scattering
in a photoionized atmosphere, $R_{\rm in}$ is the inner radius,
and $i$ is the disk inclination, defined such that 0$^{\rm \circ}$ indicates a face-on disk.  The outer radius $R_{\rm out}$ was always held fixed at 5000 $R_{\rm g}$. \\  
$^{c}$: Absolute flux calibration was not available for spectra \#1 and \#2; values of absolute line flux should be considered as estimates only. \\
}
\end{sidewaystable*}

\renewcommand{\arraystretch}{1.40}

\begin{sidewaystable*}
\caption[]{Additional best-fitting model parameters for optical spectral fits} 
\centering
\label{tab:OTHEROPTFITRESULTS}
\begin{tabular}{l|llllll|lll}  \hline\hline
                &  \multicolumn{6}{|c|}{H$\beta$ region}                                                                                                   &     \multicolumn{3}{c}{H$\alpha$ region}    \\  
                &                   &                 & [\ion{O}{iii}]  &                          & Narrow &                                     &               & Narrow H$\alpha$ & \ion{He}{i} \\
 Date           & $\chi^2/dof$      & $F_{\rm 5100}^{a}$  & flux$^{b}$      & $R_{\rm H\beta/[\ion{O}{iii}]}$ & H$\beta$ flux$^{b}$       & $R_{\rm BL/NL}^d$                       &  $\chi^2/dof$ &  flux$^{b}$        &  flux$^{b}$  \\     \hline

\#1 H/ESO 03/93   &  219.26/182=1.21  & 1.00$\pm$0.35$^{c}$ & 143$\pm$12$^{c}$ & 1.8$\pm$0.2  & 20$\pm$4$^{c}$ & $\frac{2.80\pm0.11}{1.28\pm0.21}=2.19\pm0.23$ &   --               &  --        &  --  \\
\#2 6dF 03/02   & 1271.52/820=1.55  & 4.56$\pm$0.36$^{c}$ &  81$\pm$12$^{c}$ & 3.4$\pm$0.5  & 12$\pm$3$^{c}$ & $\frac{2.89\pm0.13}{1.54\pm0.36}=1.88\pm0.19$ &  853.44/867=0.98  &  48$\pm$3$^{c}$  &   88$\pm$9$^{c}$   \\
\#3 SALT 01/22  & 1788.67/1268=1.41 & 2.81$\pm$0.08      &  83$\pm$3       & 0.8$\pm$0.1  & 10$\pm$1       & $\frac{0.47\pm0.02}{1.31\pm0.07}=0.35\pm0.02$ &  2524.13/1349=1.87 &  42$\pm$1  &    4$\pm$1      \\
\#4 VLT 02/22   &  402.44/381=1.06  & 1.72$\pm$0.08      & 108$\pm$1       & 1.2$\pm$0.4  & 11$\pm$1       & $\frac{1.17\pm0.03}{0.89\pm0.06}=1.31\pm0.06$ &   629.20/448=1.40  &  85$\pm$2  &    4$\pm$1 \\
\#5 SALT 03/22  &  859.40/1277=0.67 & 3.97$\pm$0.14      &  90$\pm$2       & 1.9$\pm$0.1  &  8$\pm$1       & $\frac{1.31\pm0.04}{1.14\pm0.13}=1.15\pm0.07$ &  1393.76/1349=1.03 &  31$\pm$1  &   41$\pm$5 \\

\#6 SALT 04/22  &  997.10/1281=0.78 & 2.70$\pm$0.02      & 112$\pm$2       & 1.8$\pm$0.6  &  6$\pm$1       & $\frac{2.26\pm0.05}{0.77\pm0.14}=2.93\pm0.20$ &  1697.43/1354=1.25 &  26$\pm$1  &   46$\pm$3 \\
\#7 SAAO 06/22  &  739.38/1078=0.69 & 9.06$\pm$0.48      & 165$\pm$5       & 2.1$\pm$0.1  &  $<$5          & --$^{e}$                                       &   132.66/850=0.16  & 106$\pm$6  &   68$\pm$17 \\
\#8 SAAO 08/22  &  767.36/1073=0.72 & 6.96$\pm$0.43      & 149$\pm$9       & 1.8$\pm$0.2  & 14$\pm$4       & $\frac{2.26\pm0.05}{0.77\pm0.14}=2.93\pm0.20$ &   324.87/850=0.38  & 104$\pm$4  &   14$\pm$14 \\
\#9 VLT 01/23   &  543.02/368=1.48  & 2.27$\pm$0.21      &  98$\pm$2       & 3.1$\pm$0.3  & 11$\pm$2       & $\frac{1.51\pm0.31}{0.93\pm0.14}=1.62\pm0.56$ &   630.04/442=1.43  &  82$\pm$10 &   91$\pm$5 \\
\#10 SAAO 03/23 &  977.38/1074=0.91 & 7.72$\pm$0.34      & 131$\pm$7       & 2.5$\pm$0.2  & 17$\pm$4       & $\frac{2.90\pm0.11}{1.15\pm0.24}=2.52\pm0.27$ &   496.98/849=0.59  &  81$\pm$4  &   47$\pm$9\\

\#11 SAAO 04/23 & 1331.23/1083=1.24 & 6.78$\pm$0.23      & 113$\pm$2       & 2.9$\pm$0.1  & 12$\pm$2       & $\frac{3.81\pm0.07}{1.01\pm0.15}=3.77\pm0.27$ &   703.17/843=0.83  &  86$\pm$3  &  101$\pm$11 \\
\#12 SALT 05/23 & 2387.77/1278=1.87 & 4.32$\pm$0.15      &  80$\pm$5       & 3.5$\pm$0.3  & 10$\pm$1       & $\frac{1.83\pm0.13}{1.46\pm0.11}=1.25\pm0.13$ &  1319.22/1345=0.98 &  35$\pm$1  &   77$\pm$3 \\
\#13 VLT 06/23  &  521.97/367=1.42  & 5.27$\pm$0.23      &  95$\pm$2       & 4.3$\pm$0.3  & 17$\pm$2       & $\frac{2.69\pm0.20}{1.51\pm0.15}=1.78\pm0.25$ &   604.27/445=1.36  & 148$\pm$10 &  149$\pm$8 \\
\#14 SALT 07/23 & 2288.66/1280=1.79 & 4.36$\pm$0.14      &  79$\pm$7       & 3.8$\pm$0.4  & 10$\pm$1       & $\frac{2.17\pm0.11}{1.32\pm0.11}=1.64\pm0.15$ &  1319.33/1346=0.98 &  35$\pm$1  &   72$\pm$3   \\
\#15 VLT 07/23  &  464.60/373/1.27  & 4.51$\pm$0.22      &  97$\pm$8       & 2.5$\pm$0.3  & 10$\pm$1       & $\frac{2.73\pm0.12}{1.08\pm0.12}=2.53\pm0.32$ &   760.49/443=1.72  & 110$\pm$10 &  140$\pm$7   \\

\#16 SAAO 07/23 & 1372.40/1059=1.30 & 7.76$\pm$0.58      &  98$\pm$10      & 3.5$\pm$0.5  & 18$\pm$4       & $\frac{2.54\pm0.27}{1.52\pm0.32}=1.67\pm0.36$ &   580.44/843=0.69  & 123$\pm$14 &   81$\pm$9  \\
\#17 VLT 01/24  &  667.94/373=1.82  & 4.77$\pm$0.24      & 100$\pm$2       & 3.9$\pm$0.2  & 11$\pm$3       & $\frac{2.12\pm0.25}{0.85\pm0.19}=2.49\pm0.75$ &   725.32/443=1.64  & 108$\pm$13 &  118$\pm$5  \\
\#18 SALT 03/24 &  833.60/1287=0.65 & 5.07$\pm$0.34      &  77$\pm$5       & 4.4$\pm$0.8  & 12$\pm$2       & $\frac{3.09\pm0.89}{1.64\pm0.30}=1.88\pm1.07$ &  1034.59/1348=0.77 &  32$\pm$2  &   94$\pm$6   \\
\#19 SALT 05/24 & 1157.03/1280=0.90 & 4.75$\pm$0.29      &  83$\pm$8       & 4.6$\pm$0.5  &  9$\pm$1       & $\frac{2.17\pm0.21}{1.29\pm0.17}=1.68\pm0.30$ &  1385.65/1346=1.03 &  44$\pm$2  &  109$\pm$5 \\
\#20 SALT 07/24 &  837.42/1277=0.66 & 3.49$\pm$0.24      &  79$\pm$6       & 4.7$\pm$0.4  & 10$\pm$1       & $\frac{2.06\pm0.15}{1.47\pm0.15}=1.40\pm0.16$ &  1023.50/1347=0.76 &  47$\pm$1  &  102$\pm$4 \\
\#21 SALT 12/24 & 1231.28/1267=0.97 & 4.64$\pm$0.29      &  86$\pm$8       & 4.7$\pm$0.5  & 10$\pm$1       & $\frac{2.16\pm0.18}{1.44\pm0.16}=1.50\pm0.20$ &  1157.24/1341=0.86 &  45$\pm$1  &  100$\pm$4 \\  \hline\hline

\end{tabular}
\tablefoot{
$^{a}$: Flux densities are in units of $10^{-16}$ erg cm$^{-2}$ s$^{-1}$ \AA$^{-1}$ \\
$^{b}$: Integrated line fluxes are in units are $10^{-16}$ erg cm$^{-2}$ s$^{-1}$ \\
$^{c}$: Absolute flux calibration was not available for spectra \#1 and \#2;
  values of $F_{\rm 5100}$ and absolute line flux should be considered as estimates only.\\
$^{d}$: Ratio of broad H$\beta$ to narrow H$\beta$ line peak flux density  (both in units of $10^{-16}$ erg cm$^{-2}$ s$^{-1}$ \AA$^{-1}$), following \citet{Runco16}. \\
$^{e}$: Ratio not calculated due to narrow H$\beta$ line not being detected. \\
}
\end{sidewaystable*}

\subsection{Energy-resolved flux ratios}  \label{sec:APPDX_SED_RATIOS}

For the purpose of exploring broadband variability properties as the
source luminosity increases, we list in
Table~\ref{tab:SED_RATIO_TABLE} values of ratios of the best-fit SED
models, derived in Sect.~\ref{sec:OUVSED}.  We note that the soft
X-ray band, dominated by contributions from the warm corona, is always
the most variable band.

\end{appendix}

\end{document}